\documentclass[11pt,a4paper]{article}

\usepackage{jheppub}

\usepackage{amssymb, amsmath, amsthm}
\usepackage{mathrsfs}
\usepackage{physics}
\usepackage{marginnote}
\usepackage{bbold}

\newcommand{\barq}{\bar{q}} 
\newcommand{\barl}{\bar{\ell}}

\newcommand{\cohav}[1]{\langle {#1} \rangle}

\usepackage{slashed}
\usepackage{subcaption}
\usepackage{float}
\usepackage{hyperref}
\usepackage{tcolorbox}
\usepackage[]{todonotes}
\usepackage{graphicx}

\usepackage{xcolor}
\usepackage{soul}

\usepackage{ytableau}

\usepackage{tikz}
\usetikzlibrary{calc}
\usetikzlibrary{arrows,decorations.pathmorphing}
\usepackage{tikz-feynman}
\tikzfeynmanset{warn luatex=false}

\newcommand{\bq}{\bar{q}} 
\newcommand{\bl}{\bar{\ell}}
\newcommand{\ii}{i} 
\newcommand{\hdelta}{\hat{\delta}} 
 
 \newcommand{\commentout}[1]{}
 \renewcommand{\v}[1]{\boldsymbol{{#1}}}
 \renewcommand{\[}{\begin{equation}\begin{aligned}}
 \renewcommand{\]}{\end{aligned}\end{equation}}

\def\C{\mathbb{C}}
\renewcommand{\d}{\mathrm{d}}

\def\dd{\hat \d}
\def\del{\hat \delta}
\def\df{\d \Phi}
\def\wn{\bar}
\def\Lexp{\biggl\langle\!\!\!\biggl\langle}
\def\Rexp{\biggr\rangle\!\!\!\biggr\rangle}

\def\eqn{equation}
\def\radCol{R^a_\textrm{col}}

\def\newT{C}

\newbox\charbox
\newbox\slabox
\def\cut#1{{      
		\setbox\charbox=\hbox{$#1$}
		\setbox\slabox=\hbox{$|$}
		\dimen\charbox=\ht\slabox
		\advance\dimen\charbox by -\dp\slabox
		\advance\dimen\charbox by -\ht\charbox
		\advance\dimen\charbox by \dp\charbox
		\divide\dimen\charbox by 2
		\raise-\dimen\charbox\hbox to \wd\charbox{\hss$|$\hss}
		\llap{$#1$}
}}

\title{Classical Yang-Mills observables from amplitudes}

\author{Leonardo de la Cruz,}
\author{Ben Maybee,}
\author{Donal O'Connell,}
\author{Alasdair Ross}
\affiliation{Higgs Centre for Theoretical Physics, School of Physics and Astronomy, The University of Edinburgh, EH9 3FD, Scotland}

\emailAdd{leonardo.delacruz@ed.ac.uk}
\emailAdd{b.maybee@ed.ac.uk}
\emailAdd{donal@staffmail.ed.ac.uk}
\emailAdd{alasdair.ross@ed.ac.uk}

\abstract{The double copy suggests that the basis of the dynamics of general relativity is Yang-Mills theory. Motivated by the importance of the
relativistic two-body problem, we study the classical dynamics of colour-charged particle scattering from the perspective of amplitudes, 
rather than equations of motion. We explain how to compute the change of colour, and the radiation of colour, during a classical collision. 
We apply our formalism at next-to-leading order for the colour change and at leading order for colour radiation. 
}

\begin{document}

\maketitle
\newpage

\section{Introduction}
\label{sec:intro}

Future prospects in gravitational wave astronomy require perturbative calculations at very high precision~\cite{Babak:2017tow}. Remarkably, ideas and methods from quantum field theory offer a promising avenue of investigation~\cite{Donoghue:1993eb,Donoghue:1994dn,Donoghue:1996mt,Holstein:2004dn,Neill:2013wsa,Bjerrum-Bohr:2013bxa,Monteiro:2014cda,Luna:2016due,Damour:2016gwp,Goldberger:2016iau,Cachazo:2017jef,Laddha:2018rle,Cheung:2018wkq,Kosower:2018adc,Bern:2019nnu,Antonelli:2019ytb}. Powerful techniques have been developed to compute scattering amplitudes, and we have learned how to extract classical physics efficiently from amplitudes~\cite{Bjerrum-Bohr:2014zsa,Bjerrum-Bohr:2017dxw,Damour:2017zjx,Luna:2017dtq,Laddha:2018myi,Laddha:2018vbn,Bjerrum-Bohr:2018xdl,Sahoo:2018lxl,KoemansCollado:2019ggb,Brandhuber:2019qpg,Cristofoli:2019neg,Laddha:2019yaj,Bern:2019crd,Damgaard:2019lfh,Kalin:2019rwq,Bjerrum-Bohr:2019kec,Kalin:2019inp,Huber:2019ugz,Saha:2019tub,Aoude:2020onz,Bern:2020gjj,Cheung:2020gyp,Cristofoli:2020uzm,Bern:2020buy,Parra-Martinez:2020dzs,Haddad:2020tvs,AccettulliHuber:2020oou,Cheung:2020sdj,A:2020lub,Kalin:2020fhe,Haddad:2020que,Kalin:2020lmz,Sahoo:2020ryf,DiVecchia:2020ymx}. A crucial insight from quantum field theory is the double copy: the observation that scattering amplitudes in gravitational theories can be computed from amplitudes in gauge theories\footnote{The double copy was thoroughly reviewed recently~\cite{Bern:2019prr}}. Since perturbation theory is far simpler in Yang-Mills (YM) theory than in standard approaches to gravity, the double copy has revolutionary potential. 

Yang-Mills theory, treated as a classical field theory, shares many of the important physical features of gravity, including non-linearity and a subtle gauge structure. In this respect the YM case has always served as an excellent toy model for gravitational dynamics. But our developing understanding
of the double copy has taught us that the connection between Yang-Mills theory and gravity is deeper than this. Detailed aspects of the perturbative dynamics of gravity, including gravitational radiation, can be deduced from Yang-Mills theory and the double copy~\cite{Goldberger:2016iau,Luna:2016hge,Luna:2017dtq,Goldberger:2017vcg}. In fact the double copy extends beyond perturbation
theory, leading to exact maps~\cite{Monteiro:2014cda,Luna:2015paa,Lee:2018gxc,Luna:2018dpt,Adamo:2018mpq,CarrilloGonzalez:2019gof,Cho:2019ype,Carrillo-Gonzalez:2019aao,Moynihan:2019bor,Bah:2019sda,Huang:2019cja,Alawadhi:2019urr,Borsten:2019prq,Kim:2019jwm,Banerjee:2019saj,Bahjat-Abbas:2020cyb,Moynihan:2020gxj,Adamo:2020syc,Alfonsi:2020lub,Luna:2020adi,Keeler:2020rcv,Elor:2020nqe,Cristofoli:2020hnk,Alawadhi:2020jrv,Casali:2020vuy,Adamo:2020qru,Easson:2020esh,Chacon:2020fmr} between classical solutions of gauge theory and gravity, even when there is gravitational radiation present~\cite{Luna:2016due}. 

So we are motivated to take another look at perturbative YM processes, particularly those that connect to gravitational wave physics.
However, the double copy is best understood as a relationship between the scattering amplitudes of
YM theory and gravity. Consequently it is useful to formulate classical YM dynamics in terms of amplitudes, rather than by solving the
equations of motion. We explain how to do so in this article.
Since our interest is in the classical theory, we systematically ignore non-perturbative quantum effects. Nevertheless
we identify observables which are well-defined in the classical theory and which can be extracted unambiguously from perturbative scattering amplitudes to any order. Our approach is complementary to an effort to understand 
the double copy from a purely classical, worldline perspective~\cite{Goldberger:2016iau,Goldberger:2017frp, Goldberger:2017vcg,Goldberger:2017ogt,
Chester:2017vcz, Shen:2018ebu,Plefka:2018dpa,Plefka:2019hmz,PV:2019uuv,Almeida:2020mrg}. Perhaps the simplest way of implementing the double copy for classical quantities will
be to compute observables using classical worldline methods; extract the relevant amplitudes by comparing to our formulae for observables in terms of amplitudes, and then double copy to gravity.

Returning briefly to our underlying interest in gravitational dynamics, it is worth emphasising the importance of spin. The spin of the individual bodies in a 
compact binary coalescence event influences the details of the outgoing gravitational radiation. This spin also contains information on the poorly-understood formation channels of
the binaries. Measurement of spin is therefore one of the primary physics outputs of gravitational wave observations. It is fortunate, then, that spin can be incorporated rather 
naturally in the formalism of quantum field 
theory~\cite{Vaidya:2014kza,Guevara:2017csg,Guevara:2018wpp,Chung:2018kqs,Bautista:2019tdr,Maybee:2019jus,Guevara:2019fsj,Arkani-Hamed:2019ymq,Bautista:2019evw, delaCruz:2016wbr, Chung:2019duq,Chung:2019yfs,Chung:2020rrz,Levi:2020uwu,Bern:2020buy,Cristofoli:2020hnk,Aoude:2020mlg}, even for very large (classical) spins. However the details of spin dynamics quickly becomes complex.

In the Yang-Mills context we can also discuss the dynamics of spin. However Yang-Mills theories always involve colour, and colour is in many respects very analogous to spin. The dynamics of colour is actually a little simpler than spin, because the latter is tied together with spacetime while colour is defined in its own vector space. We therefore pay particular attention to the dynamics of colour in Yang-Mills theory. 
We discuss in detail the change of colour of a particle during a scattering event, and the radiation of colour to null infinity. We also discuss more briefly the impulse (change of momentum) and radiation of momentum in YM theory.
The methods we use build on ideas previously described by Kosower and two of us in a previous paper~\cite{Kosower:2018adc} (KMOC). 

The classical equations of central interest in our paper describe the Yang-Mills field with some gauge group coupled to several classical point-like particles. These particles each carry colour charges $c^a$ which are time-dependent vectors in the adjoint representation of the gauge group. Often known as Wong's equations \cite{Wong:1970fu}, the equations of motion for $N$ of these particles following worldlines $x_\alpha(\tau_\alpha)$ and with velocities $v_\alpha$ and momenta $p_\alpha$ are
\begin{subequations}
\label{eq:classicalWong}
	\begin{gather}
	\frac{\d p_\alpha^\mu }{\d \tau_\alpha} = g\, c^a_\alpha(\tau_\alpha)\, F^{a\,\mu\nu}\!(x_\alpha(\tau_\alpha))\,  v_{\alpha\, \nu}(\tau_\alpha)\,,  \label{Wong-momentum} 
	\\
	\frac{\d c^a_\alpha}{\d \tau_\alpha}= g f^{abc} v^\mu_\alpha(\tau_\alpha) A_\mu^b(x_\alpha(\tau_\alpha))\,c^c_\alpha(\tau_\alpha)\,,
	\label{Wong-color}
	\\
	D^\mu F_{\mu\nu}^a(x) = J^a_\nu(x) = g \sum\limits_{\alpha= 1}^N \int\!\d\tau_\alpha\,  c^a_\alpha(\tau_\alpha)  v^\mu(\tau_\alpha)\, \delta^{(4)}(x-x(\tau_\alpha))\,, \label{YangMillsEOM}
	\end{gather}
\end{subequations}
where the Yang-Mills field is $A_\mu = A_\mu^a T^a$ and $F^a_{\mu\nu}$ is the associated field strength tensor. 

Although our motivation lies in gravitational dynamics, classical YM theory is also important in other contexts. Its asymptotic symmetry group has recently
received intense study both for its own intrinsic interest but also as a toy model for the asymptotic symmetry group of gravity~\cite{Strominger:2013lka,Pate:2017vwa,Strominger:2017zoo,Campoleoni:2017qot,He:2019pll,Gonzo:2019fai,Campoleoni:2019ptc}. The radiation of colour charge of interest to us is also important in that context.

A totally different application is to the quark-gluon plasma, a high-temperature phase of matter which is described by the classical equations of Yang-Mills theory. Indeed it is 
in this context that the classical theory has its main application, as a tool to  model transport phenomena in non-Abelian plasmas in the very
high temperature regime \cite{Heinz:1983nx,Heinz:1984my,Heinz:1984yq,Heinz:1985qe, Elze:1989un, Litim:2001db}. The equations of motion also provide a successful approximation for calculating the gluon distribution functions from deeply inelastic, ultrarelativistic ion collisions \cite{McLerran:1993ni,McLerran:1993ka,Iancu:2003xm}, and radiation from classical particles has recently been calculated in this regime \cite{Kajantie:2019hft,Kajantie:2019nse}.

Our paper is organised as follows. In section~\ref{sec:setup} we review the theoretical treatment of colour in Yang-Mills theory, emphasising the importance of coherent states in the classical limit. We will see that point-like particles emerging from an underlying matter field in a representation $R$ of the gauge group have an adjoint-valued colour charge which naturally has dimensions of angular momentum. 
In section~\ref{sec:impulse} we will set up expressions for the classical and momentum impulse when scattering states have colour, and provide an explicit computation of these observables at NLO. In section~\ref{sec:radiation} we turn to radiation, constructing the total radiated colour charge in terms of amplitudes and explicitly recovering the classical LO radiation current derived in \cite{Goldberger:2016iau}. In section~\ref{sec:ClassEOM} we reproduce our earlier results using purely classical methods. Finally, we discuss our results in section~\ref{sec:conclusions}. 

\section{Review of the theory of colour}
\label{sec:setup}
In this section we review the emergence of non-Abelian colour charges $c^a_\alpha$ --- a vector in the \emph{adjoint} representation for each particle species $\alpha$ --- from a quantum field theory with scalars $\varphi_\alpha$ in \emph{any} representation $R_\alpha$ of the gauge group, coupled to the Yang-Mills field. We specialise to the case of two different scalar fields, so that $\alpha = 1, 2$. Our action is
\begin{equation}
S = \int\!\d^4x\, \left(\sum_{\alpha=1}^2\left[ (D_\mu \varphi_\alpha^\dagger) D^\mu \varphi_\alpha - \frac{m_\alpha^2}{\hbar^2} \varphi_\alpha^\dagger \varphi_\alpha\right] - \frac14 F^a_{\mu\nu} F^{a\,\mu\nu}\right), \label{eqn:scalarAction}
\end{equation}
where  $D_\mu = \partial_\mu + i g A_\mu^a T^a_R$. We have restored factors of $\hbar$ for convenience below. The generator matrices (in a representation $R$) are $T^a_R = (T_R^a)_i{ }^j$ and satisfy the Lie algebra 
$[T_R^a, T_R^b]_i{ }^j = if^{abc} (T_R^c)_i{ }^j$. 
We take the metric signature to be mostly negative.

\subsection{Colour in classical field theory}\label{sec:ftcolour}
For simplicity we begin with the case of just a single massive scalar. At the classical level, the colour charge can be obtained from the Noether current $j^a_\mu$ associated with the global part of the gauge symmetry. 
The colour charge is explicitly given by
\begin{equation}
\int\!\d^3x\, j^a_0(t,\v{x}) = i\!\int\! \d^3x\, \Big(\varphi^\dagger T^a_R\, \partial_0 \varphi - (\partial_0 \varphi^\dagger) T^a_R\, \varphi\Big)\,.  \label{eqn:colourNoetherCharge}
\end{equation}
Notice that a direct application of the Noether procedure has led to a colour charge with dimensions of action, or equivalently, of angular momentum.

It's worth dwelling on dimensional analysis in the context of the Wong equations~\eqref{eq:classicalWong} since they motivate us
to make certain choices which may, at first, seem surprising. The Yang-Mills field strength
\begin{equation}
F^a_{\mu\nu}  =\partial_\mu A^a_\nu - \partial_\nu A^a_\mu - gf^{abc} A^b_\mu A^c_\nu\,\label{eqn:fieldStrength}
\end{equation}
is obviously an important actor in these classical equations. Classical equations should contain no factors\footnote{An equivalent point of view is that any factors of $\hbar$ appearing in an equation which has classical meaning should be absorbed into 
parameters of the classical theory.} of $\hbar$, so we choose to maintain this precise expression for the field strength
when $\hbar \neq 1$. By inspection it follows\footnote{We use the notation $[x]$ for the dimensions of the quantity $x$. The symbols $L$ and $M$ stand for dimension of length and mass, respectively.} that 
$[g A^{a}_\mu] = L^{-1}$. We can develop this further; since the action of \eqn~\eqref{eqn:scalarAction} has dimensions of angular momentum, the Yang-Mills field strength must have dimensions of 
$\sqrt{M/L^3}$. Thus, from \eqn~\eqref{eqn:fieldStrength},
\begin{equation}
[A^a_\mu] = \sqrt{\frac{M}{L}}\,, \qquad [g] = \frac1{\sqrt{ML}}\,.\label{eqn:YMdims}
\end{equation}
This conclusion about the dimensions of $g$ is in contrast to the situation in electrodynamics, where $[e] = \sqrt{ML}$. Put another way, in electrodynamics the dimensionless fine structure constant is $e^2 / 4\pi \hbar$ while in our conventions
the analogue is $\hbar g^2 / 4\pi\,$! It is possible to arrange matters such that the YM and EM cases are more similar, but we find the present conventions to be convenient in perturbative calculations.

Continuing with our discussion of dimensions, note that the Yang-Mills version of the Lorentz force, \eqn~\eqref{Wong-momentum}, demonstrates that the quantity $g c^a$ must have the same dimension as the electric charge.
This is consistent with our observation above that the colour has dimensions of angular momentum.

At first our assignment of dimensions of $g$ may seem troubling from the perspective of extracting classical physics from scattering amplitudes following the algorithm described by Kosower and some of the authors (KMOC)~\cite{Kosower:2018adc}.
The fact that $g$ has dimensions of $1/\sqrt{ML}$ implies that the dimensionless coupling at each vertex is $g \sqrt \hbar$, so factors of $\hbar$ associated with the coupling appear with the opposite power to the case of electrodynamics (and gravity).
However, because
the colour charges are dimensionful the net power of $\hbar$ turns out to be the same, and thus the KMOC approach is ultimately unaltered. We will also see below that the dimensionful nature of the colour clarifies the classical limit of this aspect of the theory.
To see how this works we must quantise. 

\subsection{Colour of single-particle quantum states}\label{sec:spqcolour}

Dimensional analysis demonstrates that $\varphi$ has dimensions of $\sqrt{M/L}$, so its mode expansion is
\begin{equation}
{\varphi}_i(x) = \frac1{\sqrt{\hbar}} \int\!\df(p)\, \left(a_i(p) e^{-ip\cdot x/\hbar} + b_i^\dagger(p) e^{ip\cdot x/\hbar}\right).
\end{equation}
The index $i$ labels the representation $R$. We have normalised the ladder operators by requiring
\begin{equation}
[a_i(p), a^{\dagger j}(q)] =  2E_p\, (2\pi)^3 \delta^{(3)}(\v{p} - \v{q})\, \delta_{i}{}^{j} \equiv \del^3_\Phi(p - q)\, \delta_{i}{ }^{j} \,,\label{eqn:ladderCommutator}
\end{equation}
and likewise for the antiparticle operators. Each ladder operator therefore has dimensions of $M^{-1}$.
We also write the Lorentz-invariant phase space measure as
\begin{equation}
\df(k)=  \frac{\d^4k}{(2\pi)^4} \, 2\pi\,\Theta(k^0)\, \delta(k^2 - m^2) \equiv  \dd^4k\, \del^{(+)}(k^2 - m^2) \,,
\end{equation}
and introduce the hat notation in 
the measure and the delta functions 
\begin{equation}
\dd^np \equiv \frac{\d^n p}{(2\pi)^n}\,,  \qquad   \del^{(n)}(x)\equiv (2\pi)^n \delta(x)\,,
\end{equation}
in order to avoid a proliferation of factors of $2\pi$.

After quantisation, the colour charge of \eqn~\eqref{eqn:colourNoetherCharge} becomes a Hilbert space operator which will be important below. To emphasise that this is an 
operator we write the quantised form as $\C^a$:
\begin{equation}
\begin{aligned}
\C^a &= i\!\int\! \d^3x\, \Big(\varphi^\dagger T^a_R\, \partial_0 \varphi - (\partial_0 \varphi^\dagger) T^a_R\, \varphi\Big) \\
&= \hbar\int\!\df(p)\, \left( a^\dagger(p) \,T^a_R \, a(p) + b^\dagger(p)\, T^a_{\bar R} \, b(p)\right),\label{eqn:colourOp}
\end{aligned}
\end{equation}
using the generators of the conjugate representation $\bar R$, $T^a_{\bar R} = - T^a_R$. The overall $\hbar$ factor guarantees that the colour has dimensions of angular momentum, as we require. It is important to note that these global colour operators inherit the usual Lie algebra of the generators, modified by factors of $\hbar$, so that
\begin{equation}
[\C^a, \C^b] = i\hbar f^{abc} \C^c\,.\label{eqn:chargeLieAlgebra}
\end{equation}

Now we turn to the action of the colour operator on the single-particle states
\begin{equation}
|p^i\rangle = 
a^{\dagger i}(p)|0\rangle\,.
\end{equation}
Acting with the colour charge operator of \eqn~\eqref{eqn:colourOp} we immediately see that
\begin{equation}
\C^a|p^i\rangle = \hbar\, (T^a_R)_j{ }^i|p^j\rangle\,, \qquad \langle p_i|\C^a = \hbar\, \langle p_j|(T^a_R)_i{ }^j\,.
\end{equation}
Thus inner products yield generators scaled by $\hbar$:
\begin{equation}
\langle p_i|\C^a|p^j\rangle \equiv (\newT^a)_i{ }^j = \hbar\, (T^a_R)_i{ }^j\,.
\end{equation}
The $(C^a)_i{ }^j$ are simply rescalings of the usual generators $T^a_R$ by a factor of $\hbar$, and thus satisfy the rescaled Lie algebra in \eqn~\eqref{eqn:chargeLieAlgebra}; since this rescaling is important for us, it is useful to make
the distinction between the two. 

We may then write a generic single particle state as
\begin{equation}
|\psi\rangle = \sum_i \int\! d\Phi(p)\, \phi(p) \chi_i\, |p^i\rangle\,,
\end{equation}
where the vector $\chi_i$ labels a general colour state, and the normalisations are chosen such that
\begin{align}
\int\!\df(p) |\phi(p)|^2 = 1\,,\qquad \sum_i \chi^{i*}\chi_i = 1\,.
\end{align}
The colour operator acts on these states as
\begin{equation}
\C^a|\psi\rangle = \int\!\df(p)\, (\newT^a)_i{ }^j\, \phi(p) \chi_j|p^i\rangle\,.
\end{equation}
Furthermore, we define the colour charge of the particle as
\begin{equation}
\langle \psi |\C^a| \psi \rangle =  \chi^{i*} (\newT^a)_{i}{ }^{j}\, \chi_j\,. \label{eqn:colourCharge}
\end{equation}
Computing this charge and extracting its classical limit is the topic of the next 
section~\ref{sec:classicalSingleParticleColour}.

As a final remark on these rescaled generators, let us write out the covariant derivative in the representation $R$. In terms of $\newT^a$, 
the $\hbar$ scaling of interactions is precisely the same as in QED (and in perturbative gravity):
\begin{equation}
D_\mu = \partial_\mu + i \, g A^a_\mu T^a = \partial_\mu + \frac{ig}{\hbar}\, A^a_\mu \newT^a \,;\label{eqn:covDerivative}
\end{equation}
for comparison, the covariant derivative used by KMOC in QED was $\partial_\mu + i e A_\mu/ \hbar$~\cite{Kosower:2018adc}. Thus we have arranged that
factors of $\hbar$ appear in the same place in YM theory as in electrodynamics, provided that
the colour is measured by $C^a$. This ensures that the basic rules for obtaining the classical limits of amplitudes will be the same as in 
KMOC~\cite{Kosower:2018adc}. In practical calculations one can thus restore $\hbar$'s in colour factors and work using
$C^a$'s everywhere. However, it is worth emphasising that unlike classical colour charges, the factors $C^a$ do not commute.

\subsection{Colour of point-like particles in the classical regime}
\label{sec:classicalSingleParticleColour}
The classical point-particle picture emerges from sharply peaked quantum wavepackets. In \cite{Kosower:2018adc}, linear exponential generalisations of Gaussian wavepackets were chosen for relativistic momentum space wavepackets. The essential property that must be satisfied was for the particle to have a sharply-defined position \emph{and} a sharply-defined momentum whenever the classical limit was taken. To understand colour, governed by the Yang-Mills-Wong equations in the classical arena, a similar picture should emerge for our quantum colour operator in \eqn~\eqref{eqn:colourOp}. Following the KMOC 
philosophy~\cite{Kosower:2018adc} we will first consider the full quantum states and then the classical limit. We define the classical limit of the colour charge to be
\begin{equation}
c^a \equiv \langle \psi |\C^a| \psi \rangle\,.
\end{equation}

Our focus in this section is on the colour structure of our particle. The full state is a tensor product of colour and kinematics: 
\begin{align}
\ket{\psi}= \sum \ket{\psi_{\text{colour}}} \otimes  \ket{\psi_{\text{kin}}},
\end{align}
but as we ignore kinematics for now, we simply write the colour part of the state $|\psi_{\text{colour}}\rangle \rightarrow |\psi\rangle$ in the remainder of this section. Then, in the classical limit, the critical requirements on the colour part of the state are that
\[
\langle \psi |\C^a | \psi \rangle &= \textrm{finite} \,, \\
\langle \psi |\C^a \C^b| \psi \rangle &= \langle \psi |\C^a | \psi \rangle\langle \psi | \C^b| \psi \rangle + \textrm{negligible} \,.
\]
Since the colour operator explicitly involves a factor of $\hbar$, another parameter 
must be large so that the colour expectation $\langle \psi |\C^a | \psi \rangle$ is much bigger than $\hbar$ in the classical region. This situation is basically the same as for the usual classical limit of angular momentum: in that case we take the spin quantum number $j$ large so that $\hbar j$ is a classical angular momentum. For colour, we similarly need the size of the representation $R$ to be large. (We will see this explicitly in the case of 
$SU(3)$ in a moment.) For the second requirement we must select appropriate colour wavefunctions $\ket{\psi}$ which are analogous to the kinematic wavepacket 
in KMOC~\cite{Kosower:2018adc}.

Coherent states are the key to the classical limit very generally~\cite{Yaffe:1981vf}, including the case of angular momentum, so we choose a coherent state to describe
the colour of our particle. The states used in KMOC to describe momenta \cite{Kosower:2018adc} can themselves be  understood as coherent states for a ``first-quantised''
particle --- more specifically they are states for the restricted Poincar\'e group \cite{Kaiser:1977ys, TwarequeAli:1988tvp, Kowalski:2018xsw}. However, not all definitions of coherent 
states are equivalent, so we need to specify in what sense our states are coherent. The definition we use was introduced by Perelomov \cite{perelomov:1972}, which 
formalises the notion of coherent state for any Lie group and hence can be utilised for both the kinematic and the colour parts. It is in this sense that the states used by KMOC are coherent for the Poincar\'e group.   We 
refer the interested reader to \cite{Perelomov:1986tf} for details of the Perelomov formalism and to  \cite{Combescure2012} for applications.

For the explicit construction of the appropriate colour states we will use the Schwinger boson formalism.  For $SU(2)$, constructing irreducible representations from Schwinger bosons is a standard textbook exercise \cite{Sakurai:2011zz}. One simply introduces the Schwinger bosons --- that is, creation $a^{\dagger i}$ and annihilation $a_i$ operators, transforming in the fundamental two-dimensional representation so that $i = 1,2$. The irreducible representations of $SU(2)$ are all symmetrised tensor powers of the fundamental, so the state
\[
a^{\dagger i_1} a^{\dagger i_2} \cdots a^{\dagger i_{2j}} \ket{0} ,
\]
which is automatically symmetric in all its indices,
transforms in the spin $j$ representation. 

For groups larger than $SU(2)$, the situation is a little more complicated because the construction of a general irreducible representation requires both symmetrisation and 
antisymmetrisation over appropriate sets of indices. This leads to expressions which are involved already for $SU(3)$ \cite{Mathur:2000sv,Mathur:2002mx}. We content 
ourselves with a  brief discussion of the $SU(3)$ case which captures all of the interesting features of the general case.

One can construct all irreducible representations from tensor products only of fundamentals \cite{Mathur:2010wc,Mathur:2010ey}; 
however, for our treatment of $SU(3)$ it is helpful to instead make use of the fundamental and antifundamental, and tensor these together to generate 
representations.  Following \cite{Mathur:2000sv}, we introduce two sets of ladder operators $a_i$ and $b^i$ , $i=1, 2, 3$, which transform in the $\mathbf{3}$ and $\mathbf{3}^*$ respectively.  The colour operator can then be written as
\begin{equation}
\C^e= \hbar \left( a^\dagger \frac{\lambda^e}{2} a -
b^\dagger \frac{\bar{\lambda}^e}{2} b  \right), \quad e=1, \dots, 8\,, \label{charge-SU3}
\end{equation}
where $\lambda^e$ are the Gell-Mann matrices and $\bar\lambda^e$ are their conjugates. The operators $a$ and $b$ satisfy the usual commutation relations
\begin{align}
 [a_i, a^{\dagger j}]= \delta_{i}{ }^{j}\,, \quad   [b^i, b^{\dagger}_j]= \delta^{i}{ }_{j}\,, \quad 
 [a_i, b^j]= 0\,, \quad  [a^{\dagger i}, b^{\dagger }_j]= 0\,.
\end{align}
By virtue of these commutators, the colour operator \eqref{charge-SU3} obeys the commutation 
relation \eqref{eqn:chargeLieAlgebra}.

There are two Casimir operators given by the number operators\footnote{Here we define $a^{\dagger}  \cdot a \equiv \sum_{i=1}^3 a^{\dagger i} a_i$ and
$|\xi|^2\equiv \sum_{i=1}^3 |\xi_i|^2$.}
\begin{equation}
 \mathcal{N}_1\equiv a^\dagger \cdot a\,, \qquad  \mathcal{N}_2\equiv b^\dagger \cdot b\,,
\end{equation}
with eigenvalues $n_1$ and $n_2$ respectively, so we label irreducible representations by $[n_1, n_2]$. 
Na\"ively, the states we are looking for are constructed by acting on the vacuum state as follows:
\begin{align}
\left(a^{\dagger i_1} \cdots  a^{\dagger i_{n_1}} \right)
\left(b_{j_1}^{\dagger} \cdots  b_{j_{n_2}}^{\dagger} \right)
\ket{0}.\label{states-reducible}
\end{align}
However, these states are $SU(3)$ reducible and thus cannot be used in our construction of coherent states. We write the irreducible states schematically by acting with a Young projector $\mathcal{P}$ which appropriately (anti-) symmetrises upper and lower indices, thereby subtracting traces:
\begin{align}
\ket{\psi}_{[n_1, n_2]} \equiv \mathcal{P} \left( \left(a^{\dagger i_1} \cdots  a^{\dagger i_{n_1}} \right)
\left(b^{\dagger}_{j_1} \cdots  b_{j_{n_2}}^{\dagger} \right)\ket{0} \right).
\label{eq:YPstate}
\end{align}

In general these operations will lead to involved expressions for the states, but we can understand them from their associated Young tableaux (Fig.~\ref{SU3-YT}). Each double box column represents an operator $b_i^{\dagger}$ and each single column box represents the operator $a^{\dagger i}$, and thus for a mixed representation we have $n_2$ double columns and $n_1$ single columns.

\begin{figure}
	\centering 
	\begin{ytableau}
		j_1 & j_2  & \dots  &j_{n_2} & i_1 & i_2 & \cdots & i_{n_1}  \cr    &  &  &  
	\end{ytableau}
	\caption{Young tableau of $SU(3)$}
	\label{SU3-YT}
\end{figure}

Having constructed the irreducible states, one can define a coherent state parametrised by two triplets of complex numbers $\xi_i$ and $ \zeta^i$, $i=1,2, 3$.  These are normalised according to
\begin{equation}
|\xi|^2 = |\zeta|^2 = 1\,, \qquad \xi \cdot \zeta=0\,.
\end{equation}

We won't require fully general coherent states, but instead their projections onto the $[n_1,n_2]$ representation, which are
\begin{equation}
\ket{\xi \,\zeta}_{[n_1,n_2]}\equiv  \frac{1}{\sqrt{(n_1! n_2!)}} \left( \zeta \cdot b^\dagger\right)^{n_2} \left(\xi \cdot a^\dagger \right)^{n_1} \ket{0}.
\label{restricted-coherent}
\end{equation}
The square roots ensure that the states are normalised to unity\footnote{Note that the Young projector in equation~\eqref{eq:YPstate} is no longer necessary since the constraint $\xi \cdot \zeta = 0$ removes all the unwanted traces.}. With this normalisation we can write the identity operator as  
\begin{equation}
\mathbb{1}_{[n_1,n_2]} = \int \d \mu(\xi,\zeta) \Big(\ket{\xi \,\zeta}\bra{\xi \,\zeta}\Big)_{[n_1,n_2]},
\end{equation}
where $\int \d \mu(\xi,\zeta)$ is the $SU(3)$ Haar measure, normalised such that $\int \d \mu(\xi ,\zeta)=1$. Its precise form is irrelevant for our purposes.

With the states in hand, we can return to the expectation value of the colour operator $\C^a$ in \eqn~\eqref{eqn:colourOp}. The size of the representation, that is $n_1$ and $n_2$, must be large compared to $\hbar$ in the classical regime so that the final result is finite. To see this let us compute this expectation value explicitly.
By definition we have 
\begin{equation}
\langle \xi \,\zeta|\mathbb{C}^e|\xi \,\zeta \rangle_{[n_1,n_2]} =  \frac{\hbar}{2} \left(\langle \xi \,\zeta|a^\dagger \lambda^e a|\xi \,\zeta \rangle_{[n_1,n_2]} -
\langle \xi \,\zeta|b^\dagger \bar{\lambda}^e b|\xi \,\zeta\rangle_{[n_1,n_2]} \right).
\end{equation}
After a little algebra we find that 
\begin{equation}
\langle \xi \,\zeta|\C^e|\xi \,\zeta \rangle = \frac{\hbar}{2} \left( n_1   \xi^{*} \lambda^e \xi - n_2    \zeta^*\bar \lambda^e \zeta\right).
\end{equation}
We see that a finite charge requires a scaling limit in which we take $n_1$, $n_2$ large  as $\hbar \to 0$, keeping the product $\hbar n_i$  fixed for at least one value of $i$. The classical charge is therefore the finite c-number
\begin{equation}
c^a =  \langle \xi \,\zeta|\C^a|\xi \,\zeta \rangle_{[n_1, n_2]}  = \frac{\hbar}{2} \left( n_1   \xi^{*}\lambda^a \xi - n_2 \zeta^{*} \bar\lambda^a \zeta\right). \label{clas-charge-SU3}
\end{equation}

The other feature we must check is the expectation value of products. A similar calculation  for two pairs  of charge operators in a large representation leads to  the important property 
\[
\langle \xi\,\zeta|\C^a \C^b|\xi\,\zeta \rangle _{[n_1,n_2]} &= \langle \xi\,\zeta|\C^a  |\xi\,\zeta\rangle _{[n_1,n_2]}  \langle \xi\,\zeta| \C^b|\xi\,\zeta 
\rangle _{[n_1,n_2]} + \mathcal O (\hbar) \\
&= c^a c^b + \mathcal O(\hbar) \,. \label{factorization-charges}
\]
This is in fact a special case of a more general construction discussed in detail by Yaffe~\cite{Yaffe:1981vf}.
In appendix~\ref{SU3-coherent-example} we prove \eqn~\eqref{factorization-charges}, and show that the correction term is $\mathcal O (\hbar)$. The same argument can be used to demonstrate an important property of the coherent states in the classical limit, which is that the overlap $\langle \chi' | \chi \rangle$ is very strongly peaked about $\chi = \chi'$~\cite{Yaffe:1981vf}. We have thus constructed explicit colour states which ensure the correct classical behaviour of the colour charges. 

In the calculation of the colour impulse and radiated colour in sections \ref{sec:impulse} and \ref{sec:radiation}, we will only need to make use of the finiteness and factorisation properties, so we will avoid further use of the explicit form of the states. Henceforth we write $\chi_i$ for the parameters of a general colour state $\ket{\chi}$ with these properties, and $\d \mu(\chi)$ for the Haar measure of the colour group, whatever it may be.

\subsection{Colour of several particles}

Now that we have reviewed the theory of colour for a single particle, it's time to consider what happens when more than one particle is present. We will shortly discuss the dynamics of colour in detail; here, we set up initial states describing more than one point-like particle. 

We take our particles to be distinguishable, so they are associated with distinct quantum fields $\varphi_\alpha$ with $\alpha = 1, 2$. We only consider two different particles explicitly, though it is no more difficult to consider the many-particle case. We also restrict to scalar fields, again for simplicity. The action is therefore as given in \eqn~\eqref{eqn:scalarAction}. Both fields $\varphi_\alpha$ must be in representations $R_\alpha$ which are large, so that a classical limit is available for the individual colours.

At some initial time in the far past, we assume that our two particles both have well-defined positions, momenta and colours. In other words, particle $\alpha$ has a wavepacket $\phi_\alpha(p_\alpha)$ describing its momentum-space distribution, and a colour wavepacket $\chi_\alpha$ as described in section~\ref{sec:classicalSingleParticleColour}. In this initial state, both particles are separated by an impact parameter $b$ (which must then be very large compared to the spatial spread of the momenta in the wavepackets~\cite{Kosower:2018adc}). We further assume that there is no incoming radiation (that is, no vector boson) in the incoming state. Thus, the state is
\[
|\Psi\rangle &= \int\!\df(p_1)\df(p_2)\, \phi_1(p_1) \phi_2(p_2)\, e^{ib\cdot p_1/\hbar}\, |{p_1}\, \chi_1 ; \, {p_2}\, \chi_2 \rangle \\
&= \int\!\df(p_1)\df(p_2) \, \phi_1(p_1) \phi_2(p_2)\, e^{ib\cdot p_1/\hbar}\, \chi_{1i}\, \chi_{2j} |{p_1}^i ; \, {p_2}^j \rangle\,.\label{eqn:inState}
\]
Notice that the state $\ket{\Psi}$ refers to a multi-particle state. We reserve the notation $\ket{\psi}$ for single particle states.

We measure the colour of multi-particle states by acting with a colour operator which is simply the sum of the individual colour operators~\eqref{eqn:colourOp} for each of the scalar fields. For example, acting on the state $|{p_1}\, \chi_1 ; \, {p_2}\, \chi_2 \rangle$ we have
\[
\C^a |{p_1}\chi_1 ;& {p_2} \chi_2 \rangle = |{p_1}^{i'} \, {p_2}^{j'} \rangle \, \left( (C^a_{1})_{i'}{}^i \delta_{j'}{}^j + \delta_{i'}{}^i (C^a_{2})_{j'}{}^j \right) \chi_{1i}\, \chi_{2j} \\
&= \int \! d\mu(\chi'_1) d\mu(\chi'_2) \, \ket{{p_1} \, \chi'_1 ; \, {p_2}\, \chi'_2} \,\langle \chi'_1\, \chi'_2| C^a_{1} \otimes 1 + 1 \otimes C^a_{2} |\chi_1\, \chi_2 \rangle\\
&= \int \! d\mu(\chi'_1) d\mu(\chi'_2) \, \ket{{p_1} \, \chi'_1 ; \, {p_2}\, \chi'_2} \,\langle\chi'_1\, \chi'_2| C^a_{1+2}  |\chi_1\, \chi_2\rangle\,,\label{eqn:charge2particleAction}
\]
where $C^a_\alpha$ is the colour in representation $R_\alpha$ and we have written $C^a_{1+2}$ for the colour operator on the tensor product of representations $R_1$ and $R_2$.
In the classical regime, using the property that the overlap between states sets $\chi'_i=\chi_i $ in the classical limit, it follows that
\[
\bra{p_1\,\chi_1; \, p_2 \,\chi_2} C^a_{1+2} \ket{p_1\, \chi_1 ; \, p_2\, \chi_2 } = c_1^a + c_2^a \,,
\]
so the colours simply add.

\section{Impulse}\label{sec:impulse}

Now we move on to the dynamics of colour. Our focus in this section will be on the colour impulse --- that is, the total change in colour during a scattering event --- leaving radiation of colour to the next section. We begin by setting up the colour impulse observable in the vein of \cite{Kosower:2018adc,Maybee:2019jus} before turning to explicit examples at LO and NLO. 

\subsection{Colourful scattering observables in quantum field theory}
\label{sec:colourfulobservables}
\newcommand{\colStructure}{\mathcal{C}}

A natural observable in Yang-Mills theory is the total change in the colour charge of one of the massive scattering particles,
\[
\langle\Delta c_1^a\rangle &= \langle \Psi |S^\dagger \C^a S|\Psi \rangle - \langle \Psi|\C^a |\Psi \rangle\\ &= i \langle \Psi|[\C_1^a, T]| \Psi \rangle + \langle \Psi |T^\dagger [\C^a_1, T]|\Psi\rangle\,,\label{eqn:ColourImpulse}
\]
where we have introduced the $S$ and $T$ matrices, related by $S = 1 + iT$, and utilised the optical theorem. We call this observable the \emph{colour impulse}, as it mirrors
the structure of the momentum impulse $\Delta p_1^\mu$ of \cite{Kosower:2018adc} and angular impulse $\Delta s_1^\mu$ of \cite{Maybee:2019jus}. An immediate novelty for
this impulse is that it is a Lorentz scalar, instead transforming in the adjoint representation of the gauge group.

Substituting the 2-particle wavepackets in \eqn~\eqref{eqn:inState} yields
\begin{multline}
\cohav{\Delta c_1^a} = \prod_{i=1,2}\int\!\df(p_i) \df(p'_i)\, \phi_i(p_i) \phi^*_i(p_i') e^{ib\cdot(p_1 - p_1')/\hbar}\\ \times \langle p_1'\,\chi_1;\,p_2'\, \chi_2| i[\C^a, T] + T^\dagger[\C^a,T]| p_1\,\chi_1;\,p_2\,\chi_2\rangle\,,
\end{multline}
which we expand in terms of amplitudes by inserting complete sets of states,
\begin{equation}
\mathbb{1} = \sum_X \int\! \df(r_1) \df(r_2)\, \d\mu(\zeta_1) \d\mu(\zeta_2)|{r_1}\,\zeta_1;\,r_2\,\zeta_2;\,X \rangle \langle {r_1}\,\zeta_1;\,r_2\,\zeta_2;\,X|\,.\label{eqn:completeStates}
\end{equation}
The set $X$ could contain any number of extra gluon or scalar states, whose phase space measures and sums over any other quantum numbers are left implicit in the summation over $X$.

It is frequently convenient to write amplitudes in Yang-Mills theory in colour-ordered form; for example, see~\cite{Ochirov:2019mtf} for an application to amplitudes with multiple different external particles.
The full amplitude $\mathcal{A}$ is decomposed
onto a basis of colour factors times partial amplitudes $A$. The colour factors are associated with some set of Feynman topologies. Once a basis of independent colour structures is chosen, the corresponding partial amplitudes must be gauge invariant. Thus,
\begin{align}
\langle p_1'\, \chi_1;\,p_2'\, \chi_2|T| p_1\,\chi_1;\,p_2\,\chi_2\rangle & = \langle \chi_1\, \chi_2|\mathcal{A}(p_1,p_2 \rightarrow p_1',p_2')|\chi_1\,\chi_2\rangle\, \del^{(4)}(p_1 + p_2 - p_1' - p_2')\nonumber\\ 
= \sum_D& \expval{\colStructure(D)}\, A_D(p_1,p_2 \rightarrow p_1',p_2')\, \del^{(4)}(p_1 + p_2 - p_1' - p_2')\,,\label{eqn:colourStripping}
\end{align}
where $\colStructure(D)$ is the colour factor of diagram $D$ and $A_D$ is the associated partial amplitude. The remaining expectation value is over the colour states $\chi_i$.
Using this notation and the action of the colour operator 
in \eqn~\eqref{eqn:charge2particleAction}, we can write the colour impulse as
\[
\cohav{\Delta c_1^{a}} =& \prod_{i=1,2} \sum_D\int\!\df(p_i) \df(p'_i)\, \phi_i(p_i) \phi^*_i(p_i') e^{ib\cdot(p_1 - p_1')/\hbar}\\ &\times \bigg[i \expval{ [\newT^a, \colStructure(D)]} A_D(p_1,p_2 \rightarrow p_1',p_2')\, \del^{(4)}(p_1 + p_2 - p_1' - p_2')\\
& + \sum_X \int\!\df(r_i)\, \expval{\colStructure(D')^\dagger [\newT^a, \colStructure(D)]} A^*_{D'}(p'_1,p'_2 \rightarrow r_1,r_2) \\  \times A_D&(p_1,p_2 \rightarrow r_1',r_2',r_X)\, \del^{(4)}(p_1 + p_2 - p_1' - p_2') \del^{(4)}(p_1 + p_2 - r_1 - r_2 - r_X) \bigg]\,.
\]
Finally, let us introduce the momentum mismatch $q_i = p'_i - p_i$, and transfer $w_i = r_i - p_i$. After integrating over the delta functions, we arrive at
\[
\langle \Delta c_1^{a} \rangle =&\, i\! \int\! \df(p_1) \df(p_2) \dd^4q\, \del(2p_1\cdot q + q^2)\del(2p_2\cdot q - q^2) \\ 
&\times \Theta(p_1^0 + q^0) \Theta(p_2^0 -q^0) \phi_1(p_1) \phi_2(p_2) \phi^*_1(p_1+q)\phi^*_2(p_2-q) e^{-ib\cdot q/\hbar} \\
& \times \bigg\{ \sum_D \expval{[\newT_1^a,\colStructure(D)]} A_D(p_1,p_2  \rightarrow p_1+q,p_2-q) \\
& -i \prod_{i=1,2} \sum_X \int\! \dd^4 w_i\, \del(2p_1\cdot w_i + w_i^2) \del^{(4)}(w_1 + w_2 - r_X) \Theta(p_i^0 + w_i^0)\\
& \times \sum_{D,D'}  \expval{\colStructure(D')^\dagger [\newT_1^a,\colStructure(D)]} A_D(p_1,p_2  \rightarrow p_1 + w_1,p_2 + w_2, r_X)\\
& \qquad\qquad \times A_{D'}^*(p_1+q,p_2-q  \rightarrow p_1+w_1,p_2+ w_2, r_X) \bigg \}\,.\label{eqn:ColourImpulse_1}
\]
It is interesting to note that factors of expectation values of colour commutators, such as $\expval{[\newT_1^a,\colStructure(D)]}$, in the colour impulse play a similar role to that of the momentum mismatch $q^\mu$ in the momentum impulse $\Delta p_1^\mu$~ \cite{Kosower:2018adc}.

The momentum impulse in QED and gravity was discussed in detail in~\cite{Kosower:2018adc}. In Yang-Mills theory, the presence of colour leads to slight modifications of those KMOC expressions. The basic difference is the colour structure of the amplitude. The observable itself is built from the (colour singlet) momentum operator $\mathbb{P}^\mu_1$, so factors of $\newT_1^a$ appearing in the colour impulse, \eqn~\eqref{eqn:ColourImpulse_1}, do not arise in the momentum case. We proceed by writing the full amplitude as a sum over colour structures, finding
\[
\langle \Delta p_1^{\mu}\rangle =&\, i \!\int\! \df(p_1)\df(p_2) \dd^4q\, \del(2p_1\cdot q + q^2)\del(2p_2\cdot q - q^2) \\  &\times \Theta(p_1^0 + q^0) \Theta(p_2^0 -q^0) \phi(p_1)\phi(p_2) \phi^*(p_1+q)\phi^*(p_2-q) e^{-ib\cdot q} \\
&\times\bigg\{ \sum_Dq^{\mu}  \expval{\colStructure(D)} A_D(p_1,p_2 \rightarrow p_1+q,p_2-q) \\
& -i \prod_{i=1,2} \sum_X \int\! \dd^4 w_i\, \del(2p_1\cdot w_i + w_i^2) \del^{(4)}(w_1 + w_2 - r_X) \Theta(p_i^0 + w_i^0)\\
&\times \sum_{D,D'}  \expval{\colStructure(D')^\dagger \colStructure(D)} A_D(p_1,p_2  \rightarrow p_1 + w_1,p_2 + w_2, r_X)\\
& \qquad\qquad \times A_{D'}^*(p_1+q,p_2-q  \rightarrow p_1+w_1,p_2+ w_2, r_X) \bigg \}\,.\label{eqn:MomentumImpulse_1}
\]

By construction, both impulse observables are well defined in the classical regime. Once wavefunctions of the types described in section~\ref{sec:setup} are used, the details of the wavefunctions will not be important. However, to extract expressions which are valid in the classical approximation, it is important to be aware that the commutators of the $\newT^a$ contain powers of $\hbar$. In particular one must take care to expand all commutators of colour factors.

All other powers of $\hbar$ appear as described by KMOC. In brief, the rescaled covariant derivative of \eqn~\eqref{eqn:covDerivative} ensures that each factor of the coupling $g$ is accompanied by a factor $\hbar^{-1/2}$; all massless external and loop momenta are products of a factor of $\hbar$ and a wavenumber; care must be taken with squares of massless momenta $q^2$ in delta functions. Finally, small shifts of order $\hbar \wn q$ to the dominant momenta of order $m$ in wavefunctions can be neglected; this is an example of a general property of coherent states in the classical limit \cite{Yaffe:1981vf}. We therefore introduce the notation
\begin{equation}
\Lexp f(p_1,p_2,\cdots) \Rexp = \int\! \df(p_1) \df(p_2) |\phi(p_1)|^2 |\phi(p_2)|^2\, \langle \chi_1\,\chi_2|f(p_1,p_2,\cdots)|\chi_1\,\chi_2\rangle\,.\label{eqn:angleBrackets}
\end{equation}
The nature of the wavepackets make evaluating these expectation values very easy in the classical limit: the momentum phase space integrals are simply evaluated by replacing massive momenta with 4-velocities, $p_i \to m_i u_i$ \cite{Kosower:2018adc}, while the  colour expectation value is guaranteed by \eqn~\eqref{factorization-charges} to behave as a product of commuting classical colour charges. We will still use single angle brackets to indicate expectation values which are only over the colour states.
\newcommand{\colKnl}{\mathcal{G}}
\newcommand{\momKnl}{\mathcal{I}}

Following this procedure, the colour impulse becomes
\begin{equation}
\langle \Delta c_1^{a} \rangle \rightarrow \Delta c_1^{a} = i \, \Lexp \int\! \dd^4\wn q\, \del(2p_1\cdot \wn q) \del(2p_2\cdot\wn q) e^{-ib\cdot\wn q}\, \colKnl^{a} \Rexp\,,\label{eqn:COLimpkernel}
\end{equation}
where we define the colour kernel $\colKnl^a$ to be
\[
\colKnl^a &= \hbar^2 \sum_D [\newT_1^a,\colStructure(D)] A_D (p_1,p_2 \rightarrow p_1 + q, p_2 - q)
\\ -  &i {\hbar^4}\sum_X \prod_{i=1,2}\int\! \dd^4\wn w_i\, \del(2p_1\cdot \wn w_i + \hbar \wn w_i^2)\, \del^{(4)}(\wn w_1 + \wn w_2 - \wn r_X) \sum_{D,D'} \colStructure(D')^\dagger [\newT^a, \colStructure(D)] \\ \times&  A^*_{D'} (p_1+ q , p_2 - q \rightarrow p_1 + w_1 , p_2+ w_2, r_X) A_D(p_1,p_2 \rightarrow p_1+ w_1,p_2 - w_2, r_X)\,.
\]
We designed this kernel so that it is of order $\hbar^0$ in the classical approximation. Clearly at LO only the first term, linear in the amplitude, contributes; the second integral contributes from NLO, where it is an integral over tree level diagrams, while the first term involves one-loop amplitudes. 

In the same notation, the momentum impulse is
\begin{equation}
\langle \Delta p_1^\mu \rangle \rightarrow \Delta p_1^\mu = i\, \Lexp  \int\! \dd^4\wn q\, \del(2p_1\cdot \wn q) \del(2p_2\cdot\wn q) e^{-ib\cdot\wn q}\, \momKnl^{\mu} \Rexp\,,\label{eqn:impkernel}
\end{equation}
with momentum kernel
\[
\label{impkernel-Mom}
\momKnl^\mu &= \hbar^3 \wn q^\mu\, \sum_D \colStructure(D) A_D(p_1,p_2 \rightarrow p_1+ q,p_2 -q)\\
&-  i{\hbar^5} \sum_X \prod_{i=1,2} \int\! \dd^4\wn w_i\, \del(2p_i\cdot \wn w_i + \hbar \wn w_i^2)\, \del^{(4)}(\wn w_1 + \wn w_2 - \wn r_X)\, \wn w_1^\mu \sum_{D,D'} \colStructure(D')^\dagger\, \colStructure(D) \\ &\times A^*_{D'} (p_1+ q , p_2 - q \rightarrow p_1 + w_1, p_2+ w_2, r_X) A_D(p_1,p_2 \rightarrow p_1+ w_1,p_2 + w_2, r_X)
\]
when the scattering particles carry colour.

\subsection{Leading order}
\newcommand{\tree}{\begin{tikzpicture}[thick, baseline={([yshift=-\the\dimexpr\fontdimen22\textfont2\relax] current bounding box.center)}, decoration={markings,mark=at position 0.6 with {\arrow{Stealth}}}]
	\begin{feynman}
	\vertex (v1);
	\vertex [right = 0.3 of v1] (v2);
	\vertex [above left = .275 and .275 of v1] (o1);
	\vertex [above right = .275 and .275 of v2] (o2);
	\vertex [below left = .275 and .275 = of v1] (i1);
	\vertex [below right = .275 and .275 of v2] (i2);
	\draw (i1) -- (v1);
	\draw (v1) -- (o1);
	\draw (i2) -- (v2);
	\draw (v2) -- (o2);
	\draw (v1) -- (v2);
	\end{feynman}
	\end{tikzpicture}}

We may now compute the colour impulse explicitly. We begin at leading order (LO) for the scalar YM theory defined by \eqn~\eqref{eqn:scalarAction}, moving to next-to-leading order (NLO) in the next subsection. We will strip coupling constants from amplitudes, writing $\bar{A}^{(n)}_D$ for the charge stripped partial amplitudes at $\mathcal{O}(g^{2n+2})$. At LO the colour kernel is
\begin{equation}
\colKnl^{a,(0)} = \hbar g^2 \sum_D [\newT_1^a, \mathcal{C}(D)]\, \bar{A}_D^{(0)}(p_1,p_2 \rightarrow p_1+\hbar \wn q, p_2- \hbar \wn q)\,.
\end{equation}
Here only the $t$-channel tree topology contributes, so the sum between colour and kinematics is trivial; we simply have\footnote{We adopt the convention that time runs vertically in Feynman diagrams.}
\begin{equation}\label{eq:treeamp}
\bar{A}_{\scalebox{0.5}{\tree}} =  \frac{4 p_1\cdot p_2 +\hbar \barq^2}{\hbar^2 \barq^2}\,, \qquad \colStructure\left(\tree\right) = \newT_1\cdot\newT_2\,,
\end{equation}
and therefore the colour impulse factor is
\begin{equation}
\big[C_1^a,\colStructure\left({\scalebox{0.75}{\tree}}\right)\big] = [\newT_1^a,\newT_1^b]\newT_2^b = i\hbar f^{abc}\newT_1^c \newT_2^b \,.
\end{equation}
Inserting these expressions into the colour kernel, all factors of $\hbar$ cancel as expected for a classical observable. The classical limit is
\begin{equation}
\begin{aligned}
\Delta c_1^{a,(0)} &= -g^2 \Lexp  f^{abc}\newT_1^c \newT_2^b \int\! \dd^4 \barq \,\del (p_1\cdot \barq) \del (p_2\cdot \barq) e^{-ib\cdot \barq}  \frac{p_1\cdot p_2}{ \barq^2}\Rexp
\\&= g^2 f^{abc}c_1^b c_2^c\, u_1\cdot u_2\! \int\! \dd  ^4 \barq\, \del (u_1\cdot \barq)\del (u_2\cdot \barq) \frac{e^{-ib \cdot \barq} }{\barq^2}\,.
\end{aligned}
\end{equation}
Notice that while evaluating the large double angle brackets we obtained classical colour charges as expectations values of the $\newT_\alpha$. 

The remaining integral is straightforward but divergent. 
While we use dimensional regulation throughout the remainder of the paper to define divergent integrals, in this case it is convenient to take a different
approach. 

The logarithmic divergence in the colour may seem surprising at first. However, the spacetime position of the particle is also logarithmically divergent
in four dimensions; this is simply the familiar divergence due to the long-range nature of $1/r^2$ forces in four dimensions. We therefore 
introduce a cutoff regulator $L$ of dimensions length as follows. Consider the following quantity
\begin{align} 
2b^2 \frac{\partial \Delta c_1^{a,(0)} }{\partial b^2}= b_\mu \frac{\partial \Delta c_1^{a,(0)}}{\partial b_\mu}  =- i g^2  f^{abc}c_1^b c_2^c  \gamma b_\mu 
 \int\! \dd  ^4 \barq\, \del (u_1\cdot \barq)\del(u_2\cdot \barq) \frac{e^{-ib \cdot \barq} \barq^\mu }{\barq^2}\,,
\end{align}
where $\gamma=u_1\cdot u_2$.  The integral on the RHS was evaluated in~\cite{Kosower:2018adc}. Using that result, it is easy to show that the solution of the differential
equation is
\begin{align}
  \Delta c_1^{a,(0)}= \frac{ \gamma g^2 f^{abc}c_1^b c_2^c }{4\pi \sqrt{\gamma^2-1}} \log \left( \frac{b^2}{L^2} \right),
\end{align}
where we have included the regulator explicitly.

\subsection{Next to leading order}
\label{NLOColor}
At NLO the classical colour kernel, with $\hbar$'s from couplings removed, is 
\[
\colKnl^{a,(1)} &= g^4 \sum_\Gamma [\newT_1^a, \mathcal{C}(\Gamma)] \bar{A}_\Gamma^{(1)}(p_1,p_2 \rightarrow p_1+\hbar\wn q,p_2- \hbar\wn q)
 \\ 
&\,\,-  i g^4 \hbar^2\! \int\! \dd ^4 \barl\, \del (2p_1\cdot \barl + \hbar\barl^2)\del (2p_2\cdot \barl - \hbar\barl^2)\, \mathcal{C}\left({\scalebox{0.75}{\tree}}\right)^\dagger[\newT^a_1, \mathcal{C}\left({\scalebox{0.75}{\tree}}\right)]
\\
&\times \bar{A}_{\scalebox{0.35}{\tree}}^*(p_1+\hbar\barq , p_2 -\hbar\barq \rightarrow p_1 + \hbar\barl , p_2  -\hbar\barl) \bar{A}_{\scalebox{0.35}{\tree}}(p_1,p_2 \rightarrow p_1+\hbar\barl,p_2-\hbar\barl)\,, \label{eqn:NLOcolKnl}
\]
where $\Gamma$ is a set of one-loop topologies which span the independent colour factors. By the analysis of \cite{Kosower:2018adc}, the topologies
relevant in the classical regime are\footnote{See appendix~\ref{app:expressions} for the evaluation of these diagrams.}
\begin{equation}
\begin{tikzpicture}[baseline={([yshift=-\the\dimexpr\fontdimen22\textfont2\relax] current bounding box.center)}, decoration={markings,mark=at position 0.6 with {\arrow{Stealth}}}]
\begin{feynman}
\vertex (v1);
\vertex [right = 0.97 of v1] (v2);
\vertex [above = 0.9 of v1] (v3);
\vertex [right = 0.97 of v3] (v4);
\vertex [above left = 0.5 and 0.5 of v3] (o1);
\vertex [below left = 0.5 and 0.5 of v1] (i1);
\vertex [above right = 0.5 and 0.5 of v4] (o2);
\vertex [below right = 0.5 and 0.5 of v2] (i2);
\draw [postaction={decorate}] (i1) -- (v1);
\draw [postaction={decorate}] (v1) -- (v3);
\draw [postaction={decorate}] (v3) -- (o1);
\draw [postaction={decorate}] (i2) -- (v2);
\draw [postaction={decorate}] (v2) -- (v4);
\draw [postaction={decorate}] (v4) -- (o2);
\diagram*{(v2) -- [gluon] (v1)};
\diagram*{(v3) -- [gluon] (v4)};
\end{feynman}	
\end{tikzpicture}\quad \begin{tikzpicture}[baseline={([yshift=-\the\dimexpr\fontdimen22\textfont2\relax] current bounding box.center)}, decoration={markings,mark=at position 0.6 with {\arrow{Stealth}}}]
\begin{feynman}
\vertex (v1);
\vertex [right = 0.9 of v1] (v2);
\vertex [above = 0.9 of v1] (v3);
\vertex [right = 0.9 of v3] (v4);
\vertex [above left = 0.5 and 0.5 of v3] (o1);
\vertex [below left = 0.5 and 0.5 of v1] (i1);
\vertex [above right = 0.5 and 0.5 of v4] (o2);
\vertex [below right = 0.5 and 0.5 of v2] (i2);
\vertex [above right = 0.4 and 0.4 of v1] (g1);
\vertex [below left = 0.4 and 0.4 of v4] (g2);
\draw [postaction={decorate}] (i1) -- (v1);
\draw [postaction={decorate}] (v1) -- (v3);
\draw [postaction={decorate}] (v3) -- (o1);
\draw [postaction={decorate}] (i2) -- (v2);
\draw [postaction={decorate}] (v2) -- (v4);
\draw [postaction={decorate}] (v4) -- (o2);
\diagram*{(v4) -- [gluon] (g2)};
\diagram*{(g1) -- [gluon] (v1)};
\diagram*{(v2) -- [gluon] (v3)};
\end{feynman}
\end{tikzpicture} \quad \begin{tikzpicture}[baseline={([yshift=-\the\dimexpr\fontdimen22\textfont2\relax] current bounding box.center)}, decoration={markings,mark=at position 0.6 with {\arrow{Stealth}}}]
\begin{feynman}
\vertex (v1);
\vertex [below left = 0.45 and 0.9 of v1] (v2);
\vertex [above left = 0.45 and 0.9 of v1] (v3);
\vertex [above right = 0.95 and 0.5 of v1] (o1);
\vertex [below right = 0.95 and 0.5 of v1] (i1);
\vertex [above left = 0.5 and 0.4 of v3] (o2);
\vertex [below left = 0.5 and 0.4 of v2] (i2);
\draw [postaction={decorate}] (i1) -- (v1);
\draw [postaction={decorate}] (v1) -- (o1);
\draw [postaction={decorate}] (i2) -- (v2);
\draw [postaction={decorate}] (v2) -- (v3);
\draw [postaction={decorate}] (v3) -- (o2);
\diagram*{(v1) -- [gluon] (v2)};
\diagram*{(v1) -- [gluon] (v3)};
\end{feynman}	
\end{tikzpicture} \quad \begin{tikzpicture}[baseline={([yshift=-\the\dimexpr\fontdimen22\textfont2\relax] current bounding box.center)}, decoration={markings,mark=at position 0.6 with {\arrow{Stealth}}}]
\begin{feynman}
\vertex (v1);
\vertex [below right = 0.45 and 0.9 of v1] (v2);
\vertex [above right = 0.45 and 0.9 of v1] (v3);
\vertex [above left = 0.95 and 0.5 of v1] (o1);
\vertex [below left = 0.95 and 0.5 of v1] (i1);
\vertex [above right = 0.5 and 0.4 of v3] (o2);
\vertex [below right = 0.5 and 0.4 of v2] (i2);
\draw [postaction={decorate}] (i1) -- (v1);
\draw [postaction={decorate}] (v1) -- (o1);
\draw [postaction={decorate}] (i2) -- (v2);
\draw [postaction={decorate}] (v2) -- (v3);
\draw [postaction={decorate}] (v3) -- (o2);
\diagram*{(v2) -- [gluon] (v1)};
\diagram*{(v3) -- [gluon] (v1)};
\end{feynman}	
\end{tikzpicture} \quad \begin{tikzpicture}[baseline={([yshift=-\the\dimexpr\fontdimen22\textfont2\relax] current bounding box.center)}, decoration={markings,mark=at position 0.6 with {\arrow{Stealth}}}]
\begin{feynman}
\vertex (v1);
\vertex [right = 0.55 of v1] (g1);
\vertex [below right = 0.45 and 0.55 of g1] (v2);
\vertex [above right = 0.45 and 0.55 of g1] (v3);
\vertex [above left = 0.95 and 0.5 of v1] (o1);
\vertex [below left = 0.95 and 0.5 of v1] (i1);
\vertex [above right = 0.5 and 0.4 of v3] (o2);
\vertex [below right = 0.5 and 0.4 of v2] (i2);
\draw [postaction={decorate}] (i1) -- (v1);
\draw [postaction={decorate}] (v1) -- (o1);
\draw [postaction={decorate}] (i2) -- (v2);
\draw [postaction={decorate}] (v2) -- (v3);
\draw [postaction={decorate}] (v3) -- (o2);
\diagram*{(v1) -- [gluon,] (g1)};
\diagram*{(g1) -- [gluon] (v2)};
\diagram*{(v3) -- [gluon] (g1)};
\filldraw [color=black] (g1) circle [radius=0.5pt];
\end{feynman}	
\end{tikzpicture} \quad	\begin{tikzpicture}[baseline={([yshift=-\the\dimexpr\fontdimen22\textfont2\relax] current bounding box.center)}, decoration={markings,mark=at position 0.6 with {\arrow{Stealth}}}]
\begin{feynman}
\vertex (v1);
\vertex [left = 0.55 of v1] (g1);
\vertex [below left = 0.45 and 0.55 of g1] (v2);
\vertex [above left = 0.45 and 0.55 of g1] (v3);
\vertex [above right = 0.95 and 0.5 of v1] (o1);
\vertex [below right = 0.95 and 0.5 of v1] (i1);
\vertex [above left = 0.5 and 0.4 of v3] (o2);
\vertex [below left = 0.5 and 0.4 of v2] (i2);
\draw [postaction={decorate}] (i1) -- (v1);
\draw [postaction={decorate}] (v1) -- (o1);
\draw [postaction={decorate}] (i2) -- (v2);
\draw [postaction={decorate}] (v2) -- (v3);
\draw [postaction={decorate}] (v3) -- (o2);
\diagram*{(v1) -- [gluon] (g1)};
\diagram*{(v2) -- [gluon] (g1)};
\diagram*{(g1) -- [gluon] (v3)};
\filldraw [color=black] (g1) circle [radius=0.5pt];
\end{feynman}	
\end{tikzpicture}.
\end{equation}
We will refer to these as the box $B$, cross box $C$, triangles $T_{ij}$ and non-Abelian diagrams $Y_{ij}$ respectively. The latter, involving the 3-gluon interaction vertex, are new to the Yang-Mills calculation. Of course these are not the only diagrams we must calculate; there is also the product of trees, which we will view as a cut box $\cut{B}$, in the non-linear part of the colour kernel. Note that we now have two distinct colour structures to calculate, one for the loops and one for the cut box. We will investigate in detail how these structures affect the cancellation of classically singular terms, but first let us work with the 1-loop, linear piece.

\subsubsection{1-loop amplitude}
\newcommand{\boxy}{\begin{tikzpicture}[thick, baseline={([yshift=-\the\dimexpr\fontdimen22\textfont2\relax] current bounding box.center)}, decoration={markings,mark=at position 0.6 with {\arrow{Stealth}}}]
	\begin{feynman}
	\vertex (v1);
	\vertex [right = 0.25 of v1] (v2);
	\vertex [above = 0.25 of v1] (v3);
	\vertex [right = 0.25 of v3] (v4);
	\vertex [above left = 0.15 and 0.15 of v3] (o1);
	\vertex [below left = 0.15 and 0.15 of v1] (i1);
	\vertex [above right = 0.15 and 0.15 of v4] (o2);
	\vertex [below right = 0.15 and 0.15 of v2] (i2);
	\draw (i1) -- (v1);
	\draw (v1) -- (v3);
	\draw (v3) -- (o1);
	\draw (i2) -- (v2);
	\draw (v2) -- (v4);
	\draw (v4) -- (o2);
	\draw (v1) -- (v2);
	\draw (v3) -- (v4);
	\end{feynman}	
	\end{tikzpicture}}
\newcommand{\crossbox}{\begin{tikzpicture}[thick, baseline={([yshift=-\the\dimexpr\fontdimen22\textfont2\relax] current bounding box.center)}, decoration={markings,mark=at position 0.6 with {\arrow{Stealth}}}]
	\begin{feynman}
	\vertex (v1);
	\vertex [right = 0.3 of v1] (v2);
	\vertex [above = 0.3 of v1] (v3);
	\vertex [right = 0.3 of v3] (v4);
	\vertex [above left = 0.125 and 0.125 of v3] (o1);
	\vertex [below left = 0.125 and 0.125 of v1] (i1);
	\vertex [above right = 0.125 and 0.125 of v4] (o2);
	\vertex [below right = 0.125 and 0.125 of v2] (i2);
	\vertex [above right = 0.1 and 0.1 of v1] (g1);
	\vertex [below left = 0.1 and 0.1 of v4] (g2);
	\draw (i1) -- (v1);
	\draw (v1) -- (v3);
	\draw (v3) -- (o1);
	\draw (i2) -- (v2);
	\draw (v2) -- (v4);
	\draw (v4) -- (o2);
	\draw (v4) -- (g2);
	\draw (g1) -- (v1);
	\draw (v2) -- (v3);
	\end{feynman}
	\end{tikzpicture}}
\newcommand{\triR}{\begin{tikzpicture}[thick, baseline={([yshift=-\the\dimexpr\fontdimen22\textfont2\relax] current bounding box.center)}, decoration={markings,mark=at position 0.6 with {\arrow{Stealth}}}]
	\begin{feynman}
	\vertex (v1);
	\vertex [below left = 0.17 and 0.25 of v1] (v2);
	\vertex [above left = 0.17 and 0.25 of v1] (v3);
	\vertex [above right = 0.275 and 0.2 of v1] (o1);
	\vertex [below right = 0.275 and 0.2 of v1] (i1);
	\vertex [above left = 0.15 and 0.1 of v3] (o2);
	\vertex [below left = 0.15 and 0.1 of v2] (i2);
	\draw (i1) -- (v1);
	\draw (v1) -- (o1);
	\draw (i2) -- (v2);
	\draw (v2) -- (v3);
	\draw (v3) -- (o2);
	\draw (v2) -- (v1);
	\draw (v3) -- (v1);
	\end{feynman}
	\end{tikzpicture}}
\newcommand{\triL}{\begin{tikzpicture}[thick, baseline={([yshift=-\the\dimexpr\fontdimen22\textfont2\relax] current bounding box.center)}, decoration={markings,mark=at position 0.6 with {\arrow{Stealth}}}]
	\begin{feynman}
	\vertex (v1);
	\vertex [below right = 0.17 and 0.25 of v1] (v2);
	\vertex [above right = 0.17 and 0.25 of v1] (v3);
	\vertex [above left = 0.275 and 0.2 of v1] (o1);
	\vertex [below left = 0.275 and 0.2 of v1] (i1);
	\vertex [above right = 0.15 and 0.1 of v3] (o2);
	\vertex [below right = 0.15 and 0.1 of v2] (i2);
	\draw (i1) -- (v1);
	\draw (v1) -- (o1);
	\draw (i2) -- (v2);
	\draw (v2) -- (v3);
	\draw (v3) -- (o2);
	\draw (v2) -- (v1);
	\draw (v3) -- (v1);
	\end{feynman}	
	\end{tikzpicture}}
\newcommand{\nonAbL}{\begin{tikzpicture}[thick, baseline={([yshift=-\the\dimexpr\fontdimen22\textfont2\relax] current bounding box.center)}, decoration={markings,mark=at position 0.6 with {\arrow{Stealth}}}]
	\begin{feynman}
	\vertex (v1);
	\vertex [right = 0.2 of v1] (g1);
	\vertex [below right = 0.175 and 0.25 of g1] (v2);
	\vertex [above right = 0.175 and 0.25 of g1] (v3);
	\vertex [above left = 0.275 and 0.2 of v1] (o1);
	\vertex [below left = 0.275 and 0.2 of v1] (i1);
	\vertex [above right = 0.15 and 0.1 of v3] (o2);
	\vertex [below right = 0.15 and 0.1 of v2] (i2);
	\draw (i1) -- (v1);
	\draw (v1) -- (o1);
	\draw (i2) -- (v2);
	\draw (v2) -- (v3);
	\draw (v3) -- (o2);
	\draw (v1) -- (g1);
	\draw (v2) -- (g1);
	\draw (v3) -- (g1);
	\end{feynman}	
	\end{tikzpicture}}
\newcommand{\nonAbR}{\begin{tikzpicture}[thick, baseline={([yshift=-\the\dimexpr\fontdimen22\textfont2\relax] current bounding box.center)}, decoration={markings,mark=at position 0.6 with {\arrow{Stealth}}}]
	\begin{feynman}
	\vertex (v1);
	\vertex [left = 0.2 of v1] (g1);
	\vertex [below left = 0.175 and 0.25 of g1] (v2);
	\vertex [above left = 0.175 and 0.25 of g1] (v3);
	\vertex [above right = 0.275 and 0.2 of v1] (o1);
	\vertex [below right = 0.275 and 0.2 of v1] (i1);
	\vertex [above left = 0.15 and 0.1 of v3] (o2);
	\vertex [below left = 0.15 and 0.1 of v2] (i2);
	\draw (i1) -- (v1);
	\draw (v1) -- (o1);
	\draw (i2) -- (v2);
	\draw (v2) -- (v3);
	\draw (v3) -- (o2);
	\draw (v1) -- (g1);
	\draw (v2) -- (g1);
	\draw (v3) -- (g1);
	\end{feynman}	
	\end{tikzpicture}}
At NLO we need to calculate the 1-loop scalar amplitude
\begin{multline}
\mathcal{A}^{(1)} = \colStructure\!\left(\boxy \right) B + \colStructure\!\left(\crossbox \right) C + \colStructure\!\left(\triR \right) T_{12} \\ + \colStructure\!\left(\triL \right) T_{21} + \colStructure\!\left(\nonAbR \right) Y_{12} + \colStructure\!\left(\nonAbL \right) Y_{21}\,.
\end{multline}
A first task is to choose a basis of independent colour structures. The complete set of colour factors can easily be calculated:
\begin{equation}
\begin{gathered}
\colStructure\!\left(\boxy \right) = \newT_1^a \newT_2^a \newT_1^b \newT_2^b\,,\\
\colStructure\!\left(\crossbox \right) = \newT_1^a \newT_2^b \newT_1^b \newT^a_2\,,\\
\colStructure\!\left(\triL \right) = \frac12\, \colStructure\!\left(\boxy \right) + \frac12\, \colStructure\!\left(\crossbox \right) = \colStructure\!\left(\triR \right),\\
\colStructure\!\left(\nonAbL \right) = \hbar\, \newT_1^a f^{abc} \newT_2^b \newT_2^c\,,\\ 
 \colStructure\!\left(\nonAbR \right) = \hbar\, \newT_1^a \newT_1^b f^{abc} \newT_2^c\,.
\end{gathered}
\end{equation}
At first sight, we appear to have a four independent colour factors: the box, cross box and the two
non-Abelian triangles. However, it is very simple to see that the latter are in fact both proportional to the tree colour factor of \eqn~\eqref{eq:treeamp}; for example, 
\[
\colStructure\!\left(\nonAbL \right) = \frac{\hbar}{2}\, \newT_1^a f^{abc} [\newT_2^b, \newT_2^c] &= \frac{i\hbar^2}{2} f^{abc} f^{bcd} \newT_1^a \newT_2^d\\
& = \frac{i\hbar^2}{2}\, \colStructure\!\left(\tree \right),
\]
where we have used \eqn~\eqref{eqn:chargeLieAlgebra}. Moreover, similar manipulations demonstrate that the cross-box colour factor is not in fact linearly independent:
\[
\colStructure\!\left(\crossbox \right) &= \newT_1^a \newT_1^b \left( \newT_2^a \newT_2^b - i\hbar f^{abc} \newT_2^c\right)\\
&= (\newT_1 \cdot \newT_2) (\newT_1 \cdot \newT_2 ) - \frac{i\hbar}{2} [\newT_1^a, \newT_2^b] f^{abc} \newT_2^c\\
& = \colStructure\!\left(\boxy \right) + \frac{\hbar^2}{2}\, \colStructure\!\left(\tree \right).
\]
\def\Aqed{\mathcal{A}^{(1, \textrm{QED})}}

Thus at 1-loop the classically significant part of the amplitude has a basis of two colour structures: the box and tree. Hence the decomposition of the 1-loop amplitude into partial amplitudes and colour structures is
\[
\mathcal{A}^{(1)} &=  \colStructure\!\left(\boxy \right) \bigg[B + C + T_{12} + T_{21}\bigg] \\ &\qquad\qquad\qquad\qquad + \frac{\hbar^2}{2}\, \colStructure\!\left(\tree \right) \bigg[C + \frac{ T_{12}}{2} + \frac{T_{21}}{2} + iY_{12} + iY_{21}\bigg]\,.\label{eqn:1loopDecomposition}
\]
This expression for the amplitude is particularly useful when taking the classical limit. The second term is proportional to two powers of $\hbar$ while the only possible singularity in $\hbar$ at one loop order is a factor $1/\hbar$ in the evaluation of the kinematic parts of the diagrams. Thus, it is clear that the second line of the expression must be a quantum 
correction, and can be dropped in calculating the classical colour impulse. Perhaps surprisingly, these terms include the sole contribution from the non-Abelian triangles $Y_{ij}$, and thus we will not need to calculate these diagrams. We learn that classically, the 1-loop scalar YM amplitude has a basis of only one colour factor, and moreover depends on the same topologies as in electrodynamics, so we have
\[
\mathcal{A}^{(1)} &=  \colStructure\!\left(\boxy \right) \Aqed + \mathcal{O}(\hbar)\,,
\]
in terms of the one-loop QED amplitude $\Aqed$.

The colour impulse factor in \eqn~\eqref{eqn:NLOcolKnl} therefore reduces to a single commutator. To calculate this we need to repeatedly apply the commutation relation in \eqn~\eqref{eqn:chargeLieAlgebra}, which yields
\[
\big[\newT_1^a, \colStructure\!\left({\scalebox{0.8}{\boxy}} \right) \big] & = [\newT_1^a, \newT_1^b \newT_1^c] \newT_2^b \newT_2^c\\
& = i\hbar f^{acd}\left( \newT_1^d \newT_1^b \newT_2^c \newT_2^b + \newT_1^b \newT_1^d \newT_2^b \newT_2^c\right)\\
&= i\hbar f^{acd}\left( \newT_1^d \newT_2^c (\newT_1\cdot\newT_2) + \left(\newT_1^d \newT_1^b + i\hbar f^{bde} \newT_1^e \right)\left( \newT_2^c \newT_2^b +i\hbar f^{bce}\newT_2^e\right)\right)\\
 & = i\hbar f^{acd}\left( 2\newT_1^d \newT_2^c (\newT_1 \cdot \newT_2) - i\hbar f^{dbe}\left(\newT_1^e \newT_2^b \newT_2^c - \newT_2^e \newT_1^b \newT_1^c\right) + \mathcal{O}(\hbar^2)\right).
\]
The colour impulse factor is itself a series in $\hbar$. The partial amplitude is also a Laurent series in $\hbar$, which is presented in appendix~\ref{app:expressions}. In brief, the leading term in this expansion is the apparently singular (enhanced by one inverse power of $\hbar$) part $\Aqed_{-1} \sim \mathcal O (\hbar^{-2})$, and the classical term $\Aqed_0\sim \mathcal O (\hbar^{-1})$. This has a very important consequence for the impulse kernel --- unlike in the QED case, the apparently singular term $\Aqed_{-1}$ in the partial amplitude now contributes classically, because of the second term in the colour impulse factor: 
\begin{multline}
\colKnl^{a,(1)}_\textrm{1-loop} = \hbar g^4 \bigg\{2 i f^{acd} \newT_1^d \newT_2^c (\newT_1\cdot \newT_2) \Big(\Aqed_{-1} + \Aqed_0\Big) \\ + \hbar f^{acd} f^{dbe}\left(\newT_1^e \newT_2^b \newT_2^c - \newT_2^e \newT_1^b \newT_1^c\right) \Aqed_{-1} \bigg\}\,.\label{eqn:colKnl1loop}
\end{multline}
However, there are still singular terms in the first line; their cancellation requires including the quadratic part of the colour  kernel.

\subsubsection{Cut box}

Rather than viewing the second term in \eqn~\eqref{eqn:NLOcolKnl} as a product of trees, we will treat the quadratic piece as a weighted cut of the box diagram, and define
\begin{multline}
\cut{B} = - i \hbar^2 \! \int\! \dd ^4 \barl\, \del (2p_1\cdot \barl + \hbar\barl^2)\del (2p_2\cdot \barl - \hbar\barl^2)\\
\times \bar{A}_{\scalebox{0.35}{\tree}}(p_1+\hbar\barq , p_2 -\hbar\barq \rightarrow p_1 + \hbar\barl , p_2  -\hbar\barl) \bar{A}_{\scalebox{0.35}{\tree}}(p_1,p_2 \rightarrow p_1+\hbar\barl,p_2-\hbar\barl)\,. \label{eqn:cutBoxFull}
\end{multline}
Using the tree in \eqn~\eqref{eq:treeamp} we can Laurent expand this expression in $\hbar$, as discussed in appendix~\ref{app:expressions}, which yields the leading terms
\[
\cut B_{-1} &= -i\frac{4 (p_1\cdot p_2)^2}{\hbar^2} \!\int\! \dd ^4 \barl \,  \frac{\del (p_1\cdot \barl)\del (p_2\cdot \barl)}{\barl^2 (\barq - \barl)^2 }\\
\cut{B}_{0} &= -i\frac{2 (p_1\cdot p_2)^2}{\hbar} \!\int\! \dd ^4 \barl  \, \frac{\barl\cdot\barq}{\barl^2 (\barq - \barl)^2 }  \left\{  \del (p_1\cdot \barl)\del '(p_2\cdot \barl) - \del (p_2\cdot \barl)\del '(p_1\cdot \barl)\right\}\label{eqn:cutBox}\,. 
\]
To determine the classical contributions we must calculate the associated colour impulse factor,
\[
\colStructure\!\left({\scalebox{0.75}{\tree}} \right)^\dagger \big[\C_1^a,\, \colStructure\!\left({\scalebox{0.75}{\tree}}\right)\big] 
&= i\hbar (\newT_1 \cdot \newT_2) f^{abc} \newT_1^c \newT_2^b\\
= i&\hbar f^{abc} \left(\newT^c_1 \newT^d_1 + i\hbar f^{dce} \newT_1^e\right) \left(\newT_2^b \newT^d_2 + i\hbar f^{dbe} \newT_2^e\right)\\
= i\hbar f^{acd}& \newT_1^d \newT_2^c (\newT_1 \cdot \newT_2) + \hbar^2 f^{acd} f^{dbe}\left(\newT_1^e \newT_2^b \newT_2^c - \newT_2^e \newT_1^b \newT_1^c\right) + \mathcal{O}(\hbar^3)\,.
\]
Clearly we have a similar situation to \eqn~\eqref{eqn:colKnl1loop}: the colour impulse factor is again an expansion in $\hbar$, and its leading term yields a classical contributions from $\cut{B}_{0}$. Meanwhile $\cut{B}_{-1}$ also contributes classically from the correction to the colour structure --- however, there is still a singular term:
\[
\colKnl^{a,(1)}_\textrm{cut box} &= i\hbar g^4 f^{acd}\Big(\newT_1^d \newT_2^c (\newT_1 \cdot \newT_2)\Big(\cut{B}_{-1} + \cut{B}_0\Big)\\ &\qquad\qquad\qquad - i\hbar f^{dbe}\left(\newT_1^e \newT_2^b \newT_2^c - \newT_2^e \newT_1^b \newT_1^c\right) \cut{B}_{-1}\Big)\\
&= 2 g^4 f^{acd} (p_1\cdot p_2)^2 \!\int\! \dd^4\barl\,\frac{1}{\barl^2 (\barq - \barl)^2 } \Bigg\{ \newT_1^d \newT_2^c (\newT_1\cdot\newT_2)\\
&\qquad \times \bigg[\frac{2\del(p_1\cdot\barl) \del(p_2\cdot\barl)}{\hbar} +  \barl\cdot\barq  \left(  \del (p_1\cdot \barl)\del '(p_2\cdot \barl) - \del (p_2\cdot \barl)\del '(p_1\cdot \barl)\right)\bigg]\\
& \qquad\qquad\qquad - 2i f^{dbe}\left(\newT_1^e \newT_2^b \newT_2^c - \newT_2^e \newT_1^b \newT_1^c\right) \del(p_1\cdot\barl) \del(p_2\cdot\barl)\Bigg\}\,.\label{eqn:colKnlcutBox}
\]
\subsubsection{Combining terms}

It now remains to calculate the full colour kernel,
\begin{equation}
\colKnl^{a,(1)} = \colKnl^{a,(1)}_\textrm{1-loop} + \colKnl^{a,(1)}_\textrm{cut box}\,.
\end{equation}
The first priority is to study the classically singular terms which sit in both parts of the kernel. Recall that each part of \eqn~\eqref{eqn:NLOcolKnl} came with a different colour kernel. The upshot of this fact is that, after explicit calculation, the singular terms now involve one common colour structure:
\begin{multline}
\colKnl^{a,(1)}_{-1} = i\hbar g^4\, f^{acd}\newT_1^d \newT_2^c (\newT_1\cdot\newT_2) \Big(2 \Aqed_{-1} + \cut{B}_{-1}\Big) \\ + \hbar^2 g^4\, f^{acd} f^{dbe}\left(\newT_1^e \newT_2^b \newT_2^c - \newT_2^e \newT_1^b \newT_1^c\right) \Big(\Aqed_{-1} + \cut{B}_{-1}\Big)\,.\label{eqn:colKnl-1}
\end{multline}
As detailed in appendix~\ref{app:expressions}, the singular term in the expansion of the QED amplitude originates entirely from box diagrams, and takes a neat form in terms of delta functions:
\begin{equation}\label{eq:b+csuper}
\Aqed_{-1} = i\frac{2 (p_1\cdot p_2)^2}{\hbar^2} \!\int\! \dd ^4 \barl\, \frac{1}{\barl^2 (\barq - \barl)^2 } \del (p_1\cdot \barl)\del (p_2\cdot \barl)\,.
\end{equation}
This expression is the same as the $\cut B _{-1}$ term in \eqn~\eqref{eqn:cutBox}. The coefficients are such that the terms in the first line of \eqn~\eqref{eqn:colKnl-1} cancel, ensuring the apparently singular part of the colour kernel vanishes.

However, an interesting new feature of the colour impulse is that the colour structure in the second line of \eqn~\eqref{eqn:colKnl-1} combines with the sum of the 1-loop singular terms to give a non-zero classical contribution:
\begin{equation}
\left[\colKnl^{a,(1)}_{-1}\right]_{\mathcal{O}(\hbar^0)} \!\!= -2 i g^4f^{acd} f^{dbe} \left( \newT_1^e \newT_2^b \newT_2^c - \newT_2^e \newT_1^b \newT_2^c\right) (p_1\cdot p_2)^2\!\int\! \dd ^4 \barl\, \frac{\del (p_1\cdot \barl)\del (p_2\cdot \barl)}{\barl^2(\barq-\barl)^2}\,.
\end{equation}

With all possible singular terms safely dealt with, it remains to combine the $\mathcal{O}(\hbar^0)$ terms in \eqn~\eqref{eqn:colKnl1loop} and \eqn~\eqref{eqn:colKnlcutBox}. Conveniently, these all have the same colour factor:
\begin{equation}
\colKnl^{a,(1)} = ig^4 \hbar f^{acd} \newT_1^d \newT_2^c (\newT_1\cdot \newT_2) \Big(2\Aqed_0 + \cut{B}_0\Big) + \left[\colKnl^{a,(1)}_{-1}\right]_{\mathcal{O}(\hbar^0)} \,.
\end{equation}
Now we can sum the diagrams in the partial amplitude, the explicit expressions for which are given in appendix~\ref{app:expressions}. The result is
\[
\colKnl^{a,(1)} &= g^4\!\int\! \dd  ^4 \barl\, \frac{1}{\barl^2 (\barl -\barq)^2}  
\Bigg\{ 4if^{acd}\newT_2^c\newT_1^d (\newT_1\cdot\newT_2) \\ &\quad\times\Bigg[ \del (p_1\cdot \barl) \bigg[ m_1^2 +  (p_1\cdot p_2)^2 \barl\cdot (\barl - \barq) \bigg(\frac{1}{(p_2\cdot \barl - i\epsilon)^2} + i \del '(p_2\cdot \barl) \bigg)\bigg]
\\ 
& \qquad\!\! +\del (p_2\cdot \barl) \bigg[m_2^2 + (p_1\cdot p_2)^2 \barl\cdot (\barl - \barq)\bigg( \frac{1}{(p_1\cdot \barl +i\epsilon)^2}  - i\del '(p_1\cdot \barl)\bigg)\bigg]\Bigg]\\
& \qquad\qquad - 2i f^{acd} f^{dbe} \left(\newT_1^e \newT_2^b \newT_2^c - \newT_2^e \newT_1^b \newT_1^c\right) (p_1\cdot p_2)^2 \del (p_1\cdot \barl) \del(p_\cdot \barl) \Bigg\}\,.\label{eqn:colKnlResult}
\]

\subsubsection{Final result}

Finally the observable, the colour impulse, is given by
\begin{equation}
\Delta c_1^{a,(1)} = \Lexp i\! \int\!\dd^4\wn q\, \del(2p_1\cdot \barq) \del(2p_2\cdot \barq) e^{-ib\cdot\barq}\, \colKnl^{a,(1)} \Rexp\,.
\end{equation}
Upon substituting the kernel in \eqn~\eqref{eqn:colKnlResult}, we can average over the momentum and colour wavefunctions implicit in the expectation value. For sharply peaked momentum wavefunctions and large $SU(N)$ representations, this merely has the effect of sending $p_i \mapsto m_i u_i$, and replacing quantum colour factors with products of commuting classical charges. Hence we finally obtain the NLO colour impulse
\[
\Delta c_1^{a,(1)} &= g^4\!\int\! \dd ^4 \barq\, \dd  ^4 \barl\, \del (u_1\cdot \barq)\del (u_2\cdot \barq) e^{-i\barq \cdot b} \frac{1}{\barl^2 (\barl -\barq)^2}  
\\ &\quad\times\Bigg\{ \del (u_1\cdot \barl) \Bigg[\frac{f^{acd}c_1^cc_2^d (c_1\cdot c_2)}{m_2}   \bigg[ 1+  (u_1\cdot u_2)^2 \barl\cdot (\barl - \barq) \bigg(\frac{1}{(u_2\cdot \barl -i\epsilon)^2} 
\\ 
& \qquad\qquad\qquad+ i \del '(u_2\cdot \barl) \bigg)\bigg]
- f^{acd}f^{dbe}c_1^b c_1^c c_2^e \frac{( u_1\cdot u_2)^2}{2} \del (u_2\cdot \barl) \Bigg]
\\ 
&\qquad+\del (u_2\cdot \barl) \Bigg[\frac{f^{acd} c_1^c c_2^d (c_1\cdot c_2)}{m_1} \bigg[ 1 + (u_1\cdot u_2)^2 \barl\cdot (\barl - \barq)\bigg( \frac{1}{(u_1\cdot \barl +i\epsilon)^2}  \\ 
& \qquad\qquad\qquad- i\del '(u_1\cdot \barl)\big)\big] + f^{acd} f^{dbe}c_1^ec_2^bc_2^c  \frac{( u_1\cdot u_2)^2}{2} \del (u_1\cdot \barl)\Bigg] \Bigg\}\,.
\]
This is found to agree with the result obtained by solving the classical equations of motion, which is discussed in section~\ref{sec:ClassEOM}.

\subsection{Momentum} \label{sec:NLOmom}

It is evident that the usual (momentum) impulse in YM theory should be similar to the QED case discussed in~\cite{Kosower:2018adc}. But it is also natural to expect some new terms in the YM impulse in view of the self-coupling of the YM field. Diagrams involving this self-coupling are present at NLO. In this subsection, we investigate the impulse in the YM case with this thought in mind. We begin with \eqn~\eqref{impkernel-Mom} for the impulse kernel $\momKnl^\mu$, which now only involves colour factors of the partial amplitudes themselves.
	
At leading order we can just reuse the expressions in \eqn~\eqref{eq:treeamp}, finding
\begin{equation}
\momKnl^{\mu,(0)} = 4 g^2 \frac{ (p_1\cdot p_2)}{\wn q^2} \wn q^\mu \newT_1\cdot\newT_2\,.
\end{equation}
Then, substituting into \eqn~\eqref{eqn:impkernel} and taking the classical limit as before we have the LO momentum impulse
\begin{equation}
\Delta p_1^{\mu,(0)}  = ig^2 c_1\cdot c_2   \int\! \dd^4 \barq\, \frac{\del (u_1\cdot \barq) \del (u_2\cdot \barq) e^{-ib\cdot \barq}}{ \barq^2}(u_1\cdot u_2 ) \barq^{\mu}\,.
\end{equation}
This expression is closely related to the NLO impulse in QED, which can be obtained from the YM case by replacing $c_1 \cdot c_2$ with the product of the electric charges of the two particles. This relationship is natural, since at leading order the gluons do not self-interact.

Just as in the colour case, at NLO the momentum kernel has linear, 1-loop and quadratic, cut box components:
\[
\momKnl^{\mu,(1)} &= \hbar g^4 \wn q^\mu \sum_\Gamma \mathcal{C}(\Gamma) \bar{A}_\Gamma^{(1)}(p_1,p_2 \rightarrow p_1+\hbar\wn q,p_2- \hbar\wn q)
\\ 
&\,-  i g^4 \hbar^3\! \int\! \dd ^4 \barl\, \del (2p_1\cdot \barl + \hbar\barl^2)\del (2p_2\cdot \barl - \hbar\barl^2)\, \barl^\mu\, \colStructure\left({\scalebox{0.75}{\tree}} \right)^\dagger \colStructure\left({\scalebox{0.75}{\tree}} \right)
\\
&\times \bar{A}_{\scalebox{0.35}{\tree}}(p_1+\hbar\barq , p_2 -\hbar\barq \rightarrow p_1 + \hbar\barl , p_2  -\hbar\barl) \bar{A}_{\scalebox{0.35}{\tree}}(p_1,p_2 \rightarrow p_1+\hbar\barl,p_2-\hbar\barl)\,. \label{eqn:NLOmomKnl}
\]

The decomposition of the 1-loop amplitude onto the colour basis in \eqn~\eqref{eqn:1loopDecomposition} makes computing the first term (linear in the one-loop amplitude) in the impulse kernel trivial; we have
\begin{equation}
\momKnl^{\mu,(1)}_\textrm{1-loop} = \hbar g^4 \wn q^\mu\, \colStructure\left(\boxy \right)\Big(\Aqed_{-1} + \Aqed_0\Big)\,.
\end{equation}
This means that the non-Abelian triangle Feynman diagrams do not contribute to the impulse: a somewhat surprising result, since it is only in these
diagrams that the self-interaction of the gluons appears.

Meanwhile we will denote the kinematic terms in the quadratic piece of the momentum kernel $\cut{B}^\mu$, which has the same definition as \eqn~\eqref{eqn:cutBoxFull} but dressed with an extra loop momentum --- explicit expressions are given in appendix~\ref{app:expressions}. Its colour factor is simply
\begin{equation}
\colStructure\!\left(\tree \right)^\dagger \colStructure\!\left(\tree \right) = (\newT_2\cdot \newT_1) (\newT_1 \cdot \newT_2) = \colStructure\!\left(\boxy \right).
\end{equation}
Thus there is only one relevant colour structure in the NLO momentum impulse, that of the box. The momentum kernel factorises accordingly:
\begin{equation}
\momKnl^{\mu,(1)} = g^4 (\newT_1\cdot\newT_2)^2 \bigg[ \hbar \wn q^\mu \Big(\Aqed_{-1} + \Aqed_0 \Big) + \cut{B}_{-1}^\mu + \cut{B}_0^\mu\bigg]\,.
\end{equation}
This is just a colour dressing of the NLO momentum impulse in QED --- in particular, the cancellation of singular terms between the cut box and 1-loop diagrams is guaranteed \cite{Kosower:2018adc}. Gathering all the terms from triangles, boxes and the cut box in appendix~\ref{app:expressions} and inserting into \eqn~\eqref{eqn:impkernel}, upon taking the classical limit in the now familiar way we find
\[
\Delta p_1 ^{\mu,(1)} &= \frac{g^4  (c_1\cdot c_2)^2 }{2} \int \!\dd  ^4 \barl\,\dd  ^4 \barq\,   \frac{ \del (u_1\cdot \barq)\del (u_2\cdot\barq) }{\barl^2 (\barl- \barq)^2}  \Bigg[ \barq^{\mu} \left\{  \frac{ \del (u_2\cdot \barl) }{m_1}  +  \frac{\del (u_1\cdot \barl)}{m_2} \right.
\\
&\qquad\qquad +  \left.  (u_1\cdot u_2)^2 \barl\cdot (\barl -\barq) \left(   \frac{\del (u_1\cdot \barl)}{m_2(u_2\cdot \barl -i\epsilon)^2} +  \frac{\del (u_2\cdot \barl) }{m_1(u_1\cdot\barl +i\epsilon)^2} \right) \right\}
\\
&\qquad\qquad -i \barl ^{\mu} \barl\cdot (\barl -\barq)\left( \frac{  \delta '(u_1\cdot \barl) \delta (u_2 \cdot \barl)}{m_2} - \frac{\del (u_1 \cdot \barl) \delta ' (u_2 \cdot \barl)}{m_ 1} \right)    \Bigg ] \,.
\label{eq:colimpfull}
\]
We have found that in the non-Abelian theory the final result for the impulse is identical to QED \cite{Kosower:2018adc} with the charge to colour replacement $Q_1Q_2 \to c_1\cdot c_2$. In fact this result follows from the colour basis decomposition in \eqn~\eqref{eqn:1loopDecomposition} and in particular the fact that the non-Abelian triangle diagrams only contribute to the $\hbar^2$ suppressed second colour structure. 

\section{Radiation}
\label{sec:radiation}
\def\asympA{\mathsf{A}}
\def\radC{\mathbb{F}}

One of the strengths of studying impulse-like observables is that radiative phenomena are naturally included, as explored in 
depth in \cite{Kosower:2018adc}. 
Moreover, the double copy makes radiation in Yang-Mills theory a powerful tool for studying its gravitational counterpart~\cite{Goldberger:2016iau,Luna:2017dtq,Shen:2018ebu}. The general discussion of radiation of momentum in KMOC~\cite{Kosower:2018adc} applies directly to the Yang-Mills case, but weakly-coupled YM theory contains another interesting observable:~the total colour radiated to infinity. In this section, we study this radiation of colour in the quantum formalism. As an explicit example we compute the leading order classical colour current found in \cite{Goldberger:2016iau} from scattering amplitudes.

\subsection{Total radiated colour}
The construction and calculation of the colour impulse relied on the adjoint-valued colour operator in \eqn~\eqref{eqn:colourOp} for a scalar field in 
representation $R$. To study the total radiated colour we need a similar operator for the gluon radiation field. This can easily be obtained by restricting to the
adjoint representation, namely by taking  $(T^a_\textrm{adj})_b{ }^c = i f^{bac}$. Since the gluon field is real and has two helicity eigenstates, its colour operator is
\begin{equation}
\radC^a = i\hbar f^{bac} \sum_{\sigma=\pm}\! \int\!\df(k)\, a_\sigma^{b\dagger}(k)\, a_\sigma^c(k)\,,\label{eqn:radColourOp}
\end{equation}
where $\sigma$ labels the helicity. Of course, it is also possible to derive this expression directly from the Noether charge for vector fields in the adjoint representation, given in \eqn~\eqref{eqn:adjointCurrent}, in close analogy to our discussion in sections~\ref{sec:ftcolour} and~\ref{sec:spqcolour}.

This adjoint colour charge is of interest elsewhere in the literature since it plays a role in the physics of YM theory at asymptotic infinity \cite{Strominger:2013lka,Pate:2017vwa,Strominger:2017zoo,Campoleoni:2017qot,He:2019pll,Gonzo:2019fai,Campoleoni:2019ptc}. In this connection, the nature of the final state of the radiation is relevant~\cite{Gonzo:2019fai}. Here, we do not compute this final state explicitly. Instead, we compute expectation values of operators on the final state. 

Following the KMOC route~\cite{Kosower:2018adc} to obtain an expression for the total colour charge radiated from a scattering event leads to
\begin{equation}
\langle \radCol \rangle = \langle \Psi| T^\dagger \radC^a T|\Psi\rangle\,,\label{eqn:rad}
\end{equation}
where we made use of the fact that there are no gauge bosons in the incoming state of \eqn~\eqref{eqn:inState}. 
Before expanding in terms of on-shell scattering amplitudes, it is worth demonstrating colour conservation in our
formalism. At the operator level, assuming only the quantum fields corresponding to particles 1 and 2 are present in addition to the Yang-Mills field, the statement that colour is conserved is 
\begin{equation}
[\C_1^a + \C_2^a + \radC^a, T] = 0\,.
\end{equation}
It then immediately follows that 
\begin{multline}
\langle \Delta c_1^a \rangle + \langle \Delta c_2^a \rangle = \langle \Psi |T^\dagger [\C_1^a , T]|\Psi\rangle + \langle \Psi|T^\dagger [\C_2^a, T]|\Psi\rangle\\
= -\langle \Psi| T^\dagger [\radC^a, T]|\Psi\rangle = -\langle \Psi| T^\dagger \radC^a T|\Psi\rangle = -\langle \radCol \rangle\,,
\label{eqn:conservation}
\end{multline}
where the second line holds from the absence of gluon radiation in the incoming state. Total colour is therefore conserved in the quantum theory, as it must be given the
associated global symmetry.

Let us proceed in expanding \eqn~\eqref{eqn:rad} by inserting complete sets of states; for leading order radiation we need to consider an extra explicit gluon with
momentum $k$ and a colour index, so we will take the set $X$ in the resolution of identity in \eqn~\eqref{eqn:completeStates} to just include the contribution
\begin{equation}
\sum_{b, \sigma}\! \int\!\df(k)\, |k^b, \sigma \rangle \langle k^b, \sigma|\,.
\end{equation}
Note that higher order corrections could also be obtained by adding further states, but we will just be interested in the lowest order case here. Using 
\eqn~\eqref{eqn:radColourOp} and integrating over intermediate delta functions we find
\[
\langle \radCol \rangle &= \sum_{b,c, \sigma} \int\!\df(k) \df(\tilde k) \df(r_1) \df(r_2)\, \d\mu(\zeta_1) \d\mu(\zeta_2)\, \\ & \qquad\qquad\qquad \times \langle \Psi|T^\dagger|{r_1}\,{r_2}\,k^b, \sigma; \zeta_1\, \zeta_2 \rangle \langle k^b, \sigma| \radC^a|\tilde k^c, \sigma \rangle \langle {r_1}\,{r_2}\,\tilde k^c, \sigma; \zeta_1 \zeta_2|T|\Psi\rangle \\
= -i\hbar & \sum_{b,c, \sigma} \int\! \df(k) \df(r_1) \df(r_2) \d\mu(\zeta_1) \d\mu(\zeta_2)\, f^{abc} \varUpsilon^{*\,b}(r_1,r_2;k, \sigma) \, \varUpsilon^c(r_1, r_2; k, \sigma)\,,\label{eqn:quantumRadiatedColour}
\]
where
\begin{multline}
\varUpsilon^a(r_1,r_2;k, \sigma) = \int\!\df(p_1) \df(p_2)\, \phi_1(p_1) \phi_2(p_2)\, e^{ib\cdot p_1/\hbar} \del^{(4)}(p_1 + p_2 - r_1 - r_2 - k) \\ \times \sum_D \langle \zeta_1\, \zeta_2| \colStructure^a(D) |\chi_1\, \chi_2 \rangle A_D(p_1,p_2 \rightarrow r_1, r_2; k, \sigma)\,.
\end{multline}
We have factorised the amplitude into colour structures $\colStructure(D)$ and partial amplitudes $A_D$, as in \eqn~\eqref{eqn:colourStripping}. However, here the colour factor gains a free index from the external gluon state.

To take the classical limit of \eqn~\eqref{eqn:quantumRadiatedColour} and introduce radiation kernels we follow \cite{Kosower:2018adc}, finding
\begin{equation}
\radCol =  -i f^{abc} \sum_\sigma \Lexp \hbar^{-2}\! \int\! \df(k)\, \mathcal{R}^{*\,b}(k, \sigma) \mathcal{R}^c(k, \sigma)\Rexp\,.\label{eqn:radiatedColour}
\end{equation}
The large angle brackets, defined in \eqn~\eqref{eqn:angleBrackets}, are the expectation value over the incoming scalar wavepackets, and thus include the colour states. The radiation kernels inherit the colour index of the external gluon, and take the form 
\begin{multline}
\mathcal{R}^a(k, \sigma) = \hbar^\frac32\! \int\!\dd^4 q_1 \dd^4  q_2\, \del(2p_1\cdot q_1 + q_1^2) \del(2p_2\cdot q_2 + q_2^2)\, 
\del^{(4)}( k -  q_1 -  q_2)\, e^{ib\cdot q_1/\hbar}\\\times \sum_D \colStructure^a(D)\, A_D(p_1 +  q_1, p_2 +  q_2 \rightarrow p_1, p_2; k, \sigma)\,.
\label{Rkernelcolour}
\end{multline}
The powers of $\hbar$ are organised such that the radiation kernel will be $\mathcal{O}(\hbar^0)$ and therefore classical in the limit. Note that because the colour
charge has dimensions of angular momentum, the $\hbar$ scaling here works out the same way as in the total radiated momentum.

\subsection{Leading order evaluation}

Let us explicitly compute the leading order radiation kernel for the scattering of two massive scalar particles, described by the action in \eqn~\eqref{eqn:scalarAction}. In the classical limit the LO kernel is given in terms of coupling constant stripped amplitudes by 
\begin{multline}
\mathcal{R}^{a,(0)}(\wn k) =  \hbar^2 g^3\! \int\!\dd^4 \wn q_1 \dd^4 \wn q_2\, \del(2p_1\cdot\wn q_1 + \hbar\wn q_1^2)  \del(2p_2\cdot\wn q_2 + \hbar\wn q_2^2) \\ \times \del^{(4)}(\wn k - \wn q_1 - \wn q_2)\, e^{ib\cdot\wn q_1}  \sum_D \colStructure^a(D) \bar{A}^{(0)}_D(p_1 + q_1, p_2 + q_2\rightarrow p_1, p_2; k, \sigma)\,.\label{eqn:radKnlLo}
\end{multline}
Clearly the terms in the amplitude which contribute to the classical radiation are those at $\mathcal{O}(\hbar^{-2})$. The shifts in the delta functions are important for obtaining this accurately. The relevant amplitude is the non-Abelian extension of the 5-point tree studied in \cite{Kosower:2018adc}; this was used in \cite{Luna:2017dtq} to take the double copy and calculate radiation in Einstein gravity. The relevant Feynman topologies for emission
from particle 1 are\footnote{The momentum routing is as indicated in \eqn~\eqref{eqn:radKnlLo}.} 
\begin{equation}
\scalebox{0.9}{
\begin{tikzpicture}[scale=0.7, baseline={([yshift=-\the\dimexpr\fontdimen22\textfont2\relax] current bounding box.center)}, decoration={markings,mark=at position 0.6 with {\arrow{Stealth}}}]
\begin{feynman}
\vertex (v1);
\vertex [right = 1.11 of v1] (v2);
\vertex [above left = 1 and 1 of v1] (o1);
\vertex [below left = 1 and 1 of v1] (i1);
\vertex [above right = 1 and 1 of v2] (o2);
\vertex [below right = 1 and 1 of v2] (i2);
\vertex [above left = 0.5 and 0.5 of v1] (g1);
\vertex [above right = 0.47 and 0.47 of g1] (g2);
\draw [postaction={decorate}] (i1) -- (v1);
\draw (v1) -- (g1);
\draw [postaction={decorate}] (g1) -- (o1);
\draw [postaction={decorate}] (i2) -- (v2);
\draw [postaction={decorate}] (v2) -- (o2);
\diagram*{(g1) -- [gluon] (g2)};
\diagram*{(v1) -- [gluon] (v2)};
\end{feynman}	
\end{tikzpicture} \quad\, \begin{tikzpicture}[scale=0.7, baseline={([yshift=-\the\dimexpr\fontdimen22\textfont2\relax] current bounding box.center)}, decoration={markings,mark=at position 0.6 with {\arrow{Stealth}}}]
\begin{feynman}
\vertex (v1);
\vertex [right = 1.11 of v1] (v2);
\vertex [above left = 1 and 1 of v1] (o1);
\vertex [below left = 1 and 1 of v1] (i1);
\vertex [above right = 1 and 1 of v2] (o2);
\vertex [below right = 1 and 1 of v2] (i2);
\vertex [below left = 0.5 and 0.5 of v1] (g1);
\vertex [below right = 0.47 and 0.47 of g1] (g2);
\draw [postaction={decorate}] (i1) -- (g1);
\draw (g1) -- (v1);
\draw [postaction={decorate}] (v1) -- (o1);
\draw [postaction={decorate}] (i2) -- (v2);
\draw [postaction={decorate}] (v2) -- (o2);
\diagram*{(g2) -- [gluon] (g1)};
\diagram*{(v1) -- [gluon] (v2)};
\end{feynman}	
\end{tikzpicture} \quad\, \begin{tikzpicture}[scale=0.7, baseline={([yshift=-\the\dimexpr\fontdimen22\textfont2\relax] current bounding box.center)}, decoration={markings,mark=at position 0.6 with {\arrow{Stealth}}}]
\begin{feynman}
\vertex (v1);
\vertex [right = 1.11 of v1] (v2);
\vertex [above left = 1 and 1 of v1] (o1);
\vertex [below left = 1 and 1 of v1] (i1);
\vertex [above right = 1 and 1 of v2] (o2);
\vertex [below right = 1 and 1 of v2] (i2);
\vertex [above right = 1 and 0.5 of v1] (g2);
\draw [postaction={decorate}] (i1) -- (v1);
\draw [postaction={decorate}] (v1) -- (o1);
\draw [postaction={decorate}] (i2) -- (v2);
\draw [postaction={decorate}] (v2) -- (o2);
\diagram*{(g2) -- [gluon] (v1)};
\diagram*{(v1) -- [gluon] (v2)};
\end{feynman}	
\end{tikzpicture} \quad\, \begin{tikzpicture}[scale=0.7, baseline={([yshift=-\the\dimexpr\fontdimen22\textfont2\relax] current bounding box.center)}, decoration={markings,mark=at position 0.6 with {\arrow{Stealth}}}]
\begin{feynman}
\vertex (v1);
\vertex [right = 1.32 of v1] (v2);
\vertex [above left = 1 and 1 of v1] (o1);
\vertex [below left = 1 and 1 of v1] (i1);
\vertex [above right = 1 and 1 of v2] (o2);
\vertex [below right = 1 and 1 of v2] (i2);
\vertex [right = 0.66 of v1] (g1);
\vertex [above = 0.96 of g1] (g2);
\draw [postaction={decorate}] (i1) -- (v1);
\draw [postaction={decorate}] (v1) -- (o1);
\draw [postaction={decorate}] (i2) -- (v2);
\draw [postaction={decorate}] (v2) -- (o2);
\diagram*{(g1) -- [gluon] (v1)};
\diagram*{(g1) -- [gluon] (v2)};
\diagram*{(g1) -- [gluon] (g2)};
\filldraw [color=black] (g1) circle [radius=0.5pt];
\end{feynman}	
\end{tikzpicture}}
\end{equation}
so we need to calculate the classical terms in the 5-point amplitude 
\newcommand{\treeLa}{\begin{tikzpicture}[thick, baseline={([yshift=-\the\dimexpr\fontdimen22\textfont2\relax] current bounding box.center)}, decoration={markings,mark=at position 0.6 with {\arrow{Stealth}}}]
	\begin{feynman}
	\vertex (v1);
	\vertex [right = 0.3 of v1] (v2);
	\vertex [above left = .25 and .25 of v1] (o1);
	\vertex [below left = .25 and .25 of v1] (i1);
	\vertex [above right = .25 and .25 of v2] (o2);
	\vertex [below right = .25 and .25 of v2] (i2);
	\vertex [above left = .1 and .1 of v1] (g1);
	\vertex [above right = 0.15 and 0.15 of g1] (g2);
	\draw (i1) -- (v1);
	\draw (v1) -- (g1);
	\draw (g1) -- (o1);
	\draw (i2) -- (v2);
	\draw (v2) -- (o2);
	\draw (g1) -- (g2);
	\draw (v1) -- (v2);
	\end{feynman}	
	\end{tikzpicture}}
\newcommand{\treeLb}{\begin{tikzpicture}[thick, baseline={([yshift=-\the\dimexpr\fontdimen22\textfont2\relax] current bounding box.center)}, decoration={markings,mark=at position 0.6 with {\arrow{Stealth}}}]
	\begin{feynman}
	\vertex (v1);
	\vertex [right = 0.3 of v1] (v2);
	\vertex [above left = .25 and .25 of v1] (o1);
	\vertex [below left = .25 and .25 of v1] (i1);
	\vertex [above right = .25 and .25 of v2] (o2);
	\vertex [below right = .25 and .25 of v2] (i2);
	\vertex [below left = .1 and .1 of v1] (g1);
	\vertex [below right = 0.15 and 0.15 of g1] (g2);
	\draw (i1) -- (g1);
	\draw (g1) -- (v1);
	\draw (v1) -- (o1);
	\draw (i2) -- (v2);
	\draw (v2) -- (o2);
	\draw (g1) -- (g2);
	\draw (v1) -- (v2);
	\end{feynman}	
	\end{tikzpicture}}
\newcommand{\treeLc}{\begin{tikzpicture}[thick, baseline={([yshift=-\the\dimexpr\fontdimen22\textfont2\relax] current bounding box.center)}, decoration={markings,mark=at position 0.6 with {\arrow{Stealth}}}]
	\begin{feynman}
	\vertex (v1);
	\vertex [right = 0.3 of v1] (v2);
	\vertex [above left = .25 and .25 of v1] (o1);
	\vertex [below left = .25 and .25 of v1] (i1);
	\vertex [above right = .25 and .25 of v2] (o2);
	\vertex [below right = .25 and .25 of v2] (i2);
	\vertex [above right = 0.25 and 0.15 of v1] (g2);
	\draw (i1) -- (v1);
	\draw (v1) -- (o1);
	\draw (i2) -- (v2);
	\draw (v2) -- (o2);
	\draw (v1) -- (g2);
	\draw (v1) -- (v2);
	\end{feynman}	
	\end{tikzpicture}}
\newcommand{\treeYM}{\begin{tikzpicture}[thick, baseline={([yshift=-\the\dimexpr\fontdimen22\textfont2\relax] current bounding box.center)}, decoration={markings,mark=at position 0.6 with {\arrow{Stealth}}}]
	\begin{feynman}
	\vertex (v1);
	\vertex [right = 0.3 of v1] (v2);
	\vertex [above left = .25 and .25 of v1] (o1);
	\vertex [below left = .25 and .25 of v1] (i1);
	\vertex [above right = .25 and .25 of v2] (o2);
	\vertex [below right = .25 and .25 of v2] (i2);
	\vertex [right = .15 of v1] (g1);
	\vertex [above = 0.275 of g1] (g2);
	\draw (i1) -- (v1);
	\draw (v1) -- (o1);
	\draw (i2) -- (v2);
	\draw (v2) -- (o2);
	\draw (g1) -- (g2);
	\draw (v1) -- (v2);
	\end{feynman}	
	\end{tikzpicture}}
\begin{align}
\mathcal{A}^{(0)}(\wn k^a) &=
\sum_D \colStructure^a(D) \bar{A}^{(0)}_D(p_1 + q_1, p_2 + q_2\rightarrow p_1, p_2; k, \sigma) \\
&=
 \left[\colStructure^a\!\left(\treeLa \right)\!A_{\scalebox{0.5}{\treeLa}} 
+ \colStructure^a\!\left(\treeLb \right)\! A_{\scalebox{0.5}{\treeLb}} 
+ \colStructure^a\!\left(\treeLc \right)\! A_{\scalebox{0.5}{\treeLc}}
 + (1\leftrightarrow 2) \right]
+ \colStructure^a\!\left(\treeYM \right)\! A_{\scalebox{0.5}{\treeYM}}\,. \nonumber
\end{align}
Explicitly, the colour factors are given by
\begin{equation}
\begin{gathered}
\colStructure^a\!\left(\treeLa \right) = (\newT_1^a \cdot \newT_1^b) \newT_2^b\,, \qquad
\colStructure^a\!\left(\treeLb \right) = (\newT_1^b \cdot \newT_1^a) \newT_2^b\,,\\
\colStructure^a\!\left(\treeLc \right) = \frac12\colStructure^a\!\left(\treeLa \right) + \frac12\colStructure^a\!\left(\treeLb \right), \qquad
\colStructure^a\!\left(\treeYM \right) = \hbar f^{abc} \newT_1^b \newT_2^c\,,
\end{gathered}
\end{equation}
with the replacement $1\leftrightarrow2$ for diagrams with gluon emission from particle 2. Just as in the 4-point case at  1-loop, this is an overcomplete set for specifying a basis, because 
\begin{equation}
\begin{aligned}
\colStructure^a\!\left(\treeLb \right) = (\newT_1^a\cdot \newT_1^b) \newT_2^b +  i\hbar f^{bac} \newT_1^c \newT_2^b = \colStructure^a\!\left(\treeLa \right) + i\, \colStructure^a\!\left(\treeYM \right).
\end{aligned}
\end{equation} 
Hence the full basis of colour factors is only 3 dimensional, and the colour decomposition of the 5-point tree is
\begin{multline}
\mathcal{A}^{(0)}(\wn k^a) = \colStructure^a\!\left(\treeLa \right)\Big(A_{\scalebox{0.5}{\treeLa}} + A_{\scalebox{0.5}{\treeLb}} + A_{\scalebox{0.5}{\treeLc}}\Big) \\ 
+ \frac12\colStructure^a\!\left(\treeYM \right)\Big(A_{\scalebox{0.5}{\treeYM}} + 2 i A_{\scalebox{0.5}{\treeLb}} + i A_{\scalebox{0.5}{\treeLc}}\Big) + (1\leftrightarrow2)\,.
\end{multline}
Given that the second structure is $\mathcal{O}(\hbar)$, it would appear that we could again neglect the second term as a quantum correction. However, this intuition is not quite correct, as calculating the associated partial amplitude shows:
\begin{multline}
A_{\scalebox{0.5}{\treeYM}} + 2 i A_{\scalebox{0.5}{\treeLb}} + i A_{\scalebox{0.5}{\treeLc}} = -\frac{4i\,\varepsilon^{h}_\mu(\wn k)}{\hbar^2} \bigg[\frac{2p_1\cdot p_2}{{\wn q_2^2\, p_1\cdot\wn k}}\, \frac{p_1^\mu}{\hbar} + \frac{1}{\hbar\, \wn q_1^2 \wn q_2^2}\Big(2 p_2\cdot\wn k\, p_1^\mu \\- p_1\cdot p_2 \,(\wn q_1^\mu - \wn q_2^\mu) - 2p_1\cdot\wn k\, p_2^\mu\Big) + \mathcal{O}(\hbar^0) \bigg]\,,
\end{multline}
where we have used $p_1\cdot\wn q_2 = p_1\cdot\wn k - \hbar\wn q_1^2/2$ on the support of the on-shell delta functions in the kernel \eqn~\eqref{eqn:radKnlLo}. The partial amplitude appears to be singular, as there is an extra power of $\hbar$ downstairs. However, this will cancel against the extra power in the colour structure, yielding a classical contribution. Meanwhile in the other partial amplitude the potentially singular terms cancel trivially, as in QED, and the contribution is classical:
\begin{multline}
A_{\scalebox{0.5}{\treeLa}} + A_{\scalebox{0.5}{\treeLb}} + A_{\scalebox{0.5}{\treeLc}} = \frac{2}{\hbar^2} \frac{\varepsilon^{h}_\mu(\wn k)}{\wn q_2^2\, p_1\cdot\wn k}\bigg[ 2p_1\cdot p_2\,\wn q_2^\mu +  \frac{p_1\cdot p_2}{p_1\cdot\wn k} p_1^\mu(\wn q_1^2 - \wn q_2^2) \\ - 2p_1\cdot\wn k\, p_2^\mu + 2p_2\cdot \wn k\, p_1^\mu + \mathcal{O}(\hbar)\bigg]\,.
\end{multline}
Summing all colour factors and partial amplitudes, the classically significant part of the 5-point amplitude is
\[
\bar{\mathcal{A}}^{(0)}&(\wn k^a) = \sum_D \colStructure^a(D) \bar{A}^{(0)}_D(\wn k) \\
= &- \frac{4\varepsilon_\mu^h(\wn k)}{\hbar^2} \bigg\{ \frac{\newT_1^a (\newT_1\cdot \newT_2)}{\wn q_2^2 \, \wn k \cdot p_1} \left[-(p_1\cdot p_2)\left(\wn q_2^\mu - \frac{\wn k\cdot\wn q_2}{\wn k\cdot p_1} p_1^\mu\right) + \wn k\cdot p_1 \, p_2^\mu - \wn k\cdot p_2\, p_1^\mu\right] \\
&\qquad\quad + \frac{if^{abc}\,\newT_1^b \newT_2^c}{\wn q_1^2 \wn q_2^2}\left[2\wn k\cdot p_2\, p_1^\mu - p_1\cdot p_2\, \wn q_1^\mu + p_1\cdot p_2 \frac{\wn q_1^2}{\wn k\cdot p_1}p_1^\mu\right] + (1\leftrightarrow 2)\bigg\}\,,
\]
where we have used that $\wn q_1^2 - \wn q_2^2 = -2\wn k\cdot \wn q_2$ since the outgoing radiation is on-shell. Finally, we can substitute into the radiation kernel in \eqn~\eqref{eqn:radKnlLo} and take the classical limit. Averaging over the wavepackets sets $p_i = m_i u_i$ and replaces quantum colour charges with their classical counterparts, yielding
\[
\mathcal{R}^{a,(0)}(\wn k) &= -g^3 \!\int\,\dd^4\wn q_1 \dd^4 \wn q_2 \, \del^{(4)}(\wn k - \wn q_1 - \wn q_2) \del(u_1\cdot\wn q_1) \del(u_2\cdot\wn q_2)\, e^{ib\cdot\wn q_1} \varepsilon_\mu^{h}\\ 
&\times \bigg\{ \frac{c_1\cdot c_2}{m_1} \frac{c_1^a}{\wn q_2^2 \, \wn k \cdot u_1} \left[-(u_1\cdot u_2)\left(\wn q_2^\mu - \frac{\wn k\cdot\wn q_2}{\wn k\cdot u_1} u_1^\mu\right) + \wn k\cdot u_1 \, u_2^\mu - \wn k\cdot u_2\, u_1^\mu\right]\\
& \qquad + \frac{if^{abc}\,c_1^b c_2^c}{\wn q_1^2 \wn q_2^2}\left[2\wn k\cdot u_2\, u_1^\mu - u_1\cdot u_2\, \wn q_1^\mu + u_1\cdot u_2 \frac{\wn q_1^2}{\wn k\cdot u_1}\,u_1^\mu\right] + (1\leftrightarrow 2)\bigg\}\,.\label{eqn:LOradKernel}
\]
Our result is equal to the leading order current $\tilde{K}^{a,(0)}$ obtained in 
\cite{Goldberger:2016iau} by iteratively solving the Wong equations in \eqn~\eqref{Wong-momentum} and \eqn~\eqref{Wong-color} for timelike particle worldlines. We will show this explicitly in the next section.

\section{Classical perspectives}\label{sec:ClassEOM}
In this section we compute the same classical observables, impulse and radiation, using purely classical techniques. These calculations are not too complex, and 
serve to verify the results we obtained using scattering amplitudes.  This gives confidence in applying these quantum methods to gravity, for example, where the 
classical calculations can become significantly more involved. We start with the colour and momentum impulses before moving to the total radiated colour charge,
discussing its relation to asymptotic symmetries.

\subsection{Impulses from equations of motion}
We start with the NLO colour impulse, initially in the more general case of a system of $N$ interacting particles, and later restrict to the $N=2$ case for comparison with earlier sections of the paper.

As discussed previously, the appropriate equations of motion for each particle's worldline are the Yang-Mills-Wong equations in \eqn~\eqref{Wong-momentum} and \eqn~\eqref{Wong-color}. We are seeking perturbative solutions, and therefore expand worldline quantities in the coupling:
\[
x_\alpha^\mu(\tau_\alpha) &= b_\alpha^\mu+u_\alpha\tau_\alpha + \Delta^{(1)}x_\alpha^\mu(\tau_\alpha) + \Delta^{(2)} x_\alpha^\mu(\tau_\alpha) +\cdots\,,\\
v_\alpha^\mu(\tau_\alpha) &= u_\alpha + \Delta^{(1)}v_\alpha^\mu(\tau_\alpha) + \Delta^{(2)} v_\alpha^\mu(\tau_\alpha) +\cdots \,, \\ 
c_\alpha^a(\tau_\alpha) &= c_\alpha^a + \Delta^{(1)}c_\alpha^a(\tau_\alpha) + \Delta^{(2)} c_\alpha^a(\tau_\alpha) +\cdots\,.
\]
Here $\Delta^{(i)}x_\alpha^\mu$ indicates quantities entering at $\mathcal{O}(g^{2i})$. Calculating higher order corrections requires solving for the gauge field $A^a_\mu(x)$, using \eqn~\eqref{YangMillsEOM}. Provided the particle worldlines remain well separated, we can find perturbative solutions using the field equation in the form \cite{Goldberger:2016iau}
\begin{equation}
\begin{gathered}
\partial^2 A_\mu^a(x) = K^a_\mu(x)\,,\\ K_\mu^a(x) \equiv J_\mu^a(x) + gf^{abc} {A}^{b\,\nu}(x) \left(\partial_\nu A_\mu^c(x) - F^c_{\mu\nu}(x)\right)\label{eqn:modifiedCurrent}.
\end{gathered}
\end{equation}
The current $K^a_\mu$ is conserved but gauge dependent; for simplicity we have chosen Lorenz gauge. Writing  $\Delta^{(i)}\!  A^{a}_{\mu}(\bl)$ for perturbative corrections to gauge-field quantities at order $\mathcal{O}(g^{2i-1})$, the LO and NLO gauge fields are solutions to the equations
\[
\partial ^2 [\Delta^{(1)}\!A^{a}_{\mu}(x)] &= \Delta^{(1)}\! J^{a}_{\mu} (x) \,,\\
\partial ^2 [\Delta^{(2)}\!A^{a}_{\mu}(x)] &= \Delta^{(2)}\! J^{a}_{\mu} (x) + gf^{abc}  \left[\Delta^{(1)}\!A^{b\,\nu}(x) \left(  \partial_{\nu} \Delta^{(1)}\!A^{c}_{\mu}(x) - \Delta^{(1)}\!F^{c}_{\mu \nu}(x) \right) \right ] \,,
\]
respectively --- note that the LO equation is the same as for Abelian electrodynamics. We solve by Fourier transforming\footnote{Our conventions for the Fourier transform  are  
$g(x)=\!\int\!\dd^4 \bl \, e^{-\ii \bl \cdot x} \tilde{g}(\bl) $ and $\tilde{g}(\bl )= \!\int\! \d^4 x \, e^{\ii \bl \cdot x} g(x)$ .}, using tildes to represent the Fourier transformed quantity. Solving the LO equation, it is easy to show that the leading order field is
\begin{align}
 \Delta^{(1)}\!  \tilde{A}^{a\mu}(\bl) = - g \sum_\alpha \hdelta(u_\alpha\cdot \bl) e^{\ii \bl\cdot b_\alpha} \frac{c_\alpha^a u^\mu_\alpha}{\bl^2}\,.
\end{align}
It will be useful to define the straight line trajectory $y_\alpha^\mu\equiv b_\alpha^\mu+u_\alpha^\mu\tau_\alpha$ corresponding to the initial unperturbed 
worldlines. With this definition the LO  equations of motion become
\[
m_\alpha \frac{\d^2\Delta^{(1)}x_\alpha^\mu}{\d\tau_\alpha^2}&= g\, c_\alpha^a \!\int\! \dd^4 \bl\, e^{-\ii \bl\cdot y_\alpha}\Delta^{(1)}\! \tilde{F}^{a\,\mu\nu}(\bl){u_{\alpha}}_{\nu}\,,\label{Dpos1}\\
\frac{\d \Delta^{(1)}c_\alpha^a}{\d\tau_\alpha}&= g f^{abc} \! \int\! \dd^4 \bl\, e^{-\ii l\cdot y_\alpha} u_{\alpha}^{\mu} \Delta^{(1)}\! \tilde{A}^{b}_\mu(\bl) c_\alpha^c\,,
\]
while at NLO, $\mathcal{O}(g^4)$,  we have
\[ 
\frac{\d^2\Delta^{(2)}x_\alpha^\mu}{\d\tau_\alpha^2}&= \frac{g}{m_\alpha} \!\int\! \dd^4 \bl\, e^{-\ii \bl\cdot y_\alpha} \Big[\Delta^{(1)}\! \tilde{F}^{a\,\mu\nu}(\bl)  \Delta^{(1)}{v_{\alpha}}_{\nu} c_\alpha^a + \Delta^{(1)}\! \tilde{F}^{a\,\mu\nu}(\bl)\\
& \times {u_{\alpha}}_{\nu} \Delta^{(1)}c_\alpha^a +  \Delta^{(2)}\! \tilde{F}^{a\,\mu\nu}(\bl)\, {u_{\alpha}}_{\nu} c_\alpha^a -  \ii \bl\cdot \Delta^{(1)}x_{\alpha}  \Delta^{(1)}\! \tilde{F}^{a\mu\nu}(\bl)\, {u_{\alpha}}_{\nu}
c_\alpha^a\Big]\,,\label{Dpos2}\\
\frac{\d \Delta^{(2)}c_\alpha^a}{\d\tau_\alpha} &= g f^{abc} \!\int\! \dd^4 \bl\, e^{-\ii \bl\cdot y_\alpha} \Big[u_\alpha\cdot \Delta^{(1)}\!\tilde{A}^{b}(\bl)\, \Delta^{(1)}c_\alpha^c + \Delta^{(1)} v_\alpha\cdot \Delta^{(1)}A^{b}(\bl)\, c^c_\alpha\\
& \qquad\qquad  + u_\alpha \cdot \Delta^{(2)}\!\tilde{A}^{b}\,c_\alpha^c -\ii \bl\cdot \Delta^{(1)}x_\alpha \ u_\alpha\cdot \Delta^{(1)}\!\tilde{A}^b(\bl)\,c_\alpha^c \Big]\,.
\]
These NLO equations involve the LO corrections to the fields and the particles' colours, positions and velocities, so although they are not the main quantities of
interest we will need to integrate the expressions in \eqn~\eqref{Dpos1}; for example,
\begin{equation}
\Delta c_{\alpha}^a(\tau_\alpha) = \int_{-\infty}^{\tau_{\alpha}} \d\tau '_\alpha \frac { \d c_\alpha^a}{\d\tau '_\alpha}\,. 
\end{equation}
In performing these integrals one must include an $i\epsilon$ convergence factor, so the definition of $y_\alpha^\mu$ is modified such that $\barl\cdot y_\alpha = \barl\cdot b_\alpha + (u_\alpha\cdot\barl + i\epsilon)\tau_\alpha$. This yields
\[
\Delta^{(1)}x^\mu_\alpha(\tau_\alpha) &= \ii g^2 \sum\limits_{\beta\ne \alpha} \frac{c_\alpha\cdot c_\beta }{m_\alpha} \!\int\!\dd^4 \bl\, e^{\ii \left(\bl\cdot b_{\beta }-\bl\cdot y_{\alpha }\right)} \hdelta(\bl\cdot u_\beta)\frac{\bl\cdot u_{\alpha } u_{\beta }^{\mu }-\bl^{\mu } u_{\alpha }\cdot u_{\beta }}{\bl^2 (\bl\cdot u_{\alpha } + i\epsilon)^2}\,, \label{defle1}\\
\Delta^{(1)}c_\alpha^a(\tau_\alpha) &=\ii g^2 \sum\limits_{\beta\ne \alpha}  f^{abc} c_\alpha^b c_\beta^c\,  u_{\alpha }\cdot u_{\beta }\! \int\!\dd^4 \bl\,  e^{\ii (\bl\cdot b_{\beta }-\bl\cdot y_{\alpha })} \frac{\hdelta(\bl\cdot u_\beta)}{\bl^2} \frac{1}{(\bl\cdot u_{\alpha } + i\epsilon)}\,.
\]
We now have the information to determine the NLO field, which is
\[
\Delta^{(2)}\!\tilde{A}^{a}_{\mu}(\bl_1) & =-g^3\frac{1}{\bl_1^2}  \sum_{\beta\neq\alpha} \int\! \dd^4\bl_2\,  e^{\ii(\bl_{1}-\bl_2) \cdot b_{\alpha}} e^{\ii \bl_2\cdot b_{\beta}}\, \hat{\delta}((\bl_1-\bl_2)\cdot u_{\alpha}) \hat{\delta}(\bl_2\cdot u_{\beta}) \\
&\times \Bigg\{\frac{c_{\alpha}^a c_{\alpha}\cdot c_{\beta}}{m_{\alpha}\bl_2^2}\left(-\frac{\bl_2^{\mu } u_{\alpha}\cdot u_{\beta}}{\bl_2\cdot u_{\alpha}}+\frac{\bl_1\cdot \bl_2 u_{\alpha}\cdot u_{\beta}\, u_{\alpha}^{\mu}}{\left(\bl_2\cdot u_{\alpha}\right)^2}-\frac{\bl_1\cdot u_{\beta}\, u_{\alpha}^{\mu }}{\bl_2\cdot u_{\alpha}} + u_{\beta}^{\mu}\right)\\
& \qquad + \frac{\ii f^{abc} c_{\alpha}^b c_{\beta}^c}{\bl_2^2} \left(\frac{\bl_2^{\mu } u_{\alpha}\cdot u_{\beta}}{(\bl_1-\bl_2)^2} - \frac{2 \bl_2\cdot u_{\alpha} u_{\beta}^{\mu }}{(\bl_1-\bl_2)^2} + \frac{ u_{\alpha}\cdot u_{\beta}\, u_{\alpha}^{\mu }}{\bl_2\cdot u_{\alpha}} \right)\!\Bigg\}\,.\label{eqn:LOradiation}
\]
It is now very simple to use the Fourier transform of the Yang-Mills equation in \eqn~\eqref{eqn:modifiedCurrent} to calculate the LO momentum space current. In this context it is useful to rename the momentum of the field $\wn k$, relabel $\barl_2 = \wn q_2$, and introduce $\wn q_1 = \barl_1 - \barl_2$. Then we find that
\[
\Delta^{(2)} \tilde K^{a\,\mu} (\wn k) &= g^3 \sum_{\beta\neq\alpha} \int\!\dd^4\wn q_1 \dd^4\wn q_2\, \del^{(4)}(\wn k - \wn q_1 - \wn q_2) \del(u_{\alpha}\cdot\wn q_1) \del(u_{\beta}\cdot\wn q_2)\, e^{ib_{\alpha}\cdot\wn q_1} e^{ib_\beta\cdot\wn q_2} \\ 
& \times \bigg\{ \frac{c_{\alpha}\cdot c_{\beta}}{\wn q_2^2 \, \wn k \cdot u_{\alpha}} \frac{c_{\alpha}^a}{m_{\alpha}} \bigg[-(u_{\alpha}\cdot u_{\beta})\left(\wn q_2^\mu - \frac{\wn k\cdot\wn q_2}{\wn k\cdot u_{\alpha}} u_{\alpha}^\mu\right) + \wn k\cdot u_{\alpha} u_{\beta}^\mu - \wn k\cdot u_{\beta}\, u_{\alpha}^\mu\bigg]\\
& \qquad + \frac{if^{abc}\,c_{\alpha}^b c_{\beta}^c}{\wn q_1^2 \wn q_2^2}\left[2\wn k\cdot u_{\beta}\, u_{\alpha}^\mu - u_{\alpha}\cdot u_{\beta}\, \wn q_1^\mu + u_{\alpha}\cdot u_{\beta} \frac{\wn q_1^2}{\wn k\cdot u_{\alpha}}\,u_{\alpha}^\mu\right]\bigg\}\,.\label{eqn:LOcurrent}
\]
This result was first obtained in Ref.~\cite{Goldberger:2016iau}. Comparing against \eqn~\eqref{eqn:LOradKernel}, we can see that 
(up to an irrelevant overall sign) the two particle restriction of the current is equal to the LO radiation kernel
calculated using amplitudes.\footnote{Notice that we set $b_2 = 0$ in sections~\ref{sec:impulse} and~\ref{sec:radiation} using translation symmetry.}

Returning to the impulse, we can skip over the current and substitute the NLO field into \eqn~\eqref{Dpos2}. A straightforward but tedious calculation then yields
the results for the NLO corrections given in appendix~\ref{app:Nparticles}. The observable quantities, the impulses, are defined by
\begin{equation}
\Delta c^a_\alpha \equiv \int_{-\infty}^\infty\! \d\tau_\alpha\, \frac { \d c_\alpha^a}{\d\tau_\alpha}\,,
\qquad 
\Delta p_\alpha^\mu \equiv m_\alpha\!  \int_{-\infty}^\infty\!\d\tau_\alpha\, \frac{\d v_\alpha^\mu}{\d \tau_\alpha}\,.
\end{equation}
Using the results for $\Delta^{(2)}c_\alpha^a$ it is  straightforward to show, after a redefinition of the integration variables, that the NLO colour impulse 
takes the form
\[
\label{ClassColImp}
&\Delta c^{a,(2)}_\alpha = f^{abc}\! \int\! \dd^4 \bq\, e^{\ii b_{2}\cdot \bq}  e^{-\ii b_{1}\cdot \bq}\, \hdelta(\bq\cdot u_1) \hdelta (\bq \cdot u_2) 
\int \frac{ \dd^4 \bl}{\bl^2 (\bl-\bq)^2}  \\
& \times \Bigg\{ \hdelta(\bl\cdot u_2) \left[\frac{c_1\cdot c_2\,c_1^b c_2^c}{m_1} \left(1+\frac{\left(u_1\cdot u_2\right)^2 \left(\bar{\ell}^2-\bar{\ell}\cdot \bar{q}\right)}{(\bar{\ell}\cdot u_1 -
i\epsilon)^2} \right)
+ i f^{bde} c_2^c c_2^d c_1^e  \frac{\left(u_1\cdot u_2\right)^2}{\bar{\ell}\cdot u_1 - i\epsilon}\right]\\
& \quad +\hdelta(\bl\cdot u_1) \Bigg[\frac{c_1\cdot c_2\, c_1^b c_2^c}{m_2} \left(1+ \frac{\left(u_1\cdot u_2\right)^2 \left(\bar{\ell}^2-\bar{\ell}\cdot \bar{q}\right)}{\left(\barl\cdot u_2 + i\epsilon\right)^2} \right) + i f^{bde} c_1^c c_1^d c_2^e \frac{ \left(u_1\cdot u_2\right)^2}{\bar{\ell}\cdot u_2 + i\epsilon}\Bigg]\\
&\qquad \qquad \qquad \qquad \qquad \qquad \qquad \qquad \qquad - 2 i f^{bde} c_1^c c_1^d c_2^e\, \del(\bl\cdot u_1) \frac{\bar{\ell}\cdot u_2}{\bar{q}^2} \Bigg\}\,.
\]
Notice that the signs of the $\ii \epsilon$ on the second and third lines in the above equation are different. This is a simple consequence of a change of variables required to associate the loop momenta of the classical calculation to that derived from amplitudes (see appendix \ref{app:Nparticles}). To see that this indeed is the same as our earlier result we must manipulate the $i\epsilon$ factors in the denominators further. For the quadratic denominator we can replace
\begin{equation}
\frac{1}{(\barl \cdot u_\alpha + i\epsilon)^2}  = i\del '(\barl \cdot u_\alpha) + \frac{1}{(\barl \cdot u_\alpha - i\epsilon)^2}\,,
\end{equation}
and for the linear denominator we make the shift $\barl \to \barq - \barl$, which simply has the effect of changing the sign of the
$i\epsilon$\footnote{The $\barq \cdot u_\alpha$ term can be ignored due to the delta function $\delta(\barq \cdot u_\alpha)$.}. 
Then these terms can be averaged and combined to form a delta function. 
These procedures are the same as for the momentum impulse  \cite{Kosower:2018adc} and are briefly reviewed  in appendix~\ref{app:expressions}. 
After these manipulations, the result from amplitudes in \eqn~\eqref{eq:colimpfull} matches the colour impulse computed classically in \eqn~\eqref{ClassColImp}, up to the term on the last line. 

This final term is spurious, and can be traced back to the non-Abelian correction to the gauge field at NLO, shown in the bottom line of \eqn~\eqref{eqn:LOradiation}. We observe that the term is proportional to the integral
\begin{align}
u_2\cdot I = \int\! \dd^4 \bl\, \hdelta(\bl\cdot u_1) \frac{ \bar{\ell}\cdot u_2}{\bl^2 (\bl-\bq)^2}\,,
\end{align}
which vanishes on the support of $\hdelta(\bq\cdot u_1)$ and $\hdelta (\bq \cdot u_2)$. This can easily be seen by writing\footnote{Any additional vectorial dependence arising by regulating the divergent integral will have vanishing dot product with $u_2$.} $I^\mu=A u_1^\mu+ B \bq^\mu$; then we have that $A=0$, and hence $u_2 \cdot I=0$.

The momentum impulse follows through similarly, and for the case of two particles we find that the result is the same as in Abelian electrodynamics, calculated by KMOC \cite{Kosower:2018adc}, but with the replacement $Q_1Q_2 \to c_1 \cdot c_2$. This is in agreement with the quantum calculation of section~\ref{sec:NLOmom}.

\subsection{Total radiated colour and asymptotic symmetries}
\label{sec:asympSyms}

Our classical calculation of the NLO impulse relied on the order $g^3$ gauge field in \eqn~\eqref{eqn:LOradiation}. This gauge field in isolation is also of interest, because it is the leading radiation field generated by the scattering of the two particles. It therefore describes the transport of momentum and colour by the classical YM field itself. This classical transport of colour deserves more discussion.

The Noether current for a vector field transforming in the adjoint of the colour group is
\[
j^a_\mu(x) = - f^{abc} A^{b\nu}(x) \left(\partial_\mu A^c_\nu(x) - \partial_\nu A^c_\mu(x) \right).\label{eqn:adjointCurrent}
\]
To measure the instantaneous rate of colour radiation at a time $t$ during a scattering event, we surround the particles by a large sphere
and measure the flux of $j^a_\mu$ across its surface, taking the limit that the radius of the sphere goes to infinity. More specifically, we are
interested in outgoing radiation from our scattering event, so we take this large radius limit at fixed retarded time $u = t - r$. We then integrate
over all retarded times. Thus the surface of integration --- namely the null future boundary $\mathscr{I}^+$ of Minkowski space --- is three-dimensional, parameterised by $u$ and the coordinates on the two-sphere. The colour radiated to $\mathscr{I}^+$ is\footnote{We use Bondi coordinates $u$, $r$, $\theta$ and $\phi$.}
\[
\radCol &= \int_{\mathscr{I}^+} \!* \, j^a \\
&= - \int_{-\infty}^\infty \! \d u \lim_{r\rightarrow\infty} \int \!\d\Omega_2\, r^2 j^a_r \,,
\]
where $j^a = j^a_\mu \, \d x^\mu$ and $* \, j^a$ is its Hodge dual.

To evaluate this integral, we need an expression for the asymptotic field. This satisfies the Yang-Mills equation, given in linearised form in 
\eqn~\eqref{eqn:modifiedCurrent}. Since there is no incoming radiation in our situation, we impose retarded boundary conditions. 
Using the standard large-distance expansion of the retarded Green's function we readily find
\begin{equation}
A_\mu^a(x) = \frac1{4\pi r}\int\!\frac{\d\omega}{2\pi}\, e^{-i\omega u} \tilde{K}^a_\mu(\wn k)\bigg|_{\wn k^\nu=(\omega, \omega\hat{\v{x}})} + 
\mathcal{O}\left(\frac1{r^2}\right)\,,\label{eqn:asymptoticGaugeField}
\end{equation} 
where $r = |\v{x}|$ and $t=x^0$. 
It may also help the reader to record the derivative of the field, which is
\begin{equation}
\partial_\mu A_\nu^a(x) = -\frac{i}{4\pi r}\int\! \frac{\d\omega}{2\pi}\, \wn k_\mu \tilde{K}^a_\nu(\wn k) e^{-i\omega u}\bigg|_{\wn k^\rho=(\omega, \omega\hat{\v{x}})} + \mathcal{O}\left(\frac1{r^2}\right) \,.
\label{eqn:radiationField}
\end{equation}
Hence upon integrating over delta functions the total radiated charge is
\begin{equation}
\radCol = \frac{i}{(4\pi)^2} \int_{-\infty}^{\infty} \!\frac{\d\omega}{2\pi} \int \d\Omega_2\, f^{abc} \tilde{K}^{b\,\nu}(-\wn k)\left(k_\nu \tilde{K}^{c}_{r}(k) - \wn k_r \tilde{K}^{c}_{\nu}(\wn k)\right)\bigg|_{\wn k=(\omega, \omega\hat{\v{x}})} \,.
\end{equation}
This expression can be considerably simplified. Current conservation is $k^\nu \tilde{K}^a_\nu(k)~=~0$, so the first term in parentheses vanishes. In the second term, we have $k_r = - \omega$.  We can further exploit the symmetry of the integral, and reality of the current $K^a_\mu(x)$ to show
\[
\radCol &= \frac{i}{(2\pi)^3} \int_0^\infty\! \d\omega\, \omega^2 \int\! \d\Omega_2\, \frac1{2\omega} f^{abc} \tilde{K}^{b}_{\nu}(-\wn k) \tilde{K}^{c\,\nu}(\wn k)\bigg|_{\wn k=(\omega, \omega\hat{\v{x}})}\\
&= \frac{i}{(2\pi)^3}\int\! \d^3\wn k\, \d\wn k^0\, \frac{\delta(\wn k^0 - |\v{\wn k}|)}{2|\v{\wn k}|} f^{abc} \tilde{K}^{b}_{\nu}(-\wn k) \tilde{K}^{c\,\nu}(\wn k)\\
&= \int\! \df(\wn k)\, if^{abc} \tilde{K}^{b\,*}_\nu (\wn k) \tilde{K}^{c\,\nu}(\wn k)\,.
\]
Finally, we use completeness of the polarisation vectors to write the classical total radiated colour in 
precisely the same form as the quantum expression:
\begin{equation}
\radCol = -if^{abc} \sum_{\sigma=\pm} \int\! \df(\wn k)\, \Big(\varepsilon^*_\sigma\cdot \tilde{K}^*(\wn k)\Big)^b \Big(\varepsilon_\sigma\cdot\tilde{K}(\wn k) \Big)^c\,.\label{eqn:classicalRadiatedCharge}
\end{equation}
Comparing with \eqn~\eqref{eqn:radiatedColour} confirms that, in the classical limit, the radiation kernel coincides (up to a possible sign) with $\varepsilon_\sigma \cdot \tilde{K}^a(\wn k)$ at large distances.

Let us finish with a few additional remarks on this radiated colour. In ordinary electrodynamics it is elementary that charge is connected with the current appearing in the equation of motion. Although we made use of the Noether current in our discussion above, it remains the case that the radiated charge is connected to the current $K^a$ in the linearised form of equation~\eqref{eqn:modifiedCurrent}. It is easy to check, using the explicit asymptotic field of \eqn~\eqref{eqn:asymptoticGaugeField}, that the radiated charge is
\[
\radCol = \frac{1}{g} \int_{\mathscr{I}^+} \!*\, K^a \,. 
\]
We may now make use of equation~\eqref{eqn:modifiedCurrent} in the form
\[
\d * F^a = - * K^a 
\]
to write
\[
\radCol = -\frac{1}{g} \int_{\mathscr{I}^+} \!\d * F^a \,,
\]
where $F^a$ is the linearised field strength. The radiated charge may therefore also be reconstructed by integration over the boundaries $\mathscr{I}^+_\pm$ of $\mathscr{I}^+$ as
\[
g \radCol = \int_{\mathscr{I}^+_-} \!* \, F^a - \int_{\mathscr{I}^+_+} \!* \, F^a \, = g c_\textrm{initial} - g c_\textrm{final} \,.
\]
In other words, the radiated charge is the difference between initial and final charges, as measured by integrating the electric fields over large spheres in the far past and the far future: total colour charge is conserved, as we also saw using quantum mechanical methods in \eqn~\eqref{eqn:conservation}.

Although our focus was on radiation of global charge, some of the expressions above are also relevant in the discussion of the larger asymptotic symmetry group of Yang-Mills theory, see for example~\cite{Luscher:1978ir,Strominger:2013lka,Adamo:2015fwa,Pate:2017vwa,Strominger:2017zoo,Campoleoni:2017qot,He:2019pll,Campoleoni:2019ptc}. It would be interesting to broaden our analysis to this context, particularly in the context of the infrared structure of loop amplitudes.

\section{Discussion}
\label{sec:conclusions}

In this article, we developed methods for computing classical observables in Yang-Mills theories from scattering amplitudes. This amounts to an extension of the scope of the KMOC formalism~\cite{Kosower:2018adc} to encompass perturbative Yang-Mills theory. In addition to the 
observables familiar from electrodynamics and gravity, namely the momentum impulse and the total radiated momentum, we constructed two new 
observables: the colour impulse and total radiated colour charge. 

Our underlying motivation is to understand the dynamics of classical general relativity through the double copy. In particular, we are
interested in the relativistic two-body problem which is so central to the physics of the compact binary coalescence events observed
by LIGO and Virgo. Consequently, we focused on observables in two-body events. Although we only considered unbound (scattering)
events, it is possible to determine the physics of bound states from our observables. This can be done concretely using effective
theories~\cite{Cheung:2018wkq}. We also hope that it may be possible to connect our observables more directly to bound states using analytic continuation,
in a manner similar to the work of K\"alin and Porto~\cite{Kalin:2019rwq,Kalin:2019inp}.

The emergence of the classical theory from an underlying perturbative quantum field theory is surprisingly intricate. Coherent states play an important
role in this story, as emphasised by Yaffe~\cite{Yaffe:1981vf} in the context of large $N$ theories. In section~\ref{sec:setup} we emphasised the role of coherent
states in describing the colour structure of particles in the classical approximation. It is also important that the representation of the corresponding quantum
field is large. This is in exact analogy with the emergence of a classical spin from a quantum system, and indeed the states we used for colour can
equally be used to describe spin. Furthermore the physics of the colour impulse in YM theory is closely analogous to the physics of angular momentum and the 
associated angular impulse~\cite{Maybee:2019jus,Guevara:2019fsj}. Since the story for colour is a little simpler, we expect it to be a useful toy model
for spin in gravity.

We studied the impulse and its colourful counterpart at NLO in YM theory. One surprise in our work was that the part of the (four-point) amplitude which is 
relevant in the classical theory is exactly proportional to the classical part of the QED four-point amplitude. Indeed the impulse at next-to-leading order in 
the YM case is basically equal to the QED case; the only difference is a charge-to-colour replacement. This is a little peculiar because it is natural to expect the non-linearity of the Yang-Mills field to enter at this order (and it does so in the quantum theory). Nevertheless the colour impulse, which is
intrinsically non-Abelian by definition, is non-vanishing. Although it is constructed from the same one-loop amplitude, an interplay of colour commutators
and classically singular terms in the amplitude results in an expression for the colour impulse which involves various different colour factors.
We confirmed the results of our calculations by a direct classical computation using the Yang-Mills-Wong worldline formalism.
It is interesting to compare our methods to those of Shen~\cite{Shen:2018ebu}, who implemented the double copy at NLO wholly within the classical 
worldline formalism following ground-breaking work of Goldberger and Ridgway~\cite{Goldberger:2016iau}. Shen found it necessary to include
vanishing terms involving structure constants in his work. Similarly, in our context, some colour factors are paired with kinematic numerators proportional
to $\hbar$. It would be interesting to use the tools developed in this paper to explore the double copy construction of Shen~\cite{Shen:2018ebu} from the
perspective of amplitudes.

Throughout our paper, we emphasised that scattering amplitudes can be used to determine classical YM observables. But so do Wong's equations. We have not addressed the question of whether it is \emph{easier} to find a particular observable from amplitudes or from the Wong equations. This question isn't really of interest to us since our goal is to understand gravity, where amplitudes are much easier to compute than any (known) classical procedure. But possibly our work offers a way to combine the advantages of classical equations and the double copy. We provided explicit expressions
for YM observables in terms of amplitudes; given a determination of these observables from the Wong equations, then it is possible to
solve for the (classical part of the) amplitude. If it is possible to compute a corresponding
gravitational amplitude from the double copy unambiguously from the classical parts of a Yang-Mills amplitude, then this method would allow
for the computation of observables in gravity from the Wong equations. Compared to the worldline double copy of Goldberger, Ridgway~\cite{Goldberger:2016iau} and Shen~\cite{Shen:2018ebu}
this suggestion would implement the double copy in a more standard manner. 
Our methods may also shine light on the difficulty implementing the double copy off-shell in the worldline theory discussed in~\cite{Plefka:2019hmz}, since 
one could check proposals for implementing a worldline double copy against our formulae.

Our expression for the colour impulse is in many ways similar to the KMOC expression for the ordinary impulse. In essence the impulse describes
a transfer of a small amount of momentum $\hbar \bar{q}$, weighted by an amplitude of order $1/\hbar$. Thus the momentum transferred by many
gluons leads to a macroscopic impulse. In the colour case, the small momentum transferred is replaced by a colour commutator. This is reminiscent
of the transition from fuzzy spaces to the continuum (see, for example~\cite{Alexanian:2001qj,Grosse:2004wm}): the momentum transfer in the impulse is the Fourier transform of a derivative, corresponding
to the commutator in a fuzzy space. Perhaps there is a clue here to how the double copy works.

Turning to radiation of colour, a first comment is that the relevant amplitude is no longer proportional to the QED case. This means that at NNLO the
impulse will no longer be proportional to the QED impulse, because the radiated momenta are genuinely different in the two cases. One motivation
for studying impulse and radiation together is that they are related by conservation of momentum, so the five-point radiation terms capture dissipative
effects in the impulse. The physics of momentum conservation and dissipation is rich, so we look forward to further work in this area. 

Colour radiation is also interesting from the point of view of asymptotic symmetry groups. Yang-Mills theory is an interesting toy model for gravity in this 
context, as pointed out in an early paper by L\"uscher~\cite{Luscher:1978ir}. It would be very interesting to study colour radiation at NLO, in particular
to understand what becomes of the infrared divergences of the loop amplitudes, and their impact on soft theorems. 
The Yang-Mills case is particularly subtle in view of the presence of 
collinear divergences. In electromagnetism and gravity a first step in these directions has recently been made \cite{A:2020lub}, where the connection between quantum and classical soft theorems in electromagnetism and gravity was studied using radiation kernels. We look forward to future progress on these fronts.

\section*{Acknowledgements}

We thank Roger Horsley, David Kosower, Se\'an Mee, Alexander Ochirov and Siddharth Pandey for useful discussions, and Ingrid Holm for correcting some typos in our equations. BM and AR are supported by STFC studentships ST/R504737/1 and ST/T506060/1 respectively. LDLC and DOC are supported by the STFC grant ST/P0000630/1.
This research was supported by the Munich Institute for Astro- and Particle Physics (MIAPP) which is funded by the Deutsche Forschungsgemeinschaft (DFG, German Research Foundation) under Germany's Excellence Strategy -- EXC-2094 -- 390783311. Some of our figures were produced with the
help of TikZ-Feynman~\cite{Ellis:2016jkw}.

\appendix

\section{Charge factorisation in SU(3)}
\label{SU3-coherent-example}
In this appendix we prove \eqn~\eqref{factorization-charges}. Using the coherent states restricted to the $SU(3)$ irreducible representation $[n_1,n_2]$ in \eqn~\eqref{restricted-coherent}, we have
\[ 
\langle \xi\,\zeta|\C^a\C^b | \xi\,\zeta\rangle_{[n_1,n_2]} &= \frac{1}{(n_1!\, n_2!)}  \bra{0} \left( \zeta^* \cdot b\right)^{n_2} \left(\xi^* \cdot a \right)^{n_1}( a^{\dagger}\lambda^a a - b^{\dagger}\bar \lambda^a b) \\& \qquad\qquad\qquad \times ( a^{\dagger}\lambda^b a - b^{\dagger}\bar \lambda^b b)\left( \zeta \cdot b^\dagger\right)^{n_2} \left(\xi \cdot a^\dagger \right)^{n_1} \ket{0}
\\
&=\frac{1}{(n_1!\, n_2!)}  \bra{0} \left( \zeta^* \cdot b\right)^{n_2} \left(\xi^* \cdot a \right)^{n_1}\Big( a^{\dagger}\lambda^a a \,  a^{\dagger}\bar \lambda^b a - b^{\dagger}\bar \lambda^a b\, a^{\dagger}\lambda^b a 
\\ & \qquad \qquad  - 
a^{\dagger} \lambda^a  a \, b^{\dagger}\bar \lambda^b b - b^{\dagger}\bar \lambda^a b \, b^{\dagger}\bar \lambda^b b \Big)\left( \zeta \cdot b^\dagger\right)^{n_2} \left(\xi \cdot a^\dagger \right)^{n_1} \ket{0} .
\]
Note that we can consider the $a$ and $b$ terms separately since they commute. The terms with only two $a$ operators (or $b$'s) reduce to products of the form
\[
\langle \xi\,\zeta|a^{\dagger i} a_j | \xi\,\zeta\rangle_{[n_1,n_2]}  &= \frac{1}{(n_1!\, n_2!)}  \bra{0} \left( \zeta^* \cdot b\right)^{n_2} \left(\xi^* \cdot a \right)^{n_1} a^{\dagger i}a_j \left( \zeta \cdot b^\dagger\right)^{n_2} \left(\xi \cdot a^\dagger \right)^{n_1} \ket{0} 
\\
&= n_1 \xi^{*i} \xi_j\,,\label{eqn:singleOpExpressions}
\]
and similarly for expressions involving $b^i b_j^{\dagger}$. We made use of the fact that the states are normalised. Next we have the term involving four $a$'s (or $b$'s), which yields
\[
\langle \xi\,\zeta|a^{\dagger i} a_j a^{\dagger k} a_l| \xi\,\zeta\rangle_{[n_1,n_2]}  =  n_1\xi^{* i} \xi_l\, \delta_j{ }^k + n_1(n_1-1) \xi^{* i} \xi_l\, \xi^{* k} \xi_j \,.
\]
In the limit where $n_1$ is large we can replace\footnote{The correction term vanishes in the classical $\hbar \to 0$ limit regardless.} the $n(n-1)$ factor with $n^2$. Using this and returning all $\lambda$ and $\hbar$ factors we find
\[
\hbar^2 (\lambda^a)_i{ }^j (\lambda^b)_k{ }^l \langle \xi\,\zeta|a^{\dagger i} a_j a^{\dagger k} a_l| \xi\,\zeta\rangle_{[n_1,n_2]}
&= \hbar^2 n_1^2\, \xi^*\lambda ^a \xi\, \xi^*\lambda^b \xi + \hbar^2 n_1\, \xi^*\lambda^a\cdot \lambda^b \xi \,.
\]

Now, gathering all the terms with a pair of $a$'s and a pair of $b$'s, which are simply products of the expressions in \eqn~\eqref{eqn:singleOpExpressions} contracted with Gell-Mann matrices, we have that
\[
\langle \xi\,\zeta|\C^a\C^b | \xi\,\zeta\rangle_{[n_1, n_2]} &= \hbar^2 \Big( n_1^2\, \xi^*\lambda^a\xi\, \xi^*\lambda^b \xi + n_2^2 \,\zeta^*\bar\lambda^a\zeta\, \zeta^* \bar\lambda^b \zeta - n_1n_2\xi^*\lambda^a \xi\, \zeta^*\bar\lambda^b\zeta \\
& \qquad- n_1n_2\, \xi^*\lambda^b \xi\, \zeta^*\bar\lambda^a\zeta \Big) + \hbar \Big(\hbar n_1\, \xi^*\lambda^a \cdot \lambda^b \xi - \hbar n_2\, \zeta^* \bar\lambda^a \cdot \bar\lambda^b \zeta \Big)\,.
\]
Recognising the charge expectation values $\langle \xi\,\zeta|\C^a\C^b | \xi\,\zeta\rangle_{[n_1,n_2]}$ from \eqn~\eqref{clas-charge-SU3}, this can be written as 
\begin{multline}
\langle \xi\,\zeta|\C^a\C^b | \xi\,\zeta\rangle_{[n_1, n_2]}  = \langle \xi\,\zeta|\C^a| \xi\,\zeta\rangle_{[n_1,n_2]}  \langle \xi\,\zeta|\C^b | \xi\,\zeta\rangle_{[n_1,n_2]} \\ + \hbar \left( \hbar n_1 \, \xi^*\lambda^a\cdot \lambda ^b \xi - \hbar n_2\, \zeta^* \bar\lambda^a\cdot \bar\lambda^b \zeta \right).
\end{multline}
The finite quantity in the classical limit $\hbar \to 0, \, n_i \to \infty$ is the product $\hbar n_i$. The term inside the brackets on the second line is itself finite, but comes with a lone $\hbar$ coefficient, and thus vanishes in the classical limit. This then proves the factorisation property in \eqn~\eqref{factorization-charges}.

\section{Diagrams and amplitude expressions}
\label{app:expressions}
In this appendix we gather the expressions necessary for calculating the 1-loop partial amplitudes introduced in section~\ref{NLOColor}. We are only interested in the leading classical terms of the relevant topologies, and will not list quantum corrections.

From \eqn~\eqref{eqn:1loopDecomposition} we know that the relevant 1-loop topologies for NLO observables in YM theory are those which contributed to the analogous QED calculation in \cite{Kosower:2018adc}. The QED amplitude
\begin{equation}
\Aqed = B + C + T_{12} + T_{21}\label{eqn:Aqed}
\end{equation}
is constructed from the triangle $T_{ij}$, box $B$, and cross box $C$ diagrams. Beginning with the triangles, the leading classical terms are
\[\label{eqn:triangles}
iT_{12} &= \hspace{-10pt} \scalebox{0.75}{\begin{tikzpicture}[baseline={([yshift=-\the\dimexpr\fontdimen22\textfont2\relax] current bounding box.center)}, decoration={markings,mark=at position 0.6 with {\arrow{Stealth}}}]
	\begin{feynman}
	\vertex (v1);
	\vertex [below left = 0.7 and 1.4 of v1] (v2);
	\vertex [above left = 0.7 and 1.4 of v1] (v3);
	\vertex [above right = 1.6 and 1.2 of v1] (o1);
	\vertex [above right = 1.6 and 0.7 of v1] (o11) {$p_2 - q$};
	\vertex [below right = 1.6 and 1.2 of v1] (i1) {$p_2$};
	\vertex [above left = 1 and 0.8 of v3] (o2);
	\vertex [above left = 1 and 0.4 of v3] (o22) {$p_1 + q$};
	\vertex [below left = 1 and 0.8 of v2] (i2) {$p_1$};
	\draw [postaction={decorate}] (i1) -- (v1);
	\draw [postaction={decorate}] (v1) -- (o1);
	\draw [postaction={decorate}] (i2) -- (v2);
	\draw [postaction={decorate}] (v2) -- (v3);
	\draw [postaction={decorate}] (v3) -- (o2);
	\diagram*{(v2) -- [white, scalar, momentum'={[arrow style = black]\(\ell\)}] (v1)};
	\diagram*{(v1) -- [gluon] (v2)};
	\diagram*{(v1) -- [gluon] (v3)};
	\end{feynman}	
	\end{tikzpicture}} \hspace{-10pt} = i\frac{2m_1^2}{\hbar} \! \int\! \dd  ^4 \barl\, \frac{ \del (p_1\cdot \barl)}{\barl^2(\barl-\barq)^2} \,+ \mathcal{O}(\hbar^0)\,,\\
iT_{21} &= \hspace{-10pt} \scalebox{0.75}{\begin{tikzpicture}[scale=1.5, baseline={([yshift=-\the\dimexpr\fontdimen22\textfont2\relax] current bounding box.center)}, decoration={markings,mark=at position 0.6 with {\arrow{Stealth}}}]
	\begin{feynman}
	\vertex (v1);
	\vertex [below right = 0.7 and 1.4 of v1] (v2);
	\vertex [above right = 0.7 and 1.4 of v1] (v3);
	\vertex [above left = 1.6 and 1.2 of v1] (o1);
	\vertex [above left = 1.6 and 0.7 of v1] (o11) {$p_1 + q$};
	\vertex [below left = 1.6 and 1.2 of v1] (i1) {$p_1$};
	\vertex [above right = 1 and 0.8 of v3] (o2);
	\vertex [above right = 1 and 0.4 of v3] (o22) {$p_2 - q$};
	\vertex [below right = 1 and 0.8 of v2] (i2) {$p_2$};
	\draw [postaction={decorate}] (i1) -- (v1);
	\draw [postaction={decorate}] (v1) -- (o1);
	\draw [postaction={decorate}] (i2) -- (v2);
	\draw [postaction={decorate}] (v2) -- (v3);
	\draw [postaction={decorate}] (v3) -- (o2);
	\diagram*{(v2) -- [gluon, momentum=\(\ell\)] (v1)};
	\diagram*{(v3) -- [gluon] (v1)};
	\end{feynman}	
	\end{tikzpicture}} \hspace{-10pt} = i\frac{2m_2^2}{\hbar}\! \int\! \dd  ^4 \barl\, \frac{ \del (p_2\cdot \barl)}{\barl^2(\barl-\barq)^2} \,+ \mathcal{O}(\hbar^0)\,.
\]
We refer the curious reader to Ref.~\cite{Kosower:2018adc} for the detailed calculations: heuristically, the $\hbar$ expansion of the diagrams is conducted by rescaling $\ell\to \hbar \barl$ and $q\to \hbar \barq$ on the support of the delta functions in \eqn~\eqref{eqn:COLimpkernel}. Propagator denominators are expanded as a series in $\hbar$. Noting that the loop integrals are symmetric under the replacement $\barl \to \barq - \barl$, this change of variables can be exploited to change the sign of the (Feynman) $i\epsilon$ in massive propagators. Then, averaging over the two expressions for the integral and applying the identities
\begin{equation}
\begin{gathered}
\label{eq:deltadef}
i\delta(x) = \frac{1}{x-i\epsilon } - \frac{1}{x+ i\epsilon }\\
-i\del '(x) = \frac{1}{(x-i\epsilon)^2 } - \frac{1}{(x+ i\epsilon)^2}
\end{gathered}
\end{equation}
leads to the expressions in \eqn~\eqref{eqn:triangles}. This symmetrisation trick is unnecessary for calculating the individual terms from box topologies, for which we choose the following momentum routing:
\begin{equation}
iB = \hspace{-10pt} \scalebox{0.85}{\begin{tikzpicture}[scale=1, baseline={([yshift=-\the\dimexpr\fontdimen22\textfont2\relax] current bounding box.center)}, decoration={markings,mark=at position 0.6 with {\arrow{Stealth}}}]
\begin{feynman}
\vertex (v1);
\vertex [right = 1.42 of v1] (v2);
\vertex [above = 1.4 of v1] (v3);
\vertex [right = 1.42 of v3] (v4);
\vertex [above left = 0.8 and 0.8 of v3] (o1);
\vertex [above left = 0.8 and 0.1 of v3] (o11) {$p_1 + q$};
\vertex [below left = 0.8 and 0.8 of v1] (i1) {$p_1$};
\vertex [above right = 0.8 and 0.8 of v4] (o2);
\vertex [above right = 0.8 and 0.1 of v4] (o22) {$p_2 - q$};
\vertex [below right = 0.8 and 0.8 of v2] (i2) {$p_2$};
\draw [postaction={decorate}] (i1) -- (v1);
\draw [postaction={decorate}] (v1) -- (v3);
\draw [postaction={decorate}] (v3) -- (o1);
\draw [postaction={decorate}] (i2) -- (v2);
\draw [postaction={decorate}] (v2) -- (v4);
\draw [postaction={decorate}] (v4) -- (o2);
\diagram*{(v2) -- [gluon, momentum=\(\ell\)] (v1)};
\diagram*{(v3) -- [gluon] (v4)};
\end{feynman}	
\end{tikzpicture}} \quad iC = \hspace{-10pt} \scalebox{0.85}{\begin{tikzpicture}[scale=1, baseline={([yshift=-\the\dimexpr\fontdimen22\textfont2\relax] current bounding box.center)}, decoration={markings,mark=at position 0.6 with {\arrow{Stealth}}}]
\begin{feynman}
\vertex (v1);
\vertex [right = 1.4 of v1] (v2);
\vertex [above = 1.4 of v1] (v3);
\vertex [right = 1.4 of v3] (v4);
\vertex [above left = 0.8 and 0.8 of v3] (o1);
\vertex [above left = 0.8 and 0.1 of v3] (o11) {$p_1 + q$};
\vertex [below left = 0.8 and 0.8 of v1] (i1) {$p_1$};
\vertex [above right = 0.8 and 0.8 of v4] (o2);
\vertex [above right = 0.8 and 0.1 of v4] (o22) {$p_2 - q$};
\vertex [below right = 0.8 and 0.8 of v2] (i2) {$p_2$};
\vertex [above right = 0.6 and 0.6 of v1] (g1);
\vertex [below left = 0.6 and 0.6 of v4] (g2);
\draw [postaction={decorate}] (i1) -- (v1);
\draw [postaction={decorate}] (v1) -- (v3);
\draw [postaction={decorate}] (v3) -- (o1);
\draw [postaction={decorate}] (i2) -- (v2);
\draw [postaction={decorate}] (v2) -- (v4);
\draw [postaction={decorate}] (v4) -- (o2);
\diagram*{(v4) -- [gluon] (g2)};
\diagram*{(g1) -- [gluon,momentum = \(\ell\)] (v1)};
\diagram*{(v2) -- [gluon] (v3)};
\end{feynman}
\end{tikzpicture}}	\quad i\cut{B} = \hspace{-10pt} \scalebox{0.85}{\begin{tikzpicture}[scale=1, baseline={([yshift=-\the\dimexpr\fontdimen22\textfont2\relax] current bounding box.center)}, decoration={markings,mark=at position 0.6 with {\arrow{Stealth}}}]
	\begin{feynman}
	\vertex (v1);
	\vertex [right = 1.42 of v1] (v2);
	\vertex [above = 1.4 of v1] (v3);
	\vertex [right = 1.42 of v3] (v4);
	\vertex [above left = 0.8 and 0.8 of v3] (o1);
	\vertex [above left = 0.8 and 0.1 of v3] (o11) {$p_1 + q$};
	\vertex [below left = 0.8 and 0.8 of v1] (i1) {$p_1$};
	\vertex [above right = 0.8 and 0.8 of v4] (o2);
	\vertex [above right = 0.8 and 0.1 of v4] (o22) {$p_2 - q$};
	\vertex [below right = 0.8 and 0.8 of v2] (i2) {$p_2$};
	\vertex [above = 0.7 of v1] (g1);
	\vertex [above = 0.7 of v2] (g2);
	\vertex [above left = 0.7 and 0.3 of v1] (g3);
	\vertex [above right = 0.7 and 0.3 of v2] (g4);
	\draw [postaction={decorate}] (i1) -- (v1);
	\draw [postaction={decorate}] (v1) -- (g1);
	\draw [postaction={decorate}] (g1) -- (v3);
	\draw [postaction={decorate}] (v3) -- (o1);
	\draw [postaction={decorate}] (i2) -- (v2);
	\draw [postaction={decorate}] (v2) -- (g2);
	\draw [postaction={decorate}] (g2) -- (v4);
	\draw [postaction={decorate}] (v4) -- (o2);
	\diagram*{(v2) -- [gluon, momentum=\(\ell\)] (v1)};
	\diagram*{(v3) -- [gluon, momentum=\(\ell - q\)] (v4)};
	\filldraw [color=white] ($  (g1) - (0, 2pt) $) rectangle ($ (g2) + (0,2pt) $) ;
	\draw [dashed] (g3) -- (g4);
	\end{feynman}	
	\end{tikzpicture}}
\end{equation}
The cut box appears outside of $\Aqed$, forming the quadratic part of the NLO colour kernel, and is defined in \eqn~\eqref{eqn:cutBox}. However, these diagrams all have kinematic coefficients of the form 
\begin{equation}
D = D_{-1} + D_0 + \mathcal{O}(\hbar^0)\,,
\end{equation}
where $D_{-1} \sim \mathcal O (\hbar^{-2})$ and $D_{0} \sim \mathcal O (\hbar^{-1})$. We choose to label the terms like this as the $\mathcal{O}(\hbar^{-1})$ terms are those which generally contribute classically, and we ignore the $ \mathcal{O}(\hbar^0)$ terms as they always act as quantum corrections.  The $\mathcal{O}(\hbar^{-2})$ terms would give rise to contributions to the impulse which are classically singular, and as shown in section~\ref{NLOColor} it is necessary to consider all three diagrams in order to see that they cancel. We have
\[
B_{-1} &= i\frac{4 (p_1\cdot p_2)^2}{\hbar^2}\! \int\! \dd^4 \barl \, \frac{1}{\barl^2 (\barq - \barl)^2 (p_2\cdot \barl - i\epsilon)(p_1\cdot \barl +i\epsilon)}\,,
\\
C_{-1} &= - i\frac{4 (p_1\cdot p_2)^2}{\hbar^2} \!\int\! \dd ^4 \barl\, \frac{1}{\barl^2 (\barq - \barl)^2 }  \frac{1}{(p_2\cdot \barl + i\epsilon)(p_1\cdot \barl + i\epsilon)}\,,
\\
\cut B_{-1} &= -i\frac{4 (p_1\cdot p_2)^2}{\hbar^2} \!\int\! \dd ^4 \barl \,  \frac{\del (p_1\cdot \barl)\del (p_2\cdot \barl)}{\barl^2 (\barq - \barl)^2 }\,,\label{eqn:singTerms}
\]
and 
\[
B_0 &= i\frac{2 p_1\cdot p_2}{\hbar} \!\int\! \dd ^4 \barl\, \frac{1}{\barl^2 (\barq - 			\barl)^2 (p_1\cdot \barl +i\epsilon)(p_2\cdot \barl - i\epsilon)} 
\\
&\qquad\qquad\times\left\{ 2(p_2-p_1)\cdot \barl +(p_1\cdot p_2) \barl^2 \left(\frac{1}{(p_2\cdot \barl - i\epsilon) } - \frac{1}{(p_1\cdot \barl + i\epsilon) } \right)\right\},
\\
C_0 &= -i\frac{2 p_1\cdot p_2}{\hbar} \!\int\! \dd ^4 \barl\, \frac{1}{\barl^2 (\barq - 			\barl)^2 (p_1\cdot \barl +i\epsilon)(p_2\cdot \barl + i\epsilon)} 
\\
&\qquad\qquad\times\left\{ 2(p_1+p_2)\cdot \barl -(p_1\cdot p_2)  \left(\frac{\barl^2}{(p_1\cdot \barl + i\epsilon) } + \frac{\barl^2  -2\barq\cdot\barl}{(p_2\cdot \barl + i\epsilon) } \right)  \right\},
\\
\cut{B}_{0} &= i\frac{2 (p_1\cdot p_2)^2}{\hbar} \!\int\! \dd ^4 \barl  \, \frac1{\barl^2 (\barq - \barl)^2 } \barl^2 \left\{  \del (p_1\cdot \barl)\del '(p_2\cdot \barl) - \del '(p_1\cdot \barl) \del (p_2\cdot \barl) \right\}.\label{eqn:clasTerms}
\]
Here the delta functions in $\cut{B}$ originate in its definition as the quadratic part of the colour kernel in \eqn~\eqref{eqn:NLOcolKnl}. Applying the symmetrisation trick and \eqn~\eqref{eq:deltadef} ensures that the sums of the box and cross box contributions can also be recast in terms of delta functions --- hence, upon including the triangle contributions, the leading terms in the expansion of the QED amplitude in \eqn~\eqref{eqn:Aqed} are
\[\label{eqn:b+cCombo}
\Aqed_{-1} &= i\frac{2 (p_1\cdot p_2)^2}{\hbar^2} \!\int\! \dd ^4 \barl\, \frac{1}{\barl^2 (\barq - \barl)^2 } \del (p_1\cdot \barl)\del (p_2\cdot \barl)\\
\Aqed_0 &= \frac{2}{\hbar} \!\int\! \dd ^4 \barl\, \frac{1}{\barl^2 (\barq - \barl)^2 } \Bigg\{i(p_1\cdot p_2)^2\bigg[\frac{1}{2}( \barl^2 - 2\barl \cdot \barq ) \Big( \del (p_1\cdot \barl) \del '(p_2\cdot \barl)
\\ &\quad - \del '(p_1\cdot \barl) \del (p_2\cdot \barl) \Big) - (\barl^2 - \barl\cdot\barq ) \left(\frac{i\del (p_1\cdot\barl)}{(p_2 \cdot \barl -i\epsilon)^2} + \frac{i\del (p_2\cdot\barl)}{(p_1 \cdot \barl +i\epsilon)^2} \right) \bigg]\\ 
& \qquad\qquad\qquad + m_1^2\, \del(p_1\cdot\barl) + m_2^2\, \del(p_2\cdot\barl)\Bigg\}\,.
\]
A similar averaging procedure can also be applied to the cut box, yielding
\begin{equation}
\cut{B}_{0} =- i\frac{2( p_1\cdot p_2)^2}{\hbar} \!\int\! \dd ^4 \barl  \, \frac{\barl\cdot\barq}{\barl^2 (\barq - \barl)^2 } \left\{  \del (p_1\cdot \barl)\del '(p_2\cdot \barl) - \del '(p_1\cdot \barl) \del (p_2\cdot \barl) \right\},
\end{equation}
which is the result listed in \eqn~\eqref{eqn:cutBox}.

Finally, for the momentum impulse the cut box $\cut{B}^{\mu}$ is dressed by a power of the loop momentum, and thus
\[
\cut B_{-1} ^{\mu} &= -i\frac{4 (p_1\cdot p_2)^2}{\hbar^2} \!\int \!\dd ^4 \barl\,   \frac{\barl^{\mu}\del (p_1\cdot \barl)\del (p_2\cdot \barl)}{\barl^2 (\barq - \barl)^2 }\,,
\\
\cut B_{0} ^{\mu} &= i\frac{2 (p_1\cdot p_2)^2}{\hbar} \!\int \!\dd ^4 \barl\,  \frac{\barl^{\mu}}{\barl^2 (\barq - \barl)^2 }  \left\{ \barl^2 \del (p_1\cdot \barl)\del '(p_2\cdot \barl) - \barl^2 \del (p_2\cdot \barl)\del '(p_1\cdot \barl)\right\}.
\]
Shifting with $\barl \to \barq - \barl$ and averaging the two expressions, these can be written equivalently as
\[
\cut B_{-1}^{\mu} &= - i\frac{2g^4 (p_1\cdot p_2)^2}{\hbar^2} \int \!\dd ^4 \barl\, \frac{\barq^{\mu} }{\barl^2 (\barq - \barl)^2 } \del (p_1\cdot \barl)\del (p_2\cdot \barl)\,,
\\ 
\cut B_{0}^{\mu} &=  i\frac{2 (p_1\cdot p_2)^2}{\hbar} \int\! \dd ^4 \barl\, \frac{\barl\cdot (\barl-\barq)}{\barl^2 (\barq - \barl)^2 } \barl^{\mu}  \left(\del (p_1\cdot \barl)\del '(p_2\cdot \barl) - \del '(p_1\cdot \barl)\del (p_2\cdot \barl) \right)
\\
& \quad - i\,\frac{ (p_1\cdot p_2)^2}{\hbar} \barq ^{\mu}\! \int \dd ^4 \barl\, \frac{ (\barl^2 -2\barq \cdot \barl)}{\barl^2 (\barq - \barl)^2 } \left(\del (p_1\cdot \barl)\del '(p_2\cdot \barl) - \del '(p_1\cdot \barl)\del (p_2\cdot \barl) \right).
\]

Applying the analysis of \cite{Kosower:2018adc}, one would expect the non-Abelian triangles
\begin{equation}
iY_{12} = \hspace{-10pt} \scalebox{0.8}{\begin{tikzpicture}[scale=1.5, baseline={([yshift=-\the\dimexpr\fontdimen22\textfont2\relax] current bounding box.center)}, decoration={markings,mark=at position 0.6 with {\arrow{Stealth}}}]
\begin{feynman}
\vertex (v1);
\vertex [left = 0.96 of v1] (g1);
\vertex [below left = 0.7 and 1.24 of g1] (v2);
\vertex [above left = 0.7 and 1.24 of g1] (v3);
\vertex [above right = 1.6 and 1.2 of v1] (o1);
\vertex [above right = 1.6 and 0.7 of v1] (o11) {$p_2 - q$};
\vertex [below right = 1.6 and 1.2 of v1] (i1) {$p_2$};
\vertex [above left = 1 and 0.8 of v3] (o2);
\vertex [above left = 1 and 0.4 of v3] (o22) {$p_1 + q$};
\vertex [below left = 1 and 0.8 of v2] (i2) {$p_1$};
\draw [postaction={decorate}] (i1) -- (v1);
\draw [postaction={decorate}] (v1) -- (o1);
\draw [postaction={decorate}] (i2) -- (v2);
\draw [postaction={decorate}] (v2) -- (v3);
\draw [postaction={decorate}] (v3) -- (o2);
\diagram*{(v1) -- [gluon,momentum=\(q\)] (g1)};
\diagram*{(v2) -- [gluon,momentum'=\(\ell\)] (g1)};
\diagram*{(g1) -- [gluon] (v3)};
\filldraw [color=black] (g1) circle [radius=0.5pt];
\end{feynman}	
\end{tikzpicture}} \qquad iY_{21} = \hspace{-10pt} \scalebox{0.8}{\begin{tikzpicture}[scale=1.5, baseline={([yshift=-\the\dimexpr\fontdimen22\textfont2\relax] current bounding box.center)}, decoration={markings,mark=at position 0.6 with {\arrow{Stealth}}}]
\begin{feynman}
\vertex (v1);
\vertex [right = 0.96 of v1] (g1);
\vertex [below right = 0.7 and 1.24 of g1] (v2);
\vertex [above right = 0.7 and 1.24 of g1] (v3);
\vertex [above left = 1.6 and 1.2 of v1] (o1);
\vertex [above left = 1.6 and 0.7 of v1] (o11) {$p_1 + q$};
\vertex [below left = 1.6 and 1.2 of v1] (i1) {$p_1$};
\vertex [above right = 1 and 0.8 of v3] (o2);
\vertex [above right = 1 and 0.4 of v3] (o22) {$p_2 - q$};
\vertex [below right = 1 and 0.8 of v2] (i2) {$p_2$};
\draw [postaction={decorate}] (i1) -- (v1);
\draw [postaction={decorate}] (v1) -- (o1);
\draw [postaction={decorate}] (i2) -- (v2);
\draw [postaction={decorate}] (v2) -- (v3);
\draw [postaction={decorate}] (v3) -- (o2);
\diagram*{(v1) -- [gluon, momentum'=\(q\)] (g1)};
\diagram*{(v2) -- [white, scalar, momentum={[arrow style = black]\(\ell\)}] (g1)};
\diagram*{(g1) -- [gluon] (v2)};
\diagram*{(v3) -- [gluon] (g1)};
\filldraw [color=black] (g1) circle [radius=0.5pt];
\end{feynman}	
\end{tikzpicture}}
\end{equation}
to contribute to the NLO observables. The series in $\hbar$ for these partial amplitudes contain terms at $\mathcal O(\hbar^{-2})$; however, the decomposition of the full amplitude onto the colour basis in \eqn~\eqref{eqn:1loopDecomposition} shows that these always act as quantum corrections, and thus we need not calculate these diagrams.

\section{Colour deflection for $N$ particles}\label{app:Nparticles}
\newcommand{\colKnlCl}{{\mathcal{H}}}
Here we give the full general $N$ particle results for the colour deflection. The strategy to  perform the calculation is 
iterative and follows  Ref.~\cite{Goldberger:2016iau}.  However, here we have not introduced an extra integration and performed a sum over integration 
labels as in \cite{Goldberger:2016iau}. As can be seen in equations~\eqref{defle1},  we are  removing self-interactions. At NLO this leads to an
important distinction between sums over particle species. Accordingly, in the following expressions we have separated the contributions 
according to the type of sum involved. Our result for the NLO colour deflection is
\[\label{ColorDef-2}
\Delta^{(2)}c_{\alpha}^a(\tau_{\alpha}) & = 
\sum_{ \substack{ \beta=1, \beta\ne \alpha \\ \gamma=1, \gamma\ne \alpha}  }^N \!\int \!\dd^4 \bq_2 
\!\int \!\dd^4 \bq_3\,  \hdelta(\bq_3\cdot u_{\gamma}) \hdelta(\bq_2\cdot u_{\beta}) e^{\ii \bq_3 \cdot b_{\gamma}} e^{\ii \bq_2 \cdot b_{\beta}} \\
&\qquad\qquad\qquad\qquad \times e^{-\ii (\bq_2+\bq_3) \cdot (b_{\alpha} + u_{\alpha} \tau_{\alpha})}
   \colKnlCl_{A}^{a,(2)}(\bq_1, \bq_3; u_{\alpha}, u_{\gamma}, u_{\beta})\\
&+ \sum\limits_{\substack{\beta=1, \beta\ne \alpha \\{\gamma}=1,  \gamma \ne {\beta}}}^N  \int \!\dd^4 \bq_2 \int \!\dd^4 \bq_3  \, \hat{\delta}(\bq_3\cdot u_{\gamma}) \hat{\delta }\left(\bq_2\cdot u_{\beta}\right) e^{\ii \bq_3 \cdot b_{\gamma}} e^{\ii \bq_2 \cdot b_{\beta}} \\
&\qquad\qquad\qquad\qquad \times e^{-\ii (\bq_2+\bq_3) \cdot (b_{\alpha} + u_{\alpha} \tau_{\alpha})}
 \colKnlCl_{B}^{a,(2)}(\bq_1, \bq_3; u_{\alpha}, u_{\gamma}, u_{\beta})\,,
\]
where $\bq_{ij\dots}=\bq_i+\bq_j+\cdots$ and 
\begin{equation}
\text{tr}(A,B,C,D)\equiv  \frac{1}{4} \text{tr}(\slashed{A} \slashed{B} \slashed{C}\slashed{D})\,.
\end{equation}
Here,
\[ \label{eqn:ColKernel_general}
&\colKnlCl_{A}^{a,(2)}(\bq_1, \bq_3; u_{\alpha}, u_{\gamma}, u_{\beta})=
 \ii \frac{f^{abc} c_{\gamma}^b c_{\alpha}^c c_{\alpha}\cdot c_{\beta}}{m_{\alpha}\, \bar{q}_2^2 \bar{q}_3^2\, \bar{q}_{23}\cdot u_{\alpha} 
 (\bar{q}_2\cdot u_{\alpha})^2}\\
 & \times \left( u_{\alpha}\cdot u_{\gamma} \text{tr}(\bar{q}_2, \bar{q}_3, u_{\alpha}, u_{\beta})+
 \bar{q}_2 \cdot u_{\alpha} \text{tr}(\bar{q}_2, u_{\beta},  u_{\alpha},  u_{\gamma}) 
 \right)\\
& \qquad\qquad\qquad\qquad - f^{abc} f^{cde} c_{\gamma}^b c_{\beta}^d c_{\alpha}^e 
\left(\frac{u_{\alpha}\cdot u_{\beta} u_{\alpha}\cdot u_{\gamma}}{\bar{q}_2^2 \bar{q}_3^2 \bar{q}_2\cdot u_{\alpha} \bar{q}_{23}\cdot u_{\alpha}}\right),\\
& \colKnlCl_{B}^{a,(2)}(\bq_1, \bq_3; u_{\alpha}, u_{\gamma}, u_{\beta})= 
\ii   \frac{f^{abc} c_{\beta}^b c_{\alpha}^c c_{\beta}\cdot c_{\gamma}}{m_{\beta}\, \bar{q}_3^2 \bar{q}_{23}^2\, \bar{q}_{23}\cdot u_{\alpha} 
(\bar{q}_3\cdot u_{\beta})^2} 
\bigg(
u_{\alpha}\cdot u_{\beta} \text{tr}(\bar{q}_{23}, u_{\beta}, \bar{q}_3, u_{\gamma})+ \\
& \bar{q}_3 \cdot u_{\beta} \text{tr}(\bar{q}_3, u_{\alpha},  u_{\beta},  u_{\gamma})
\bigg)+\!\frac{f^{abc} f^{bde} c_{\alpha}^c c_{\gamma}^d c_{\beta}^e}{\bar{q}_3^2 \bar{q}_{23}^2 \, \bar{q}_{23}\cdot u_{\alpha}}
\\& \qquad\qquad \times \left( -\frac{u_{\beta}\cdot u_{\gamma} \bar{q}_3\cdot u_{\alpha}}{\bar{q}_2^2}+\frac{2 u_{\alpha}\cdot u_{\gamma} \bar{q}_3\cdot u_{\beta}}{\bar{q}_2^2}-\frac{u_{\alpha}\cdot u_{\beta} u_{\beta}\cdot u_{\gamma}}{\bar{q}_3\cdot u_{\beta}} \right).
\]
In order to reach the form  of the colour impulse  in section \ref{sec:ClassEOM} we set $N=2$ and perform the time integration on
the support of the on-shell conditions. To recover the final observables we define the loop momentum 
as $\bl\equiv\bq_3$ and $\bl\equiv\bq_2+\bq_3$ in the first and second contributions in \eqref{eqn:ColKernel_general} respectively.

\bibliographystyle{JHEP}

\begin{thebibliography}{100}

\bibitem{Babak:2017tow}
S.~Babak, J.~Gair, A.~Sesana, E.~Barausse, C.~F. Sopuerta, C.~P. Berry,
  E.~Berti, P.~Amaro-Seoane, A.~Petiteau, and A.~Klein, {\it {Science with the
  space-based interferometer LISA. V: Extreme mass-ratio inspirals}},  {\em
  Phys. Rev. D} {\bf 95} (2017), no.~10 103012,
  [\href{http://arxiv.org/abs/1703.09722}{{\tt arXiv:1703.09722}}].

\bibitem{Donoghue:1993eb}
J.~F. Donoghue, {\it {Leading quantum correction to the Newtonian potential}},
  {\em Phys. Rev. Lett.} {\bf 72} (1994) 2996--2999,
  [\href{http://arxiv.org/abs/gr-qc/9310024}{{\tt gr-qc/9310024}}].

\bibitem{Donoghue:1994dn}
J.~F. Donoghue, {\it {General relativity as an effective field theory: The
  leading quantum corrections}},  {\em Phys. Rev.} {\bf D50} (1994) 3874--3888,
  [\href{http://arxiv.org/abs/gr-qc/9405057}{{\tt gr-qc/9405057}}].

\bibitem{Donoghue:1996mt}
J.~F. Donoghue and T.~Torma, {\it {On the power counting of loop diagrams in
  general relativity}},  {\em Phys. Rev.} {\bf D54} (1996) 4963--4972,
  [\href{http://arxiv.org/abs/hep-th/9602121}{{\tt hep-th/9602121}}].

\bibitem{Holstein:2004dn}
B.~R. Holstein and J.~F. Donoghue, {\it {Classical Physics and Quantum Loops}},
   {\em Phys. Rev. Lett.} {\bf 93} (2004) 201602,
  [\href{http://arxiv.org/abs/hep-th/0405239}{{\tt hep-th/0405239}}].

\bibitem{Neill:2013wsa}
D.~Neill and I.~Z. Rothstein, {\it {Classical Space-Times from the S Matrix}},
  {\em Nucl. Phys.} {\bf B877} (2013) 177--189,
  [\href{http://arxiv.org/abs/1304.7263}{{\tt arXiv:1304.7263}}].

\bibitem{Bjerrum-Bohr:2013bxa}
N.~E.~J. Bjerrum-Bohr, J.~F. Donoghue, and P.~Vanhove, {\it {On-shell
  Techniques and Universal Results in Quantum Gravity}},  {\em JHEP} {\bf 02}
  (2014) 111, [\href{http://arxiv.org/abs/1309.0804}{{\tt arXiv:1309.0804}}].

\bibitem{Monteiro:2014cda}
R.~Monteiro, D.~O'Connell, and C.~D. White, {\it {Black holes and the double
  copy}},  {\em JHEP} {\bf 12} (2014) 056,
  [\href{http://arxiv.org/abs/1410.0239}{{\tt arXiv:1410.0239}}].

\bibitem{Luna:2016due}
A.~Luna, R.~Monteiro, I.~Nicholson, D.~O'Connell, and C.~D. White, {\it {The
  double copy: Bremsstrahlung and accelerating black holes}},  {\em JHEP} {\bf
  06} (2016) 023, [\href{http://arxiv.org/abs/1603.05737}{{\tt
  arXiv:1603.05737}}].

\bibitem{Damour:2016gwp}
T.~Damour, {\it {Gravitational scattering, post-Minkowskian approximation and
  Effective One-Body theory}},  {\em Phys. Rev.} {\bf D94} (2016), no.~10
  104015, [\href{http://arxiv.org/abs/1609.00354}{{\tt arXiv:1609.00354}}].

\bibitem{Goldberger:2016iau}
W.~D. Goldberger and A.~K. Ridgway, {\it {Radiation and the classical double
  copy for color charges}},  {\em Phys. Rev. D} {\bf 95} (2017), no.~12 125010,
  [\href{http://arxiv.org/abs/1611.03493}{{\tt arXiv:1611.03493}}].

\bibitem{Cachazo:2017jef}
F.~Cachazo and A.~Guevara, {\it {Leading Singularities and Classical
  Gravitational Scattering}},  {\em JHEP} {\bf 02} (2020) 181,
  [\href{http://arxiv.org/abs/1705.10262}{{\tt arXiv:1705.10262}}].

\bibitem{Laddha:2018rle}
A.~Laddha and A.~Sen, {\it {Gravity Waves from Soft Theorem in General
  Dimensions}},  {\em JHEP} {\bf 09} (2018) 105,
  [\href{http://arxiv.org/abs/1801.07719}{{\tt arXiv:1801.07719}}].

\bibitem{Cheung:2018wkq}
C.~Cheung, I.~Z. Rothstein, and M.~P. Solon, {\it {From Scattering Amplitudes
  to Classical Potentials in the Post-Minkowskian Expansion}},  {\em Phys. Rev.
  Lett.} {\bf 121} (2018), no.~25 251101,
  [\href{http://arxiv.org/abs/1808.02489}{{\tt arXiv:1808.02489}}].

\bibitem{Kosower:2018adc}
D.~A. Kosower, B.~Maybee, and D.~O'Connell, {\it {Amplitudes, Observables, and
  Classical Scattering}},  {\em JHEP} {\bf 02} (2019) 137,
  [\href{http://arxiv.org/abs/1811.10950}{{\tt arXiv:1811.10950}}].

\bibitem{Bern:2019nnu}
Z.~Bern, C.~Cheung, R.~Roiban, C.-H. Shen, M.~P. Solon, and M.~Zeng, {\it
  {Scattering Amplitudes and the Conservative Hamiltonian for Binary Systems at
  Third Post-Minkowskian Order}},  {\em Phys. Rev. Lett.} {\bf 122} (2019),
  no.~20 201603, [\href{http://arxiv.org/abs/1901.04424}{{\tt
  arXiv:1901.04424}}].

\bibitem{Antonelli:2019ytb}
A.~Antonelli, A.~Buonanno, J.~Steinhoff, M.~van~de Meent, and J.~Vines, {\it
  {Energetics of two-body Hamiltonians in post-Minkowskian gravity}},  {\em
  Phys. Rev.} {\bf D99} (2019), no.~10 104004,
  [\href{http://arxiv.org/abs/1901.07102}{{\tt arXiv:1901.07102}}].

\bibitem{Bjerrum-Bohr:2014zsa}
N.~E.~J. Bjerrum-Bohr, J.~F. Donoghue, B.~R. Holstein, L.~Plant\'{e}, and
  P.~Vanhove, {\it {Bending of Light in Quantum Gravity}},  {\em Phys. Rev.
  Lett.} {\bf 114} (2015), no.~6 061301,
  [\href{http://arxiv.org/abs/1410.7590}{{\tt arXiv:1410.7590}}].

\bibitem{Bjerrum-Bohr:2017dxw}
N.~E.~J. Bjerrum-Bohr, B.~R. Holstein, J.~F. Donoghue, L.~Plant\'{e}, and
  P.~Vanhove, {\it {Illuminating Light Bending}},  {\em PoS} {\bf CORFU2016}
  (2017) 077, [\href{http://arxiv.org/abs/1704.01624}{{\tt arXiv:1704.01624}}].

\bibitem{Damour:2017zjx}
T.~Damour, {\it {High-energy gravitational scattering and the general
  relativistic two-body problem}},  {\em Phys. Rev.} {\bf D97} (2018), no.~4
  044038, [\href{http://arxiv.org/abs/1710.10599}{{\tt arXiv:1710.10599}}].

\bibitem{Luna:2017dtq}
A.~Luna, I.~Nicholson, D.~O'Connell, and C.~D. White, {\it {Inelastic Black
  Hole Scattering from Charged Scalar Amplitudes}},  {\em JHEP} {\bf 03} (2018)
  044, [\href{http://arxiv.org/abs/1711.03901}{{\tt arXiv:1711.03901}}].

\bibitem{Laddha:2018myi}
A.~Laddha and A.~Sen, {\it {Logarithmic Terms in the Soft Expansion in Four
  Dimensions}},  {\em JHEP} {\bf 10} (2018) 056,
  [\href{http://arxiv.org/abs/1804.09193}{{\tt arXiv:1804.09193}}].

\bibitem{Laddha:2018vbn}
A.~Laddha and A.~Sen, {\it {Observational Signature of the Logarithmic Terms in
  the Soft Graviton Theorem}},  {\em Phys. Rev. D} {\bf 100} (2019), no.~2
  024009, [\href{http://arxiv.org/abs/1806.01872}{{\tt arXiv:1806.01872}}].

\bibitem{Bjerrum-Bohr:2018xdl}
N.~E.~J. Bjerrum-Bohr, P.~H. Damgaard, G.~Festuccia, L.~Plant\'{e}, and
  P.~Vanhove, {\it {General Relativity from Scattering Amplitudes}},  {\em
  Phys. Rev. Lett.} {\bf 121} (2018), no.~17 171601,
  [\href{http://arxiv.org/abs/1806.04920}{{\tt arXiv:1806.04920}}].

\bibitem{Sahoo:2018lxl}
B.~Sahoo and A.~Sen, {\it {Classical and Quantum Results on Logarithmic Terms
  in the Soft Theorem in Four Dimensions}},  {\em JHEP} {\bf 02} (2019) 086,
  [\href{http://arxiv.org/abs/1808.03288}{{\tt arXiv:1808.03288}}].

\bibitem{KoemansCollado:2019ggb}
A.~Koemans~Collado, P.~Di~Vecchia, and R.~Russo, {\it {Revisiting the second
  post-Minkowskian eikonal and the dynamics of binary black holes}},  {\em
  Phys. Rev. D} {\bf 100} (2019), no.~6 066028,
  [\href{http://arxiv.org/abs/1904.02667}{{\tt arXiv:1904.02667}}].

\bibitem{Brandhuber:2019qpg}
A.~Brandhuber and G.~Travaglini, {\it {On higher-derivative effects on the
  gravitational potential and particle bending}},  {\em JHEP} {\bf 01} (2020)
  010, [\href{http://arxiv.org/abs/1905.05657}{{\tt arXiv:1905.05657}}].

\bibitem{Cristofoli:2019neg}
A.~Cristofoli, N.~Bjerrum-Bohr, P.~H. Damgaard, and P.~Vanhove, {\it
  {Post-Minkowskian Hamiltonians in general relativity}},  {\em Phys. Rev. D}
  {\bf 100} (2019), no.~8 084040, [\href{http://arxiv.org/abs/1906.01579}{{\tt
  arXiv:1906.01579}}].

\bibitem{Laddha:2019yaj}
A.~Laddha and A.~Sen, {\it {Classical proof of the classical soft graviton
  theorem in $D > 4$}},  {\em Phys. Rev. D} {\bf 101} (2020), no.~8 084011,
  [\href{http://arxiv.org/abs/1906.08288}{{\tt arXiv:1906.08288}}].

\bibitem{Bern:2019crd}
Z.~Bern, C.~Cheung, R.~Roiban, C.-H. Shen, M.~P. Solon, and M.~Zeng, {\it
  {Black Hole Binary Dynamics from the Double Copy and Effective Theory}},
  {\em JHEP} {\bf 10} (2019) 206, [\href{http://arxiv.org/abs/1908.01493}{{\tt
  arXiv:1908.01493}}].

\bibitem{Damgaard:2019lfh}
P.~H. Damgaard, K.~Haddad, and A.~Helset, {\it {Heavy Black Hole Effective
  Theory}},  {\em JHEP} {\bf 11} (2019) 070,
  [\href{http://arxiv.org/abs/1908.10308}{{\tt arXiv:1908.10308}}].

\bibitem{Kalin:2019rwq}
G.~K{\"a}lin and R.~A. Porto, {\it {From Boundary Data to Bound States}},  {\em
  JHEP} {\bf 01} (2020) 072, [\href{http://arxiv.org/abs/1910.03008}{{\tt
  arXiv:1910.03008}}].

\bibitem{Bjerrum-Bohr:2019kec}
N.~Bjerrum-Bohr, A.~Cristofoli, and P.~H. Damgaard, {\it {Post-Minkowskian
  Scattering Angle in Einstein Gravity}},  {\em JHEP} {\bf 08} (2020) 038,
  [\href{http://arxiv.org/abs/1910.09366}{{\tt arXiv:1910.09366}}].

\bibitem{Kalin:2019inp}
G.~K{\"a}lin and R.~A. Porto, {\it {From boundary data to bound states. Part
  II. Scattering angle to dynamical invariants (with twist)}},  {\em JHEP} {\bf
  02} (2020) 120, [\href{http://arxiv.org/abs/1911.09130}{{\tt
  arXiv:1911.09130}}].

\bibitem{Huber:2019ugz}
M.~Accettulli~Huber, A.~Brandhuber, S.~De~Angelis, and G.~Travaglini, {\it
  {Note on the absence of $R^2$ corrections to Newton's potential}},  {\em
  Phys. Rev. D} {\bf 101} (2020), no.~4 046011,
  [\href{http://arxiv.org/abs/1911.10108}{{\tt arXiv:1911.10108}}].

\bibitem{Saha:2019tub}
A.~P. Saha, B.~Sahoo, and A.~Sen, {\it {Proof of the classical soft graviton
  theorem in $D$ = 4}},  {\em JHEP} {\bf 06} (2020) 153,
  [\href{http://arxiv.org/abs/1912.06413}{{\tt arXiv:1912.06413}}].

\bibitem{Aoude:2020onz}
R.~Aoude, K.~Haddad, and A.~Helset, {\it {On-shell heavy particle effective
  theories}},  {\em JHEP} {\bf 05} (2020) 051,
  [\href{http://arxiv.org/abs/2001.09164}{{\tt arXiv:2001.09164}}].

\bibitem{Bern:2020gjj}
Z.~Bern, H.~Ita, J.~Parra-Martinez, and M.~S. Ruf, {\it {Universality in the
  classical limit of massless gravitational scattering}},  {\em Phys. Rev.
  Lett.} {\bf 125} (2020), no.~3 031601,
  [\href{http://arxiv.org/abs/2002.02459}{{\tt arXiv:2002.02459}}].

\bibitem{Cheung:2020gyp}
C.~Cheung and M.~P. Solon, {\it {Classical gravitational scattering at $
  \mathcal{O} $(G$^{3}$) from Feynman diagrams}},  {\em JHEP} {\bf 06} (2020)
  144, [\href{http://arxiv.org/abs/2003.08351}{{\tt arXiv:2003.08351}}].

\bibitem{Cristofoli:2020uzm}
A.~Cristofoli, P.~H. Damgaard, P.~Di~Vecchia, and C.~Heissenberg, {\it
  {Second-order Post-Minkowskian scattering in arbitrary dimensions}},  {\em
  JHEP} {\bf 07} (2020) 122, [\href{http://arxiv.org/abs/2003.10274}{{\tt
  arXiv:2003.10274}}].

\bibitem{Bern:2020buy}
Z.~Bern, A.~Luna, R.~Roiban, C.-H. Shen, and M.~Zeng, {\it {Spinning Black Hole
  Binary Dynamics, Scattering Amplitudes and Effective Field Theory}},
  \href{http://arxiv.org/abs/2005.03071}{{\tt arXiv:2005.03071}}.

\bibitem{Parra-Martinez:2020dzs}
J.~Parra-Martinez, M.~S. Ruf, and M.~Zeng, {\it {Extremal black hole scattering
  at $O(G^3)$: graviton dominance, eikonal exponentiation, and differential
  equations}},  \href{http://arxiv.org/abs/2005.04236}{{\tt arXiv:2005.04236}}.

\bibitem{Haddad:2020tvs}
K.~Haddad and A.~Helset, {\it {The double copy for heavy particles}},
  \href{http://arxiv.org/abs/2005.13897}{{\tt arXiv:2005.13897}}.

\bibitem{AccettulliHuber:2020oou}
M.~Accettulli~Huber, A.~Brandhuber, S.~De~Angelis, and G.~Travaglini, {\it
  {Eikonal phase matrix, deflection angle and time delay in effective field
  theories of gravity}},  {\em Phys. Rev. D} {\bf 102} (2020), no.~4 046014,
  [\href{http://arxiv.org/abs/2006.02375}{{\tt arXiv:2006.02375}}].

\bibitem{Cheung:2020sdj}
C.~Cheung and M.~P. Solon, {\it {Tidal Effects in the Post-Minkowskian
  Expansion}},  \href{http://arxiv.org/abs/2006.06665}{{\tt arXiv:2006.06665}}.

\bibitem{A:2020lub}
M.~A., D.~Ghosh, A.~Laddha, and A.~P.~V., {\it {Soft Radiation from Scattering
  Amplitudes Revisited}},  \href{http://arxiv.org/abs/2007.02077}{{\tt
  arXiv:2007.02077}}.

\bibitem{Kalin:2020fhe}
G.~K{\"a}lin, Z.~Liu, and R.~A. Porto, {\it {Conservative Dynamics of Binary
  Systems to Third Post-Minkowskian Order from the Effective Field Theory
  Approach}},  \href{http://arxiv.org/abs/2007.04977}{{\tt arXiv:2007.04977}}.

\bibitem{Haddad:2020que}
K.~Haddad and A.~Helset, {\it {Gravitational tidal effects in quantum field
  theory}},  \href{http://arxiv.org/abs/2008.04920}{{\tt arXiv:2008.04920}}.

\bibitem{Kalin:2020lmz}
G.~K{\"a}lin, Z.~Liu, and R.~A. Porto, {\it {Conservative Tidal Effects in
  Compact Binary Systems to Next-to-Leading Post-Minkowskian Order}},
  \href{http://arxiv.org/abs/2008.06047}{{\tt arXiv:2008.06047}}.

\bibitem{Sahoo:2020ryf}
B.~Sahoo, {\it {Classical Sub-subleading Soft Photon and Soft Graviton Theorems
  in Four Spacetime Dimensions}},  \href{http://arxiv.org/abs/2008.04376}{{\tt
  arXiv:2008.04376}}.

\bibitem{DiVecchia:2020ymx}
P.~Di~Vecchia, C.~Heissenberg, R.~Russo, and G.~Veneziano, {\it {Universality
  of ultra-relativistic gravitational scattering}},
  \href{http://arxiv.org/abs/2008.12743}{{\tt arXiv:2008.12743}}.

\bibitem{Bern:2019prr}
Z.~Bern, J.~J. Carrasco, M.~Chiodaroli, H.~Johansson, and R.~Roiban, {\it {The
  Duality Between Color and Kinematics and its Applications}},
  \href{http://arxiv.org/abs/1909.01358}{{\tt arXiv:1909.01358}}.

\bibitem{Luna:2016hge}
A.~Luna, R.~Monteiro, I.~Nicholson, A.~Ochirov, D.~O'Connell, N.~Westerberg,
  and C.~D. White, {\it {Perturbative spacetimes from Yang-Mills theory}},
  {\em JHEP} {\bf 04} (2017) 069, [\href{http://arxiv.org/abs/1611.07508}{{\tt
  arXiv:1611.07508}}].

\bibitem{Goldberger:2017vcg}
W.~D. Goldberger and A.~K. Ridgway, {\it {Bound states and the classical double
  copy}},  {\em Phys. Rev.} {\bf D97} (2018), no.~8 085019,
  [\href{http://arxiv.org/abs/1711.09493}{{\tt arXiv:1711.09493}}].

\bibitem{Luna:2015paa}
A.~Luna, R.~Monteiro, D.~O'Connell, and C.~D. White, {\it {The classical double
  copy for Taub--NUT spacetime}},  {\em Phys. Lett.} {\bf B750} (2015)
  272--277, [\href{http://arxiv.org/abs/1507.01869}{{\tt arXiv:1507.01869}}].

\bibitem{Lee:2018gxc}
K.~Lee, {\it {Kerr-Schild Double Field Theory and Classical Double Copy}},
  {\em JHEP} {\bf 10} (2018) 027, [\href{http://arxiv.org/abs/1807.08443}{{\tt
  arXiv:1807.08443}}].

\bibitem{Luna:2018dpt}
A.~Luna, R.~Monteiro, I.~Nicholson, and D.~O'Connell, {\it {Type D Spacetimes
  and the Weyl Double Copy}},  {\em Class. Quant. Grav.} {\bf 36} (2019)
  065003, [\href{http://arxiv.org/abs/1810.08183}{{\tt arXiv:1810.08183}}].

\bibitem{Adamo:2018mpq}
T.~Adamo, E.~Casali, L.~Mason, and S.~Nekovar, {\it {Plane wave backgrounds and
  colour-kinematics duality}},  {\em JHEP} {\bf 02} (2019) 198,
  [\href{http://arxiv.org/abs/1810.05115}{{\tt arXiv:1810.05115}}].

\bibitem{CarrilloGonzalez:2019gof}
M.~Carrillo~Gonz\'{a}lez, B.~Melcher, K.~Ratliff, S.~Watson, and C.~D. White,
  {\it {The classical double copy in three spacetime dimensions}},
  \href{http://arxiv.org/abs/1904.11001}{{\tt arXiv:1904.11001}}.

\bibitem{Cho:2019ype}
W.~Cho and K.~Lee, {\it {Heterotic Kerr-Schild Double Field Theory and
  Classical Double Copy}},  {\em JHEP} {\bf 07} (2019) 030,
  [\href{http://arxiv.org/abs/1904.11650}{{\tt arXiv:1904.11650}}].

\bibitem{Carrillo-Gonzalez:2019aao}
M.~Carrillo~González, R.~Penco, and M.~Trodden, {\it {Shift symmetries, soft
  limits, and the double copy beyond leading order}},
  \href{http://arxiv.org/abs/1908.07531}{{\tt arXiv:1908.07531}}.

\bibitem{Moynihan:2019bor}
N.~Moynihan, {\it {Kerr-Newman from Minimal Coupling}},  {\em JHEP} {\bf 01}
  (2020) 014, [\href{http://arxiv.org/abs/1909.05217}{{\tt arXiv:1909.05217}}].

\bibitem{Bah:2019sda}
I.~Bah, R.~Dempsey, and P.~Weck, {\it {Kerr-Schild Double Copy and Complex
  Worldlines}},  {\em JHEP} {\bf 02} (2020) 180,
  [\href{http://arxiv.org/abs/1910.04197}{{\tt arXiv:1910.04197}}].

\bibitem{Huang:2019cja}
Y.-T. Huang, U.~Kol, and D.~O'Connell, {\it {Double copy of electric-magnetic
  duality}},  {\em Phys. Rev. D} {\bf 102} (2020), no.~4 046005,
  [\href{http://arxiv.org/abs/1911.06318}{{\tt arXiv:1911.06318}}].

\bibitem{Alawadhi:2019urr}
R.~Alawadhi, D.~S. Berman, B.~Spence, and D.~Peinador~Veiga, {\it {S-duality
  and the double copy}},  {\em JHEP} {\bf 03} (2020) 059,
  [\href{http://arxiv.org/abs/1911.06797}{{\tt arXiv:1911.06797}}].

\bibitem{Borsten:2019prq}
L.~Borsten, I.~Jubb, V.~Makwana, and S.~Nagy, {\it {Gauge × gauge on
  spheres}},  {\em JHEP} {\bf 06} (2020) 096,
  [\href{http://arxiv.org/abs/1911.12324}{{\tt arXiv:1911.12324}}].

\bibitem{Kim:2019jwm}
K.~Kim, K.~Lee, R.~Monteiro, I.~Nicholson, and D.~Peinador~Veiga, {\it {The
  Classical Double Copy of a Point Charge}},  {\em JHEP} {\bf 02} (2020) 046,
  [\href{http://arxiv.org/abs/1912.02177}{{\tt arXiv:1912.02177}}].

\bibitem{Banerjee:2019saj}
A.~Banerjee, E.~Colg\'ain, J.~Rosabal, and H.~Yavartanoo, {\it {Ehlers as EM
  duality in the double copy}},  \href{http://arxiv.org/abs/1912.02597}{{\tt
  arXiv:1912.02597}}.

\bibitem{Bahjat-Abbas:2020cyb}
N.~Bahjat-Abbas, R.~Stark-Much\~ao, and C.~D. White, {\it {Monopoles,
  shockwaves and the classical double copy}},  {\em JHEP} {\bf 04} (2020) 102,
  [\href{http://arxiv.org/abs/2001.09918}{{\tt arXiv:2001.09918}}].

\bibitem{Moynihan:2020gxj}
N.~Moynihan and J.~Murugan, {\it {On-Shell Electric-Magnetic Duality and the
  Dual Graviton}},  \href{http://arxiv.org/abs/2002.11085}{{\tt
  arXiv:2002.11085}}.

\bibitem{Adamo:2020syc}
T.~Adamo, L.~Mason, and A.~Sharma, {\it {MHV scattering of gluons and gravitons
  in chiral strong fields}},  {\em Phys. Rev. Lett.} {\bf 125} (2020), no.~4
  041602, [\href{http://arxiv.org/abs/2003.13501}{{\tt arXiv:2003.13501}}].

\bibitem{Alfonsi:2020lub}
L.~Alfonsi, C.~D. White, and S.~Wikeley, {\it {Topology and Wilson lines:
  global aspects of the double copy}},  {\em JHEP} {\bf 07} (2020) 091,
  [\href{http://arxiv.org/abs/2004.07181}{{\tt arXiv:2004.07181}}].

\bibitem{Luna:2020adi}
A.~Luna, S.~Nagy, and C.~D. White, {\it {The convolutional double copy: a case
  study with a point}},  \href{http://arxiv.org/abs/2004.11254}{{\tt
  arXiv:2004.11254}}.

\bibitem{Keeler:2020rcv}
C.~Keeler, T.~Manton, and N.~Monga, {\it {From Navier-Stokes to Maxwell via
  Einstein}},  {\em JHEP} {\bf 08} (2020) 147,
  [\href{http://arxiv.org/abs/2005.04242}{{\tt arXiv:2005.04242}}].

\bibitem{Elor:2020nqe}
G.~Elor, K.~Farnsworth, M.~L. Graesser, and G.~Herczeg, {\it {The
  Newman-Penrose Map and the Classical Double Copy}},
  \href{http://arxiv.org/abs/2006.08630}{{\tt arXiv:2006.08630}}.

\bibitem{Cristofoli:2020hnk}
A.~Cristofoli, {\it {Gravitational shock waves and scattering amplitudes}},
  \href{http://arxiv.org/abs/2006.08283}{{\tt arXiv:2006.08283}}.

\bibitem{Alawadhi:2020jrv}
R.~Alawadhi, D.~S. Berman, and B.~Spence, {\it {Weyl doubling}},
  \href{http://arxiv.org/abs/2007.03264}{{\tt arXiv:2007.03264}}.

\bibitem{Casali:2020vuy}
E.~Casali and A.~Puhm, {\it {A Double Copy for Celestial Amplitudes}},
  \href{http://arxiv.org/abs/2007.15027}{{\tt arXiv:2007.15027}}.

\bibitem{Adamo:2020qru}
T.~Adamo and A.~Ilderton, {\it {Classical and quantum double copy of
  back-reaction}},  \href{http://arxiv.org/abs/2005.05807}{{\tt
  arXiv:2005.05807}}.

\bibitem{Easson:2020esh}
D.~A. Easson, C.~Keeler, and T.~Manton, {\it {The classical double copy of
  regular non-singular black holes}},
  \href{http://arxiv.org/abs/2007.16186}{{\tt arXiv:2007.16186}}.

\bibitem{Chacon:2020fmr}
E.~Chac\'on, H.~Garc\'ia-Compe\'an, A.~Luna, R.~Monteiro, and C.~D. White, {\it
  {New heavenly double copies}},  \href{http://arxiv.org/abs/2008.09603}{{\tt
  arXiv:2008.09603}}.

\bibitem{Goldberger:2017frp}
W.~D. Goldberger, S.~G. Prabhu, and J.~O. Thompson, {\it {Classical gluon and
  graviton radiation from the bi-adjoint scalar double copy}},  {\em Phys.
  Rev.} {\bf D96} (2017), no.~6 065009,
  [\href{http://arxiv.org/abs/1705.09263}{{\tt arXiv:1705.09263}}].

\bibitem{Goldberger:2017ogt}
W.~D. Goldberger, J.~Li, and S.~G. Prabhu, {\it {Spinning particles, axion
  radiation, and the classical double copy}},  {\em Phys. Rev.} {\bf D97}
  (2018), no.~10 105018, [\href{http://arxiv.org/abs/1712.09250}{{\tt
  arXiv:1712.09250}}].

\bibitem{Chester:2017vcz}
D.~Chester, {\it {Radiative double copy for Einstein-Yang-Mills theory}},  {\em
  Phys. Rev.} {\bf D97} (2018), no.~8 084025,
  [\href{http://arxiv.org/abs/1712.08684}{{\tt arXiv:1712.08684}}].

\bibitem{Shen:2018ebu}
C.-H. Shen, {\it {Gravitational Radiation from Color-Kinematics Duality}},
  {\em JHEP} {\bf 11} (2018) 162, [\href{http://arxiv.org/abs/1806.07388}{{\tt
  arXiv:1806.07388}}].

\bibitem{Plefka:2018dpa}
J.~Plefka, J.~Steinhoff, and W.~Wormsbecher, {\it {Effective action of dilaton
  gravity as the classical double copy of Yang-Mills theory}},  {\em Phys.
  Rev.} {\bf D99} (2019), no.~2 024021,
  [\href{http://arxiv.org/abs/1807.09859}{{\tt arXiv:1807.09859}}].

\bibitem{Plefka:2019hmz}
J.~Plefka, C.~Shi, J.~Steinhoff, and T.~Wang, {\it {Breakdown of the classical
  double copy for the effective action of dilaton-gravity at NNLO}},  {\em
  Phys. Rev. D} {\bf 100} (2019), no.~8 086006,
  [\href{http://arxiv.org/abs/1906.05875}{{\tt arXiv:1906.05875}}].

\bibitem{PV:2019uuv}
A.~P.V. and A.~Manu, {\it {Classical double copy from Color Kinematics duality:
  A proof in the soft limit}},  {\em Phys. Rev. D} {\bf 101} (2020), no.~4
  046014, [\href{http://arxiv.org/abs/1907.10021}{{\tt arXiv:1907.10021}}].

\bibitem{Almeida:2020mrg}
G.~L. Almeida, S.~Foffa, and R.~Sturani, {\it {Classical Gravitational
  Self-Energy from Double Copy}},  \href{http://arxiv.org/abs/2008.06195}{{\tt
  arXiv:2008.06195}}.

\bibitem{Vaidya:2014kza}
V.~Vaidya, {\it {Gravitational spin Hamiltonians from the S matrix}},  {\em
  Phys. Rev.} {\bf D91} (2015), no.~2 024017,
  [\href{http://arxiv.org/abs/1410.5348}{{\tt arXiv:1410.5348}}].

\bibitem{Guevara:2017csg}
A.~Guevara, {\it {Holomorphic Classical Limit for Spin Effects in Gravitational
  and Electromagnetic Scattering}},  {\em JHEP} {\bf 04} (2019) 033,
  [\href{http://arxiv.org/abs/1706.02314}{{\tt arXiv:1706.02314}}].

\bibitem{Guevara:2018wpp}
A.~Guevara, A.~Ochirov, and J.~Vines, {\it {Scattering of Spinning Black Holes
  from Exponentiated Soft Factors}},  {\em JHEP} {\bf 09} (2019) 056,
  [\href{http://arxiv.org/abs/1812.06895}{{\tt arXiv:1812.06895}}].

\bibitem{Chung:2018kqs}
M.-Z. Chung, Y.-T. Huang, J.-W. Kim, and S.~Lee, {\it {The simplest massive
  S-matrix: from minimal coupling to Black Holes}},  {\em JHEP} {\bf 04} (2019)
  156, [\href{http://arxiv.org/abs/1812.08752}{{\tt arXiv:1812.08752}}].

\bibitem{Bautista:2019tdr}
Y.~F. Bautista and A.~Guevara, {\it {From Scattering Amplitudes to Classical
  Physics: Universality, Double Copy and Soft Theorems}},
  \href{http://arxiv.org/abs/1903.12419}{{\tt arXiv:1903.12419}}.

\bibitem{Maybee:2019jus}
B.~Maybee, D.~O'Connell, and J.~Vines, {\it {Observables and amplitudes for
  spinning particles and black holes}},  {\em JHEP} {\bf 12} (2019) 156,
  [\href{http://arxiv.org/abs/1906.09260}{{\tt arXiv:1906.09260}}].

\bibitem{Guevara:2019fsj}
A.~Guevara, A.~Ochirov, and J.~Vines, {\it {Black-hole scattering with general
  spin directions from minimal-coupling amplitudes}},  {\em Phys. Rev. D} {\bf
  100} (2019), no.~10 104024, [\href{http://arxiv.org/abs/1906.10071}{{\tt
  arXiv:1906.10071}}].

\bibitem{Arkani-Hamed:2019ymq}
N.~Arkani-Hamed, Y.-t. Huang, and D.~O'Connell, {\it {Kerr black holes as
  elementary particles}},  {\em JHEP} {\bf 01} (2020) 046,
  [\href{http://arxiv.org/abs/1906.10100}{{\tt arXiv:1906.10100}}].

\bibitem{Bautista:2019evw}
Y.~F. Bautista and A.~Guevara, {\it {On the Double Copy for Spinning Matter}},
  \href{http://arxiv.org/abs/1908.11349}{{\tt arXiv:1908.11349}}.

\bibitem{delaCruz:2016wbr}
L.~de~la Cruz, A.~Kniss, and S.~Weinzierl, {\it {Double Copies of Fermions as
  Matter that Interacts Only Gravitationally}},  {\em Phys. Rev. Lett.} {\bf
  116} (2016), no.~20 201601, [\href{http://arxiv.org/abs/1601.04523}{{\tt
  arXiv:1601.04523}}].

\bibitem{Chung:2019duq}
M.-Z. Chung, Y.-T. Huang, and J.-W. Kim, {\it {Classical potential for general
  spinning bodies}},  \href{http://arxiv.org/abs/1908.08463}{{\tt
  arXiv:1908.08463}}.

\bibitem{Chung:2019yfs}
M.-Z. Chung, Y.-T. Huang, and J.-W. Kim, {\it {Kerr-Newman stress-tensor from
  minimal coupling to all orders in spin}},
  \href{http://arxiv.org/abs/1911.12775}{{\tt arXiv:1911.12775}}.

\bibitem{Chung:2020rrz}
M.-Z. Chung, Y.-t. Huang, J.-W. Kim, and S.~Lee, {\it {Complete Hamiltonian for
  spinning binary systems at first post-Minkowskian order}},  {\em JHEP} {\bf
  05} (2020) 105, [\href{http://arxiv.org/abs/2003.06600}{{\tt
  arXiv:2003.06600}}].

\bibitem{Levi:2020uwu}
M.~Levi, A.~J. Mcleod, and M.~Von~Hippel, {\it {NNNLO gravitational
  quadratic-in-spin interactions at the quartic order in G}},
  \href{http://arxiv.org/abs/2003.07890}{{\tt arXiv:2003.07890}}.

\bibitem{Aoude:2020mlg}
R.~Aoude, M.-Z. Chung, Y.-t. Huang, C.~S. Machado, and M.-K. Tam, {\it {The
  silence of binary Kerr}},  \href{http://arxiv.org/abs/2007.09486}{{\tt
  arXiv:2007.09486}}.

\bibitem{Wong:1970fu}
S.~Wong, {\it {Field and particle equations for the classical Yang-Mills field
  and particles with isotopic spin}},  {\em Nuovo Cim. A} {\bf 65} (1970)
  689--694.

\bibitem{Strominger:2013lka}
A.~Strominger, {\it {Asymptotic Symmetries of Yang-Mills Theory}},  {\em JHEP}
  {\bf 07} (2014) 151, [\href{http://arxiv.org/abs/1308.0589}{{\tt
  arXiv:1308.0589}}].

\bibitem{Pate:2017vwa}
M.~Pate, A.-M. Raclariu, and A.~Strominger, {\it {Color Memory: A Yang-Mills
  Analog of Gravitational Wave Memory}},  {\em Phys. Rev. Lett.} {\bf 119}
  (2017), no.~26 261602, [\href{http://arxiv.org/abs/1707.08016}{{\tt
  arXiv:1707.08016}}].

\bibitem{Strominger:2017zoo}
A.~Strominger, {\it {Lectures on the Infrared Structure of Gravity and Gauge
  Theory}},  \href{http://arxiv.org/abs/1703.05448}{{\tt arXiv:1703.05448}}.

\bibitem{Campoleoni:2017qot}
A.~Campoleoni, D.~Francia, and C.~Heissenberg, {\it {Asymptotic Charges at Null
  Infinity in Any Dimension}},  {\em Universe} {\bf 4} (2018), no.~3 47,
  [\href{http://arxiv.org/abs/1712.09591}{{\tt arXiv:1712.09591}}].

\bibitem{He:2019pll}
T.~He and P.~Mitra, {\it {Asymptotic symmetries in (d + 2)-dimensional gauge
  theories}},  {\em JHEP} {\bf 10} (2019) 277,
  [\href{http://arxiv.org/abs/1903.03607}{{\tt arXiv:1903.03607}}].

\bibitem{Gonzo:2019fai}
R.~Gonzo, T.~McLoughlin, D.~Medrano, and A.~Spiering, {\it {Asymptotic Charges
  and Coherent States in QCD}},  \href{http://arxiv.org/abs/1906.11763}{{\tt
  arXiv:1906.11763}}.

\bibitem{Campoleoni:2019ptc}
A.~Campoleoni, D.~Francia, and C.~Heissenberg, {\it {Electromagnetic and color
  memory in even dimensions}},  {\em Phys. Rev. D} {\bf 100} (2019), no.~8
  085015, [\href{http://arxiv.org/abs/1907.05187}{{\tt arXiv:1907.05187}}].

\bibitem{Heinz:1983nx}
U.~W. Heinz, {\it {Kinetic Theory for Nonabelian Plasmas}},  {\em Phys. Rev.
  Lett.} {\bf 51} (1983) 351.

\bibitem{Heinz:1984my}
U.~W. Heinz, {\it {A Relativistic Colored Spinning Particle in an External
  Color Field}},  {\em Phys. Lett. B} {\bf 144} (1984) 228--230.

\bibitem{Heinz:1984yq}
U.~W. Heinz, {\it {Quark - Gluon Transport Theory. Part 1. the Classical
  Theory}},  {\em Annals Phys.} {\bf 161} (1985) 48.

\bibitem{Heinz:1985qe}
U.~W. Heinz, {\it {Quark - Gluon Transport Theory. Part 2. Color Response and
  Color Correlations in a Quark - Gluon Plasma}},  {\em Annals Phys.} {\bf 168}
  (1986) 148.

\bibitem{Elze:1989un}
H.-T. Elze and U.~W. Heinz, {\it {Quark - Gluon Transport Theory}},  {\em Phys.
  Rept.} {\bf 183} (1989) 81--135.

\bibitem{Litim:2001db}
D.~F. Litim and C.~Manuel, {\it {Semi-classical transport theory for
  non-Abelian plasmas}},  {\em Phys. Rept.} {\bf 364} (2002) 451--539,
  [\href{http://arxiv.org/abs/hep-ph/0110104}{{\tt hep-ph/0110104}}].

\bibitem{McLerran:1993ni}
L.~D. McLerran and R.~Venugopalan, {\it {Computing quark and gluon distribution
  functions for very large nuclei}},  {\em Phys. Rev. D} {\bf 49} (1994)
  2233--2241, [\href{http://arxiv.org/abs/hep-ph/9309289}{{\tt
  hep-ph/9309289}}].

\bibitem{McLerran:1993ka}
L.~D. McLerran and R.~Venugopalan, {\it {Gluon distribution functions for very
  large nuclei at small transverse momentum}},  {\em Phys. Rev. D} {\bf 49}
  (1994) 3352--3355, [\href{http://arxiv.org/abs/hep-ph/9311205}{{\tt
  hep-ph/9311205}}].

\bibitem{Iancu:2003xm}
E.~Iancu and R.~Venugopalan, {\it {The Color glass condensate and high-energy
  scattering in QCD}},  in {\em {Quark-Gluon plasma 4}}, pp.~249--3363.
\newblock World Scientific, 2003.
\newblock \href{http://arxiv.org/abs/hep-ph/0303204}{{\tt hep-ph/0303204}}.

\bibitem{Kajantie:2019hft}
K.~Kajantie, L.~D. McLerran, and R.~Paatelainen, {\it {Gluon Radiation from a
  Classical Point Particle}},  {\em Phys. Rev. D} {\bf 100} (2019), no.~5
  054011, [\href{http://arxiv.org/abs/1903.01381}{{\tt arXiv:1903.01381}}].

\bibitem{Kajantie:2019nse}
K.~Kajantie, L.~D. McLerran, and R.~Paatelainen, {\it {Gluon Radiation from a
  classical point particle II: dense gluon fields}},  {\em Phys. Rev. D} {\bf
  101} (2020), no.~5 054012, [\href{http://arxiv.org/abs/1911.12738}{{\tt
  arXiv:1911.12738}}].

\bibitem{Yaffe:1981vf}
L.~G. Yaffe, {\it {Large N Limits as Classical Mechanics}},  {\em Rev. Mod.
  Phys.} {\bf 54} (1982) 407.

\bibitem{Kaiser:1977ys}
G.~Kaiser, {\it {Phase Space Approach to Relativistic Quantum Mechanics. 1.
  Coherent State Representation for Massive Scalar Particles}},  {\em J. Math.
  Phys.} {\bf 18} (1977) 952--959.

\bibitem{TwarequeAli:1988tvp}
S.~Twareque~Ali and J.~Antoine, {\it {Coherent states of the 1+1 dimensional
  Poincare group: Square integrability and a relativistic Weyl transform}},
  {\em Ann. Inst. H. Poincare Phys. Theor.} {\bf 51} (1989) 23--44.

\bibitem{Kowalski:2018xsw}
K.~Kowalski, J.~Rembieli\'nski, and J.-P. Gazeau, {\it {On the coherent states
  for a relativistic scalar particle}},  {\em Annals Phys.} {\bf 399} (2018)
  204--223, [\href{http://arxiv.org/abs/1903.07312}{{\tt arXiv:1903.07312}}].

\bibitem{perelomov:1972}
A.~M. Perelomov, {\it Coherent states for arbitrary {L}ie group},  {\em Commun.
  Math. Phys.} {\bf 26} (1972) 222--236,
  [\href{http://arxiv.org/abs/math-ph/0203002}{{\tt math-ph/0203002}}].

\bibitem{Perelomov:1986tf}
A.~Perelomov, {\em {Generalized coherent states and their applications}}.
\newblock Springer, 1986.

\bibitem{Combescure2012}
M.~Combescure and D.~Robert, {\em Coherent States and Applications in
  Mathematical Physics}.
\newblock Springer, 2012.

\bibitem{Sakurai:2011zz}
J.~J. Sakurai and J.~Napolitano, {\em {Modern Quantum Mechanics}}.
\newblock Cambridge University Press, Cambridge, 2017.

\bibitem{Mathur:2000sv}
M.~Mathur and D.~Sen, {\it {Coherent states for SU(3)}},  {\em J. Math. Phys.}
  {\bf 42} (2001) 4181--4196,
  [\href{http://arxiv.org/abs/quant-ph/0012099}{{\tt quant-ph/0012099}}].

\bibitem{Mathur:2002mx}
M.~Mathur and H.~Mani, {\it {SU(N) coherent states}},  {\em J. Math. Phys.}
  {\bf 43} (2002) 5351, [\href{http://arxiv.org/abs/quant-ph/0206005}{{\tt
  quant-ph/0206005}}].

\bibitem{Mathur:2010wc}
M.~Mathur, I.~Raychowdhury, and R.~Anishetty, {\it {SU(N) Irreducible Schwinger
  Bosons}},  {\em J. Math. Phys.} {\bf 51} (2010) 093504,
  [\href{http://arxiv.org/abs/1003.5487}{{\tt arXiv:1003.5487}}].

\bibitem{Mathur:2010ey}
M.~Mathur and I.~Raychowdhury, {\it {SU(N) Coherent States and Irreducible
  Schwinger Bosons}},  {\em J. Phys. A} {\bf 44} (2011) 035203,
  [\href{http://arxiv.org/abs/1007.1510}{{\tt arXiv:1007.1510}}].

\bibitem{Ochirov:2019mtf}
A.~Ochirov and B.~Page, {\it {Multi-Quark Colour Decompositions from
  Unitarity}},  {\em JHEP} {\bf 10} (2019) 058,
  [\href{http://arxiv.org/abs/1908.02695}{{\tt arXiv:1908.02695}}].

\bibitem{Luscher:1978ir}
M.~L{\"u}scher, {\it {Asymptotic behaviour of classical Yang-Mills fields in
  Minkowski space}},  {\em Nucl. Phys. B} {\bf 140} (1978) 429--448.

\bibitem{Adamo:2015fwa}
T.~Adamo and E.~Casali, {\it {Perturbative gauge theory at null infinity}},
  {\em Phys. Rev. D} {\bf 91} (2015), no.~12 125022,
  [\href{http://arxiv.org/abs/1504.02304}{{\tt arXiv:1504.02304}}].

\bibitem{Alexanian:2001qj}
G.~Alexanian, A.~Balachandran, G.~Immirzi, and B.~Ydri, {\it {Fuzzy CP**2}},
  {\em J. Geom. Phys.} {\bf 42} (2002) 28--53,
  [\href{http://arxiv.org/abs/hep-th/0103023}{{\tt hep-th/0103023}}].

\bibitem{Grosse:2004wm}
H.~Grosse and H.~Steinacker, {\it {Finite gauge theory on fuzzy CP**2}},  {\em
  Nucl. Phys. B} {\bf 707} (2005) 145--198,
  [\href{http://arxiv.org/abs/hep-th/0407089}{{\tt hep-th/0407089}}].

\bibitem{Ellis:2016jkw}
J.~Ellis, {\it {TikZ-Feynman: Feynman diagrams with TikZ}},  {\em Comput. Phys.
  Commun.} {\bf 210} (2017) 103--123,
  [\href{http://arxiv.org/abs/1601.05437}{{\tt arXiv:1601.05437}}].

\end{thebibliography}

\providecommand{\href}[2]{#2}\begingroup\raggedright\endgroup

\end{document}